%% file: main.tex
\documentclass[11pt]{article}
\PassOptionsToPackage{final}{showlabels}
\input{template}

\usepackage{placeins}
\usepackage{needspace}

\newcommand{\ub}{\mathbf{u}}
\renewcommand{\sb}{\mathbf{s}}
\newcommand{\Ks}{\mathsf{K}}
\newcommand{\M}{\mathcal{M}}
\newcommand{\Ws}{\mathsf{W}}
\newcommand{\ws}{\mathsf{w}}
\newcommand{\Ps}{\mathsf{P}}
\newcommand{\Rs}{\mathsf{R}}

\newcommand{\rs}{\mathsf{r}}

\renewcommand{\qb}{\mathbf{q}}

\newcommand{\op}{\mathrm{op}}
\newcommand{\SNR}{\mathrm{SNR}}
\newcommand{\Col}{\mathrm{Col}}

\renewcommand{\Kb}{\mathbf{K}}
\renewcommand{\Ib}{\mathbf{I}}
\renewcommand{\P}{\mathbb{P}}
\newcommand{\iid}{\stackrel{\mathrm{i.i.d}}{\sim}}
\numberwithin{figure}{section}

\date{}
\title{Adaptive Sparsity-Aware Projection-Assisted Spectral Embedding for High-Dimensional Noisy Data with Nonlinear Structures}
\author[1]{Yichen Hu  \thanks{E-mail: ethhu@ucdavis.edu. The author is partially supported by NSF DMS-2515104. } }
\author[1]{Xiucai Ding \thanks{E-mail: xcading@ucdavis.edu. The author is partially supported by NSF DMS-2515104.} }
\author[2]{Jian Kang \thanks{E-mail: jiankang@umich.edu. The author is partially supported by NSF IIS2123777.}}
\affil[1]{Department of Statistics and Data Science, University of California, Davis}
\affil[2]{Department of Biostatistics, University of Michigan, Ann Arbor}

\begin{document}
\maketitle
\begin{abstract}
In this paper, we propose SPASE, an adaptive, sparsity-aware, projection-assisted spectral embedding method for feature extraction from high-dimensional noisy data with underlying nonlinear structures and latent sparsity patterns. We first construct a data-driven projection that adapts to the hidden sparsity patterns and the signal-to-noise ratio, reducing noise-induced distortion of pairwise distances. We then apply kernel-based spectral embedding to the projected data to explore the underlying geometric structure. The key methodological contribution is an energy-based projection selection algorithm that is adaptive to the signal-to-noise ratio and sparsity of the signal representation, while mitigating the impact of noise. On the theoretical side, we show that the resulting embeddings converge to the eigenfunctions of an integral operator associated with the latent signal, and the method can remain effective when the signal-to-noise ratio is low. We demonstrate the empirical performance of the proposed method through simulation studies and real-data applications, showing improved accuracy in downstream tasks such as classification and clustering.

\end{abstract}

\section{Introduction}
Feature extraction is a fundamental problem in modern data science, particularly in the analysis of high-dimensional data. Classical linear methods such as principal component analysis (PCA)~\cite{hotelling1933analysis} are natural starting points when the dominant variability of the data can be well represented by a low-dimensional linear subspace. However, many modern datasets exhibit intrinsically nonlinear geometric structure that cannot be adequately captured by linear representations alone. To address this challenge, a large class of manifold learning and nonlinear spectral embedding methods has been developed, including kernel PCA \cite{scholkopf1998kernel}, Isomap \cite{tenenbaum2000global}, locally linear embedding \cite{roweis2000lle}, Laplacian eigenmaps \cite{belkin2003laplacian}, diffusion maps \cite{coifman2006diffusion}, vector diffusion maps \cite{singer2012vdm}, t-SNE \cite{van2008visualizing}, and UMAP \cite{mcinnes2018umap}. More recently, neural-network-based approaches such as autoencoders \cite{hinton2006reducing}, denoising autoencoders \cite{vincent2008extracting}, and variational autoencoders \cite{kingma2014autoencoding} have provided additional flexible frameworks for nonlinear representation learning. For downstream tasks such as classification, clustering, and visualization, informative low-dimensional representations are often essential for revealing meaningful structure and enabling reliable statistical analysis.

Existing manifold learning and spectral embedding methods are typically most
effective when the observed geometry still reliably reflects the underlying signal structure, such as in clean or signal-dominated settings. In many modern high-dimensional applications, however, the ambient noise can substantially distort pairwise distances, affinities, and neighborhood relations before the final embedding step is even performed. As a result, the observed geometry itself may no longer provide a reliable representation of the latent nonlinear structure. For example, in biomedical imaging such as cryo-electron microscopy, low-dose acquisition often leads to extremely low signal-to-noise ratios \cite{baxter2009snrcryoem}. At the same time, signals in such applications often possess additional low-complexity structure, such as sparse or compressible representations \cite{olshausen1996emergence}. This additional structure creates an opportunity to partially recover or stabilize the underlying geometry before applying nonlinear spectral embedding methods.

Motivated by these observations, in this paper, we study feature extraction under the
signal-plus-noise model
\begin{equation}
\xb_i = \sb_i + \zb_i \in \R^p, \qquad 1 \leq i \leq n,
\label{eq:spn}
\end{equation}
where $\sb_i$ denotes the latent signal component and $\zb_i$ denotes the
additive noise component. We focus on the setting where the latent signal has
nonlinear geometric structure and admits a sparse or compressible
representation. We propose an adaptive sparsity-aware projection-assisted spectral embedding method for this problem. The main idea is to first perform an adaptive projection step to denoise the pairwise distances and then apply spectral embedding to the projected data. We first summarize some relevant related works in Section \ref{sec:literature} and then provide an overview of our method and contribution in Section \ref{sec:contrib}.

\subsection{Some Related Works}
\label{sec:literature}

The behavior of manifold learning and spectral embedding methods has been extensively studied in fixed-dimensional settings. In the noiseless case, a substantial theoretical literature establishes the convergence of graph- and kernel-based constructions to their population or continuum counterparts; see, for example, \cite{belkin2006convergence,hein2007graph,von2008consistency,ting2010analysis,garcia2020error,trillos2021geometric,dunson2021spectral,cheng2022eigen,li2025central}. Related analyses have also been developed for specific methods and models, including continuum Isomap \cite{bernstein2007continuumisomap}, locally linear embedding \cite{wu2018think}, and vector diffusion maps and connection Laplacians \cite{singer2012vdm,singer2017connection}. In parallel, a number of robust variants have been proposed in fixed-dimensional signal-plus-noise settings, including perturbation analysis of graph-Laplacian eigenvectors for denoising \cite{meyer2014perturbation}, robust locally linear embedding \cite{chang2006robustlle}, and robust kernel Isomap \cite{kwok2007robustkernelisomap}. Collectively, these works show that moderate noise can often be accommodated when the observed geometry still reflects the latent signal structure with reasonable fidelity.

The situation becomes substantially more challenging in high-dimensional noisy settings, where the main difficulty is not only the perturbation of the final embedding step, but also the distortion of the geometry itself. High-dimensional noise may induce concentration effects and geometric artifacts, so that distances, affinities, and neighborhood relations constructed from the observations become much less informative \cite{aggarwal2001surprising,elkaroui2010spectrum,el2010information,bordenave2013euclidean}. Nevertheless, when the signal dominates the noise, recent works have shown that direct graph- and kernel-based spectral methods can still recover meaningful low-dimensional structure under suitable conditions. These include kernel spectral embedding with adaptive bandwidth selection \cite{ding2023learning}, robustness analysis for connection graph Laplacians under additive noise \cite{elkaroui2016graph}, graph-Laplacian analysis under high-dimensional noise \cite{ding2022impact}, landmark-based spectral embedding \cite{shen2022roseland}, and multiview sensor-fusion methods \cite{ding2024kernel,ding2026generalized}. Taken together, these results provide theoretical justification for nonlinear spectral embedding in certain high-dimensional signal-plus-noise regimes, provided that the underlying geometry remains sufficiently informative and
that the bandwidth is properly chosen.

Much less is understood in low-SNR regimes, where the noise dominates the signal and the observed geometry may no longer provide a reliable basis for direct graph or kernel construction. In this setting, additional low-complexity structure becomes particularly important. For linear methods, sparsity has been widely used to identify informative directions in high-dimensional noisy data through sparse PCA and related approaches. Early sparse PCA methods include penalized regression, regularized low-rank approximation, and penalized matrix decomposition \cite{zou2006sparsepca,shen2008sparsepca,witten2009penalized}. Related transform-domain reduction and feature-selection ideas before downstream low-dimensional analysis appear in \cite{wickerhauser1994large,feng2000face,nadler2005prediction}, while selection-based methods for sparse or compressible signals in prescribed bases are studied in \cite{johnstone2009consistency}. Subsequent developments include augmented sparse PCA \cite{paul2007augmented}, two-stage coordinate-selection procedures \cite{birnbaum2013minimax}, and adaptive sparse principal subspace estimation \cite{cai2013optimal}. However, these methods are primarily designed for linear covariance structures, and much less is known about how sparsity can be used to stabilize nonlinear geometric structure for manifold learning and spectral embedding in high-dimensional low-SNR settings.

To address this gap, we propose an adaptive sparsity-aware projection-assisted spectral embedding method for feature extraction from high-dimensional noisy data with nonlinear structures under sparse or compressible representations. The key idea is to use adaptive projection to denoise the pairwise geometry before applying nonlinear spectral embedding. Our approach applies across a broad range of SNR levels, while being particularly useful when the noise dominates the signal. An overview of our method and contributions is given in Section \ref{sec:contrib}.

\subsection{An Overview of Main Results and Contributions}
\label{sec:contrib}
\begin{figure}[t]
\centering
\includegraphics[width=\textwidth]{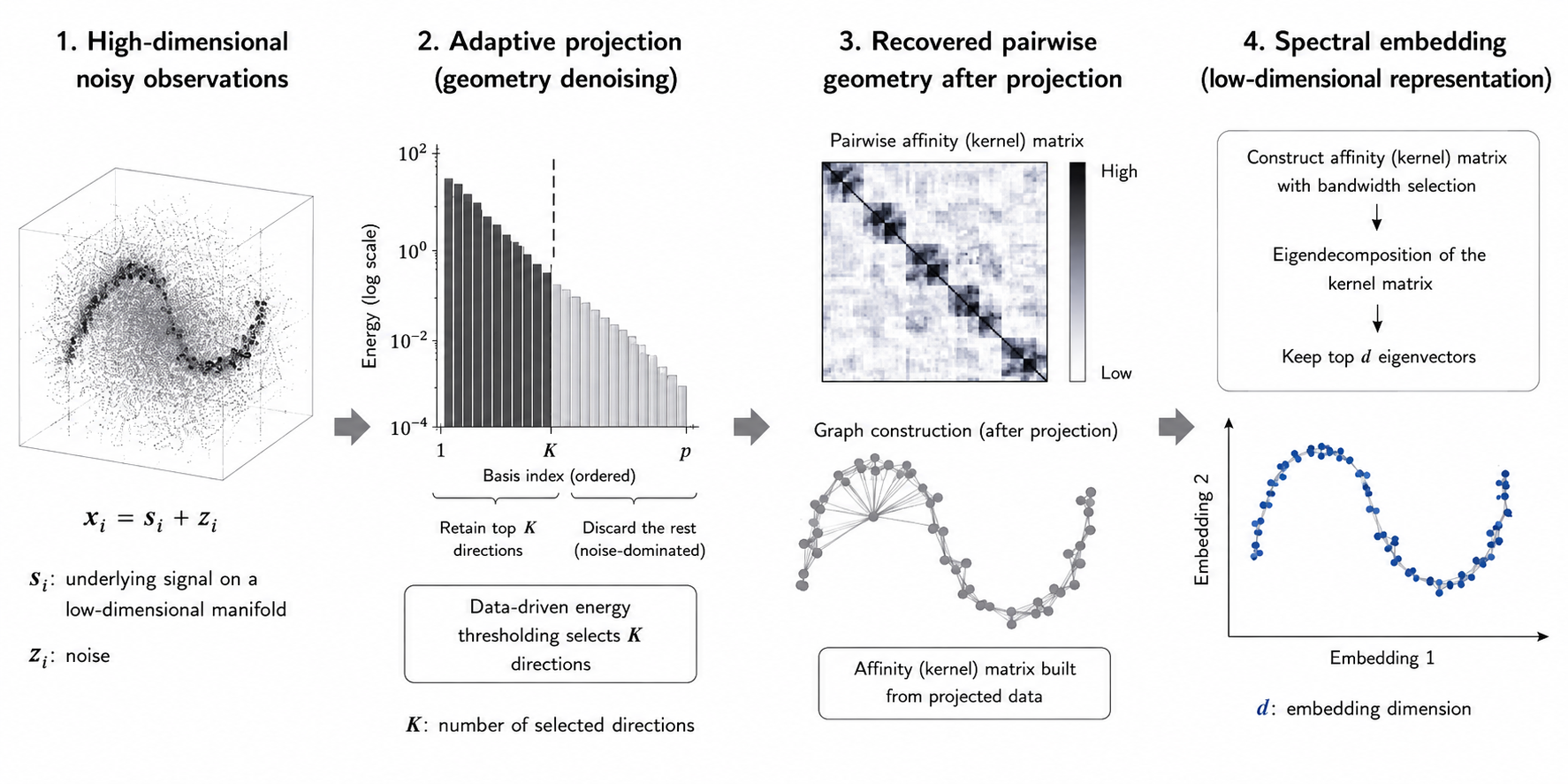}
\caption{
Overview of the proposed adaptive sparsity-aware projection-assisted spectral embedding framework. The method first performs adaptive projection to denoise pairwise geometry in high-dimensional noisy observations and then applies spectral embedding to the projected data to recover the latent nonlinear structure.
}
\label{fig:overview}
\end{figure}
Figure~\ref{fig:overview} illustrates the overall workflow of the proposed framework. As in \cite{scholkopf1998kernel,belkin2003laplacian,coifman2006diffusion,singer2012vdm,ding2023learning}, our algorithm relies on pairwise distances to construct an affinity matrix for spectral embedding. In high-dimensional noisy settings, however, the main challenge is that these pairwise distances may already become unreliable before the final eigendecomposition is performed. In particular, high-dimensional asymptotics reveal rigid geometric effects in which observations tend to become nearly equidistant or nearly orthogonal \cite{hall2005geometric,ahn2007mild}, while nearest-neighbor structure may become increasingly unreliable through hubness and related distortions \cite{radovanovic2010hubs,angiulli2018behavior}. Moreover, high-dimensional noise can substantially alter the effective kernel geometry extracted from the observations \cite{el2010information}. Consequently, the main issue is not only whether the spectral embedding step itself is stable, but also whether the input geometry still faithfully reflects the latent signal structure. The goal of our method is thus to denoise the pairwise geometry before applying spectral embedding.

To address this difficulty, we exploit additional sparsity structure in the latent signal. When the signal admits a sparse representation in a prescribed basis, most of its energy is concentrated on a relatively small number of directions, while the ambient noise is more diffusely spread across coordinates. This separation between concentrated signal energy and diffuse noise motivates an adaptive geometry-denoising step prior to manifold learning. Specifically, we perform a preliminary data-adaptive projection step in which empirically energetic directions are retained and noise-dominated ones are discarded, similar in spirit to earlier transform-domain reduction and feature-screening ideas; see, for example, \cite{wickerhauser1994large,feng2000face,nadler2005prediction,johnstone2009consistency}. The resulting projection stabilizes the pairwise geometry before subsequent spectral analysis.

Based on this idea, we propose an adaptive sparsity-aware projection-assisted spectral embedding method in Section~\ref{sec_methodology}. The overall procedure consists of two main steps. First, we construct a data-driven projection by selecting directions in a prescribed basis so as to reduce the effect of noise before pairwise distances are computed. The key methodological contribution lies in this step: we introduce an empirical-energy-based thresholding rule (Algorithm~\ref{alg:Kselect}) that is fully data-driven and, as shown in Proposition~\ref{prop:K}, adapts to the unknown underlying SNR and sparsity structure of the data. Intuitively, the procedure retains directions whose empirical energies remain distinguishable from the ambient noise level, while filtering out directions dominated by noise. As shown in Theorem~\ref{thm:pairwise}, the projection denoises the pairwise geometry of the data and makes subsequent spectral analysis more reliable.

Second, after this denoising projection step, we apply a spectral embedding method with bandwidth selection in the spirit of \cite{ding2023learning} to the projected data. This produces a unified project-then-embed framework that combines adaptive geometry recovery with nonlinear spectral embedding. We show in Theorem~\ref{thm:total_convergence} that, after normalization, the resulting embeddings converge to the eigenfunctions of an integral operator associated with the clean latent signal. Although we adopt \cite{ding2023learning} here for concreteness and tractability, the second step can in principle be replaced by other spectral embedding methods, for example, diffusion-based procedures such as Laplacian eigenmaps \cite{belkin2003laplacian}, diffusion maps \cite{coifman2006diffusion}, and vector diffusion maps \cite{singer2012vdm}; we leave these extensions for future work.

Our theoretical analysis in Section~\ref{sec:main_results} shows that the proposed framework applies across a broad range of SNR levels and remains effective even in low-SNR regimes where the noise dominates the signal and direct geometric construction may fail. In addition to the theoretical analysis, Sections~\ref{sec:simulation_analysis} and \ref{sec:real_data} compare the empirical performance of the proposed method with \cite{roweis2000lle,balasubramanian2002isomap,coifman2006diffusion,hinton2006reducing,vincent2008extracting,ding2023learning} through simulation studies and real data examples. The results demonstrate that the proposed geometry-denoising framework substantially improves the robustness of nonlinear feature extraction in high-dimensional noisy settings, particularly in low-SNR regimes.

\subsection{Organization and Notations}
The rest of the paper is organized as follows. In Section~\ref{sec_methodology}, we introduce the proposed method. In Section~\ref{sec:main_results}, we state the main assumptions and establish the theoretical guarantees. Sections~\ref{sec:simulation_analysis} and \ref{sec:real_data} present simulation studies and real data analyses. Additional simulation and real data results are provided in Appendices~\ref{app:additional_simulation} and \ref{app:additional_real_data}, respectively. Supplementary algorithms are given in Appendix~\ref{sec:supp_algo}, and technical proofs are deferred to Appendix~\ref{sec:proofs}.

\noindent{\bf Conventions.}
We write $C$, $C_0$, etc.\ for generic large positive constants whose values may change from line to line, and $\epsilon$, $\delta$, etc.\ for generic small positive constants. We use $\floor{a}$ for the greatest integer less than or equal to $a$. When a constant depends on a parameter $a$, we indicate this by $C(a)$ or $C_a$. For sequences $\{a_n\}$ and $\{b_n\}$ depending on $n$, $a_n=\OO(b_n)$ means $|a_n|\le C|b_n|$ for some constant $C>0$, while $a_n=\oo(b_n)$ means $|a_n|\le c_n|b_n|$ for some positive sequence $c_n\downarrow 0$ as $n\to\infty$. We also use $a_n\lesssim b_n$ to mean $a_n=\OO(b_n)$, $a_n\gtrsim b_n$ to mean $b_n=\OO(a_n)$, and $a_n\asymp b_n$ to mean both $a_n=\OO(b_n)$ and $b_n=\OO(a_n)$. The probabilistic analogues are $a_n=\OOp(b_n)$ and $a_n=\oop(b_n)$.

Bold lowercase letters such as $\xb,\yb$ denote vectors, while bold uppercase letters such as $\mathbf{K}$ denote matrices. Calligraphic letters such as $\mathcal{K}$ and $\mathcal{L}$ denote operators or function spaces, and plain letters such as $f,g$ denote generic functions. For norms, $\norm{\xb}_2=\big(\sum_{j=1}^p |x_j|^2\big)^{1/2}$ denotes the $\ell_2$-norm of $\xb=(x_1,\cdots,x_p)^\top\in\R^p$. We denote by $\norm{\mathcal{L}}$ and $\norm{\mathbf{A}}$ the operator norms of an operator $\mathcal{L}$ and a matrix $\mathbf{A}$, respectively. For a matrix $\mathbf A$, let $\Pi_{\mathbf A}$ denote the orthogonal projection onto $\Col(\mathbf A)$.

\section{Methodology}
\label{sec_methodology}

In this section, we introduce the proposed methodology. The overall framework consists of two main stages: an adaptive projection step for denoising pairwise geometry, followed by nonlinear spectral embedding on the projected data. We first construct an adaptive sparse projection to denoise the pairwise geometry of the observed high-dimensional data in Section~\ref{sec:projection}. We then apply a spectral embedding method to the projected samples in Section~\ref{sec:embedding}. Finally, in Section~\ref{sec:downstream}, we describe how the resulting embeddings can be used for downstream tasks such as clustering and prediction. For notational convenience, throughout the paper we adopt the convention that for a positive integer $n$, $[n]=\{1,2,\ldots,n\}$.

\subsection{Adaptive Sparse Projection}
\label{sec:projection}

To denoise the pairwise geometry, we first construct a data-driven projection from the observed samples $\{\xb_i\}_{i\in[n]}\subset\R^p$, which is the key step in our methodology. Let $\Ws=\{\ws_j\}_{j\in[p]}$ be a prescribed orthonormal basis of $\R^p$, for example, the Haar wavelet basis \cite{daubechies1992ten}. In this paper, we may take $\Ws$ to be a wavelet basis, since many structured signals have local features and are therefore represented by a relatively small number of large coefficients in wavelet bases \cite{mallat1989theory,donoho1994ideal,donoho1995adapting}. For each $1\le i\le n$ and $1\le j\le p$, define
\begin{align}
    y_{ij}=(\xb_i-\bar\xb)^\top \ws_j,\label{eq:y}
\end{align}
where $\bar\xb=n^{-1}\sum_i\xb_i$, and compute the empirical energy of the $j$-th direction by
\begin{align}
    E_j=\frac{1}{n}\sum_{i=1}^n y_{ij}^2.
    \label{eq:Ej}
\end{align}
We then order $\{E_j\}_{j\in[p]}$ as $E_{(1)}\ge E_{(2)}\ge\cdots\ge E_{(p)}.$ The basic idea is to retain the basis directions corresponding to the $\Ks$ largest empirical energies.

A central issue in this step is the choice of the retained dimension $\Ks$. If $\Ks$ is too large, we retain too many noise-dominated directions; if $\Ks$ is too small, we may discard directions that still contain useful signal information. To address this issue, we choose $\Ks$ in a data-driven way based on the ordered empirical energies. More specifically, we use $\hat\sigma^2$, computed as the average of the smallest $s$ empirical energies, as an estimate of the noise variance, and choose $\Ks$ as the largest $j$ such that $E_{(j)}\ge (1+\tau)\hat\sigma^2$, where $\tau$ is a prespecified small positive constant. The role of $s$ is to ensure that $\hat\sigma^2$ provides a reliable estimate of the noise level. In practice, $s$ can be chosen as a small fixed integer. As long as $s$ is fixed, it does not affect the theoretical order of $\Ks$ under the conditions of Proposition~\ref{prop:K}. The parameter $\tau$ controls how conservative the cutoff is: a larger $\tau$ leads to a smaller $\Ks$, while a smaller $\tau$ leads to a larger $\Ks$. However, changing $\tau$ mainly affects directions whose signal energies are comparable to the noise level, which typically carry only a small fraction of the signal information. By contrast, clearly signal-dominated directions remain selected and clearly noise-dominated directions remain excluded. Theoretically, each fixed $\tau>0$ affects the order of $\Ks$ only through multiplicative constants under the conditions of Proposition~\ref{prop:K}. In this paper, we fix $\tau=0.5$ in our experiments. The selection of $\Ks$ is summarized in Algorithm~\ref{alg:Kselect}.

\begin{algorithm}[!ht]
    \caption{\bf Adaptive selection of the projection dimension $\Ks$}
    \label{alg:Kselect}
    \normalsize
    \begin{flushleft}
        \vspace{0pt}
        \noindent{\bf Inputs:} Ordered empirical energies $\{E_{(j)}\}_{j=1}^p$, a small positive fixed number $\tau$, and a small fixed integer $s$.

        \noindent{\bf Step one:}
        Compute the estimated noise variance
        $$
        \widehat\sigma^2=\frac{1}{s}\sum_{j=1}^{s}E_{(p+1-j)}.
        $$

        \noindent{\bf Step two:}
        Compute
        \begin{align}
            \Ks=\max\left\{j:\,E_{(j)} \ge (1+\tau)\hat\sigma^2\right\}.
            \label{eq:Ks}
        \end{align}

        \noindent{\bf Output:} The selected projection dimension $\Ks$.
    \end{flushleft}
\end{algorithm}
\begin{remark}
    The selected projection dimension $\Ks$ is controlled jointly by the relative strength between the signal and the noise and by how sparse the signal is in the chosen basis. When the relative strength between the signal and the noise is larger, more basis directions have empirical energies that remain clearly above the noise level, and the selected dimension becomes larger. By contrast, when relative strength between the signal and the noise is smaller, fewer directions can be reliably distinguished and the selected dimension becomes smaller. Likewise, when the signal is more sparse in the basis, its energy is concentrated on fewer directions, which creates a sharper separation between informative and uninformative directions and makes the choice of $\Ks$ smaller. We will illustrate these effects on $\Ks$ in Figure~\ref{fig:sim_k_vs_q} in Section~\ref{sec:simulation_analysis}.
\end{remark}

Once $\Ks$ is selected, we retain the $\Ks$ basis directions corresponding to the $\Ks$ largest empirical energies, and denote the set of their indices by $\mathcal{I}_{\Ks}$. We construct the projection matrix as
\begin{align}
    \Ps=(\ws_j)_{j\in\mathcal{I}_{\Ks}}\in\R^{p\times \Ks}.\label{eq:P}
\end{align}
This projection step produces the lower-dimensional representation that will be used in the subsequent spectral embedding procedure.

\subsection{Spectral Embedding after Projection}
\label{sec:embedding}
Once the projection matrix $\Ps$ in \eqref{eq:P} is constructed, we project the original observations into the selected lower-dimensional subspace and then apply a kernel spectral embedding procedure, in the spirit of \cite{ding2023learning}, to the projected data. More specifically, for each $i=1,\dots,n$, define
\begin{align}
    \yb_i=\Ps^\top(\xb_i-\bar\xb)\in\R^\Ks.
    \label{eq:projected_y}
\end{align}
Instead of constructing affinities directly from the original high-dimensional observations, we construct them from the projected samples $\{\yb_i\}_{i=1}^n$, whose pairwise geometry is expected to be more robust after the adaptive projection step.

The embedding procedure is then applied to $\yb_i$ in \eqref{eq:projected_y}. Given a bandwidth parameter $h_n$, we construct the kernel matrix $\Kb_n = (K(i,j))$ as
\begin{align}
K(i,j) = \exp\left(-\frac{\norm{\yb_i - \yb_j}_2^2}{h_n}\right),
\quad 1 \le i,j \le n, \label{eq:Kn}
\end{align}
and compute its eigendecomposition
\begin{align}
\frac{1}{n}\Kb_n = \mathbf U \mathbf\Lambda \mathbf U^\top, \label{eq:eigendecomposition}
\end{align}
where $\mathbf\Lambda = \mathrm{diag}(\lambda_1, \dots, \lambda_n)$ with $\lambda_1 \ge \cdots \ge \lambda_n$, and $\mathbf U = (\ub_1, \dots, \ub_n) \in \mathbb{R}^{n \times n}$. For the embedding step, given an embedding dimension $\mathsf{d}$, we define the spectral embeddings as
\begin{align}
\qb_i = (\ub_{1i}, \dots, \ub_{\mathsf{d}i})^\top \in \mathbb{R}^{\mathsf{d}}, \quad i = 1, \dots, n. \label{eq:qb}
\end{align}
In practice, the embedding dimension $\mathsf{d}$ is selected based on the specific underlying task. Algorithm~\ref{alg:d} in Appendix~\ref{sec:supp_algo} provides a data-driven option for choosing $\mathsf d$, following \cite{ding2022impact}.

Finally, we note that the construction of the kernel matrix in \eqref{eq:Kn} depends on the choice of a key parameter $h_n$. Inspired by \cite{ding2022impact}, in this paper we select the bandwidth based on pairwise distances, allowing it to adapt to the underlying geometry, sparsity, and signal-to-noise ratio. Specifically, we compute all pairwise squared Euclidean distances of $\{\yb_i\}_{i \in [n]}$, denoted by $d_{ij}=\|\yb_i-\yb_j \|_2^2, 1 \leq i \neq j \leq  n$, and define their empirical cumulative distribution function as
\begin{align}
G_n(t) = \frac{2}{n(n-1)} \sum_{1 \le i < j \le n} \mathbf{1}_{\{d_{ij} \le t\}}. \label{eq:Gn}
\end{align}
We then set the bandwidth $h_n$ as the $\omega$-quantile of $G_n(t)$, given by
\begin{align}
h_n \coloneqq \inf\{t \ge 0 : G_n(t) \ge \omega\}, \label{eq:hn-alg}
\end{align}
where $\omega \in (0,1)$ is a prespecified quantile parameter. Practical guidelines for selecting $\omega$ are discussed in \cite{ding2022impact}. For completeness, we provide Algorithm~\ref{alg:omega} in Appendix~\ref{sec:supp_algo} for its selection.

\begin{remark}
Two remarks are in order. First, the spectral embedding step above is to provide a concrete nonlinear embedding procedure after the adaptive projection has stabilized the pairwise geometry. For this reason, this step may be replaced by other spectral embedding methods, for example, diffusion based methods such as Laplacian eigenmaps \cite{belkin2003laplacian}, diffusion maps \cite{coifman2006diffusion}, and vector diffusion maps \cite{singer2012vdm}. Second,     although we focus on the Gaussian kernel function in \eqref{eq:Kn}, our results extend to a broader class of kernel functions under suitable regularity conditions. We will discuss the general kernel assumption and the corresponding extension of our results in Appendix \ref{sec:proof_main}.
\end{remark}

\subsection{Downstream analysis}
\label{sec:downstream}

The learned embeddings can be used as low-dimensional features for downstream tasks such as clustering and prediction. In this paper, we focus on spectral clustering and classification. For spectral clustering, the setting is relatively direct, since one may apply a standard clustering algorithm directly to the embedded points $\{\qb_i\}_{i\in[n]}$ in \eqref{eq:qb}. In practice, this can be combined with commonly used methods such as K-means \cite{macqueen1967some}. The main point is that clustering is carried out in the learned low-dimensional space rather than in the original noisy ambient space, so the representation produced by the project--then--embed pipeline serves as a denoised feature map for unsupervised analysis.

For the downstream classification, the situation is more delicate, since one typically works with separate training and test sets. In particular, once the embedding $\{\qb_i\}_{i\in[n]}$ is learned from the training data, one cannot simply obtain the test embeddings by rerunning the whole spectral procedure on the test set alone, because this would place the test samples in a different coordinate system and make them no longer directly comparable with the geometry learned from the training set. On the other hand, one also cannot first embed the combined training and test samples and then split them afterward, since this would allow information from the test set to enter the construction of the training representation.

This motivates an out-of-sample extension within the same framework. Suppose the projection matrix $\Ps$, the bandwidth $h_n$, and the leading eigenpairs $\{(\lambda_i,\ub_i)\}_{i=1}^{\mathsf d}$ have already been obtained from the training samples. Given a new observation $\xb_j^{\mathrm{new}}\in\R^p$, we first project it using the same matrix $\Ps$ learned from the training stage, and then embed it by a Nystr\"om-type extension \cite{williams2001using} based on its kernel affinities with the projected training samples in \eqref{eq:projected_y}. In this way, the new observation is mapped into the same low-dimensional coordinate system as the training data, so that the training embeddings $\{\qb_i\}_{i\in[n]}$ and test embeddings $\{\qb_j^{\mathrm{new}}\}_{j\in[m]}$ remain directly comparable. The out-of-sample extension procedure is summarized in Algorithm~\ref{alg:pred}.

\begin{algorithm}[!ht]
    \caption{\bf Out-of-sample extension}
    \label{alg:pred}
    \normalsize
    \begin{flushleft}
        \vspace{0pt}
        \noindent{\bf Inputs:} New samples
        $\{\xb_j^{\mathrm{new}}\}_{j\in[m]}$, projection matrix $\Ps$, training mean $\bar\xb$, projected training samples $\{\yb_i\}_{i=1}^n$, bandwidth $h_n$, eigenpairs
        $\{(\lambda_i,\ub_i)\}_{i=1}^{\mathsf d}$ with $\lambda_{\mathsf d}>0$.

        \noindent{\bf Step one: projection\\}
        Compute the projected new samples
        $$
        \yb_j^{\mathrm{new}}=\Ps^\top(\xb_j^{\mathrm{new}}-\bar\xb)
        \in\R^\Ks,\qquad j=1,\dots,m.
        $$

        \noindent{\bf Step two: Nystr\"om extension\\}
        For each $j=1,\dots,m$, form the kernel vector
        $$
        \mathbf{k}_j^{\mathrm{new}}
        =
        \left(
        \exp\left(-\frac{\norm{\yb_j^{\mathrm{new}}-\yb_1}_2^2}{h_n}\right),
        \cdots,
        \exp\left(-\frac{\norm{\yb_j^{\mathrm{new}}-\yb_n}_2^2}{h_n}\right)
        \right)^\top\in\R^n.
        $$
        Let
        $$
         {\mathbf\Lambda}_{\mathsf d}
        =
        \mathrm{diag}(\lambda_1,\dots,\lambda_{\mathsf d})
        \quad\text{and}\quad
         {\mathbf U}_{\mathsf d}
        =
        (\ub_1,\dots,\ub_{\mathsf d})\in\R^{n\times \mathsf d}.
        $$
        Define
        $$
        \qb_j^{\mathrm{new}}
        =
        \frac{1}{n} {\mathbf\Lambda}_{\mathsf d}^{-1}
         {\mathbf U}_{\mathsf d}^\top
        \mathbf{k}_j^{\mathrm{new}}
        \in\R^\mathsf d.
        $$

        \noindent{\bf Output:} Out-of-sample embeddings
        $\{\qb_j^{\mathrm{new}}\}_{j\in[m]}$.
    \end{flushleft}
\end{algorithm}

    The out-of-sample extension uses exactly the same projection matrix $\Ps$, bandwidth $h_n$, and leading eigenpairs learned from the training data. As a result, new observations can be mapped into the same embedding space without repeating the full training procedure.

After this extension step, one may train a standard classifier on the training embeddings $\{\qb_i\}_{i\in[n]}$ obtained in \eqref{eq:qb} together with their labels, and then evaluate it on the out-of-sample test embeddings $\{\qb_j^{\mathrm{new}}\}_{j\in[m]}$ obtained in Algorithm~\ref{alg:pred}. One can use simple classifiers such as a multilayer perceptron (MLP) \cite{rumelhart1986learning} and linear discriminant analysis (LDA) \cite{fisher1936use}, though other classifiers can also be used in exactly the same way.

\section{Main Results}
\label{sec:main_results}
In this section, we develop the theoretical framework for the proposed method and present its main convergence results. Our goal is to quantify how accurately the spectral embeddings in \eqref{eq:qb}, after normalization, approximate the eigenfunctions of an integral operator defined by the clean signal distribution. We begin by introducing the structural assumptions on the signal, the noise, and the sparsity representation in Section~\ref{sec:ass}. We then study the recovery of pairwise distances after adaptive projection in Section \ref{sec:recover}, followed by the spectral convergence of the resulting kernel embedding constructed from the projected data in \eqref{eq:qb} in Section \ref{sec:spectral}.

\subsection{Notations and assumptions}
\label{sec:ass}
Throughout the paper, we follow the setup of \cite{elkaroui2016graph,ding2022impact,shen2022roseland} and consider the signal-plus-noise model \eqref{eq:spn}.
For the signal components $\sb_i$, we follow \cite{cheng2013local,shen2022roseland,wu2018think,ding2023learning} and impose the following assumption.

\begin{assumption}
    \label{ass:model}
    The signals $\{\sb_i\}_{i\in[n]}$ are independent $\R^p$-valued random vectors with common distribution $\mu_s$ on the Borel $\sigma$-algebra $\mathcal B(\R^p)$. $\mu_s$ is supported on $\mathcal S\coloneq\iota(\M)$, where $\M$ is a fixed $d$-dimensional Riemannian manifold with $d\ge1$, and	$\iota:\M\to\R^p$ is a smooth isometric embedding. The linear span of $\iota(\M)$ has dimension $r$ for some fixed integer $r<p$. Let $V$ denote the Riemannian volume measure on $\M$. Assume that $\mu_s$ is absolutely continuous with respect to $\iota_\# V$, the pushforward of $V$ under $\iota$, defined by $(\iota_\# V)(A)=V(\iota^{-1}(A))$ for $A\in\mathcal B(\R^p)$.
\end{assumption}

Assumption~\ref{ass:model} distinguishes the intrinsic manifold dimension $d$ from the dimension $r$ of the linear signal space. The role of the projection step is to exploit the latter before constructing any distance-based affinity. Based on Assumption~\ref{ass:model}, there exists a rotation matrix $\Rs=(\rs_1,\dots,\rs_p)\in\R^{p\times p}$ whose first $r$ columns form an orthonormal basis of $\operatorname{span}\{\iota(\M)\}$, such that, for each $\sb_i\in\iota(\M)$,
\begin{align}
    \Rs^\top \sb_i=\bigl((\sb_i^0)^\top,0,\dots,0\bigr)^\top
    \in\R^p,\quad i=1,\cdots,n,
    \label{eq:rotate}
\end{align}
where $\sb_i^0=(s_{i1},\dots,s_{ir})^\top\in\R^r$ denotes the reduced coordinates.
Hence, we write $\Rs_r=(\rs_1,\dots,\rs_r)\in\R^{p\times r}$ and define
\begin{equation}\label{eq:transform}
T:\R^p\to\R^r,\qquad T(\sb)=\Rs_r^\top\sb.
\end{equation}
Let $\mu\coloneq T_\#\mu_s$ be the pushforward probability measure induced by the reduced signal coordinates. Thus, $\{\sb_i^0\}_{i\in[n]}$ are i.i.d.\ from $\mu$. Moreover, we define the reduced signal manifold
$$
\M^0\coloneq T\bigl(\iota(\M)\bigr)\subset\R^r.
$$
Since $T$ restricts to a linear isometry on $\operatorname{span}(\iota(\M))$, the composition $T\circ\iota:\M\to\R^r$ is also a smooth isometric embedding. Thus $\M^0$ is a $d$-dimensional Riemannian submanifold of $\R^r$, with induced volume measure $V_0=(T\circ\iota)_\#V$. Assumption~\ref{ass:model} implies $\mu$ is absolutely continuous with respect to $V_0$, and we can therefore write
\begin{equation}
    f_\mu\coloneq\frac{\dd\mu}{\dd V_0}.\label{eq:f_mu}
\end{equation}

We next impose regularity conditions on the reduced signal distribution $\mu$ and the additive noise in \eqref{eq:spn}. Those assumptions are commonly used in the literature; see, for example, \cite{cheng2013local,ding2023learning}.
\begin{assumption}
    \label{ass:signal}
    We assume that $\M^0$ is compact and that the reduced signal measure $\mu$ admits a continuous density $f_\mu$, bounded above and below by some positive constants on $\M^0$ with respect to $V_0$. We also assume that $\Sigma=\Cov(\sb_i^0)$ is positive definite. We further assume that the noise vectors $\{\zb_i\}_{i\in[n]}$ are independent of the signals $\{\sb_i\}_{i\in[n]}$, and are independent sub-Gaussian random vectors with independent entries, and
    \begin{align}
        \E(\zb_i)=0,\quad\Cov(\zb_i)=\sigma^2\Ib_p.
        \label{eq:noise-cov}
    \end{align}
    Moreover, for some $ 0<\alpha<\infty$, we assume that $p\asymp n^\alpha$.
\end{assumption}

Let $\lambda_1^0\ge\cdots\ge\lambda_r^0>0$ denote the eigenvalues of $\Sigma$. We define the signal strength by
\begin{align}
    \Theta=\operatorname{tr}(\Sigma)=\sum_{j=1}^r\lambda_j^0.\label{eq:signal_strength}
\end{align}
Based on the model (\ref{eq:spn}), we see that the overall signal-to-noise ratio of noisy samples $\{\xb_i\}_{i\in[n]}$ can be defined as
\begin{align}
    \SNR\equiv\SNR(n)\coloneq\frac{\Theta}{p\sigma^2}.\label{eq:SNR}
\end{align}

We then impose a sparsity assumption motivated by \cite{johnstone2009consistency,paul2007augmented}, which encodes the sparsity structure that makes adaptive projection possible. Let $\Ws=(\ws_1,\dots,\ws_p)\in\R^{p\times p}$ be the orthonormal basis used in \eqref{eq:y}. For each $k=1,\dots,r$, write the $k$-th signal direction $\rs_k$ of $\Rs$ in this basis as
\begin{align}
    \rs_k=\sum_{j=1}^p a_{jk}\ws_j \in \mathbb{R}^p.\label{eq:proj}
\end{align}

\begin{assumption}
    \label{ass:lq}
    Assume that for some $q\in(0,2)$ and constants
    $0<c_k\le C_k<\infty$,
    \begin{align}
        c_k j^{-1/q}\le|a_{(j)k}|\le C_k j^{-1/q},
        \qquad j=1,\dots,p,\label{eq:lq}
    \end{align}
    where $|a_{(1)k}|\ge\cdots\ge|a_{(p)k}|$ denotes the decreasing
    rearrangement of $\{|a_{jk}|\}_{j\in[p]}$.
\end{assumption}

Smaller values of $q$ in (\ref{eq:lq}) correspond to faster energy decay and stronger concentration of the signal on a few basis directions. Larger values of $q$ spread the signal more evenly across the basis and therefore make the selection of directions less effective.

Finally, to streamline the statements of our results, we adopt the notion of stochastic domination \cite{erdos2017dynamical} throughout the remainder of the paper. Let $X=\{X^{(n)}(u): n\in\N,\ u\in U^{(n)}\}$ and $Y=\{Y^{(n)}(u): n\in\N,\ u\in U^{(n)}\}$ be two families of nonnegative random variables, where $U^{(n)}$ is a possibly $n$-dependent parameter set. We say that $X$ is stochastically dominated by $Y$, uniformly over $u\in U^{(n)}$, if for every small $\epsilon>0$ and large $D>0$, there exists $n_0(\epsilon,D)\in\N$ such that
$$
\sup_{u\in U^{(n)}}\P\Bigl(X^{(n)}(u)>n^\epsilon Y^{(n)}(u)\Bigr)\le n^{-D},
\qquad n\ge n_0(\epsilon,D).
$$
We then write $X\prec Y$ or equivalently $X=\OO_{\prec}(Y)$.

\subsection{Recovery of Pairwise Distances}
\label{sec:recover}
Recall the signal strength $\Theta$, signal-to-noise ratio $\SNR$, and the sparsity parameter $q$ in \eqref{eq:signal_strength}, \eqref{eq:SNR} and \eqref{eq:lq}. We begin by quantifying the projection dimension $\Ks$ selected by Algorithm~\ref{alg:Kselect}.

\begin{proposition}[Selected projection dimension]
    \label{prop:K}
    Suppose Assumptions~\ref{ass:model}, \ref{ass:signal}, and \ref{ass:lq} hold and $\sigma^2/\Theta=\oo(1)$. Then, with high probability, for sufficiently large $n$, for any fixed constants $s$ and $\tau$ in Algorithm \ref{alg:Kselect}
    \begin{align}
        \Ks
        \asymp
        \begin{cases}
            (\Theta/\sigma^2)^{q/2}, & \text{if  }\SNR\lesssim p^{2/q-1}, \\[0.5ex]
            p, & \text{if  } \SNR\gtrsim p^{2/q-1}.
        \end{cases}
        \label{eq:K-cases}
    \end{align}
\end{proposition}

\begin{remark}
	Proposition~\ref{prop:K} shows how the selected projection dimension depends on the $\SNR$ and the sparsity parameter $q$. When $\SNR\lesssim p^{2/q-1}$, the energy-based criterion screens out prescribed basis directions carrying small signal variances relative to the noise variance $\sigma^2$, leading to $\Ks\asymp(\Theta/\sigma^2)^{q/2}$. When $\SNR\gtrsim p^{2/q-1}$, the signal variance in every prescribed basis direction is of order $\sigma^2$ or larger, and the procedure retains a number of directions proportional to $p$. The condition $\sigma^2/\Theta=\oo(1)$ provides a sufficient condition for a nonempty selection of directions with high probability, allowing the two-sided characterization of $\Ks$ in Proposition~\ref{prop:K}.
\end{remark}

Once $\Ks$ and the associated directions are chosen, we next study how accurately the selected projection preserves the latent signal pairwise geometry. The projection $\Ps$ in \eqref{eq:P} obtained by Algorithm~\ref{alg:Kselect} reduces the number of noise coordinates from $p$ to $\Ks$, motivating the retained-noise scale $\Ks\sigma^2$. We define the effective SNR after projection as follows
\begin{align}
    \SNR_{\mathrm{eff}}
    \coloneq
    \begin{cases}
        \left(\dfrac{\Theta}{\sigma^2}\right)^{1-q/2},
        & \text{if } \SNR\lesssim p^{2/q-1}, \\[1.2ex]
        \dfrac{\Theta}{\sigma^2 p},
        & \text{if } \SNR\gtrsim p^{2/q-1}.
    \end{cases}
    \label{eq:SNR-eff}
\end{align}
Recall \eqref{eq:projected_y} that $\yb_i=\Ps^\top(\xb_i-\bar\xb)$ is the corresponding projected data. The resulting pairwise-distance error between $\norm{\yb_i-\yb_j}_2^2$ and $\norm{\sb_i-\sb_j}_2^2$ naturally decomposes into two sources: a bias caused by truncating weak signal directions, and a stochastic error caused by the projected noise.

Next, we define two rates for the two sources of error as
\begin{align}
    \psi_{n,\mathrm{bias}}
    \coloneq
    \begin{cases}
        \frac{1}{\SNR_{\mathrm{eff}}},
        & \text{if } \SNR\lesssim p^{2/q-1}, \\[1.2ex]
        p^{1-2/q},
        & \text{if } \SNR\gtrsim p^{2/q-1},
    \end{cases}
    \label{eq:psi-proj-bias}
\end{align}
and
\begin{align}
    \psi_{n,\mathrm{noise}}
    \coloneq
    \begin{cases}
        \frac{1}{\SNR_{\mathrm{eff}}}
        +
        \sigma\left(\dfrac{\Theta}{\sigma^2}\right)^{((q-1)_+)/2},
        & \text{if } \SNR\lesssim p^{2/q-1}, \\[2ex]
        \frac{1}{\SNR_{\mathrm{eff}}}
        +
        \sigma p^{(1-1/q)_+},
        & \text{if } \SNR\gtrsim p^{2/q-1},
    \end{cases}
    \label{eq:psi-proj-noise}
\end{align}
Moreover, we define the combined rate
\begin{align}
    \psi_n
    \coloneq
    \psi_{n,\mathrm{bias}}+\psi_{n,\mathrm{noise}}.
    \label{eq:psi-proj}
\end{align}

\begin{theorem}[Uniform recovery of pairwise distances]
    \label{thm:pairwise}
    Suppose Assumptions~\ref{ass:model}, \ref{ass:signal}, and \ref{ass:lq} hold. Then
    \begin{align}
        \max_{1\le i<j\le n}
        \left|\|\yb_i-\yb_j\|_2^2-\|\sb_i-\sb_j\|_2^2\right|
        =
        \OO_\prec(\psi_n).
        \label{eq:pairwise-uniform}
    \end{align}
\end{theorem}

\begin{remark}
    \label{rmk:pairwise}
    Theorem~\ref{thm:pairwise} shows that pairwise-distance recovery behaves differently in different $\SNR$ regimes. When $\SNR\lesssim p^{2/q-1}$, the selected projection dimension is smaller, and the resulting error is controlled by the effective post-projection signal-to-noise ratio. When $\SNR\gtrsim p^{2/q-1}$, the selected projection dimension is of order $p$, so the rate reflects the full scale after projection.

    The sparsity parameter $q$ determines how favorable this recovery is. For $0<q\le 1$,
    \begin{align*}
        \psi_{n}
        \asymp
        \begin{cases}
            \frac{1}{\SNR_{\mathrm{eff}}}+\sigma,
            & \text{if } \SNR\lesssim p^{2/q-1}, \\[1.2ex]
            p^{1-2/q}+\sigma,
            & \text{if } \SNR\gtrsim p^{2/q-1},
        \end{cases}
    \end{align*}
    and for $1<q<2$,
    \begin{align*}
        \psi_{n}
        \asymp
        \begin{cases}
            \frac{1}{\SNR_{\mathrm{eff}}}+\sigma\left(\dfrac{\Theta}{\sigma^2}\right)^{(q-1)/2},
            & \text{if } \SNR\lesssim p^{2/q-1}, \\[1.2ex]
            p^{1-2/q}+\sigma p^{1-1/q},
            & \text{if } \SNR\gtrsim p^{2/q-1}.
        \end{cases}
    \end{align*}
    Smaller values of $q$ lead to stronger concentration of the signal on a few directions and hence better recovery, whereas larger values of $q$ spread the signal more broadly and lead to a larger error rate.

        We point out that the rate bound in Theorem~\ref{thm:pairwise} does not require the condition $\sigma^2/\Theta=\oo(1)$ imposed in Proposition~\ref{prop:K}. However, if $\sigma\lesssim n^{-\mathsf c}$ for some fixed $\mathsf c>0$, consistency holds across all the $\SNR$ regimes and sparsity levels discussed above. Specifically, for any fixed $q\in(0,2)$ and $p\asymp n^\alpha$ as in Assumptions~\ref{ass:lq} and~\ref{ass:signal}, respectively, Theorem~\ref{thm:pairwise} implies that there exists a fixed $\mathsf c'>0$ such that
    	$$
    	\max_{1\le i<j\le n}
    	\left|
    	\|\yb_i-\yb_j\|_2^2
    	-
    	\|\sb_i-\sb_j\|_2^2
    	\right|
    	=
    	\OO_\prec(n^{-\mathsf c'}).
    	$$
\end{remark}

Theorem~\ref{thm:pairwise} provides the algorithmic justification for the adaptive projection step. It shows that, after selecting the basis directions with large enough empirical energy, the projected observations retain the essential pairwise geometry of the underlying signals. In this sense, the projection step serves as a geometric denoising procedure: it suppresses noise-dominated directions while preserving the distance information needed for the subsequent kernel construction. This is precisely why the projected data, rather than the original observations, form a more reliable input for the spectral embedding algorithm.

\subsection{Convergence of Spectral Embedding}
\label{sec:spectral}

We now state the spectral convergence results for $\Kb_n$ in \eqref{eq:Kn}.
Recall that $\mu$ is the distribution of the reduced signal coordinates $\sb_i^0$ on $\M^0\subset\R^r$ as in Assumption~\ref{ass:signal}.
Let $\bm{U}\in\R^r$ and $\bm{V}\in\R^r$ be i.i.d.\ random vectors with distribution $\mu$, and set $\bm{D}=\|\bm{U}-\bm{V}\|_2^2.$ Denote the cumulative distribution function of $\bm{D}$ by
\begin{align}
	G(t)&\coloneq\P(\bm D\le t),\label{eq:cdf}
\end{align}
and define the population bandwidth $\mathsf h$ by
\begin{align}
	\mathsf h\coloneq \inf\{t\ge 0:G(t)\ge \omega\},\label{eq:h}
\end{align}
with $\omega$ as in \eqref{eq:hn-alg}.

Using this bandwidth $\mathsf h$, we define the population kernel operator $\mathcal K:L^2(\R^r,\mu)\to L^2(\R^r,\mu)$ by
\begin{align}
    (\mathcal K g)(\bm{u})
    \coloneq
    \int_{\M^0} \exp\!\left(-\frac{\|\bm{u}-\bm{v}\|_2^2}{\mathsf{h}}\right)g(\bm{v})\,\mu(\dd\bm{v}),
    \qquad \bm{u}\in\R^r.\notag
\end{align}
Let $\{(\gamma_i,\phi_i)\}_{i\ge 1}$ denote the eigenvalue-eigenfunction pairs of $\mathcal K$, with $\{\phi_i\}$ orthonormal in $L^2(\R^r,\mu)$, ordered so that $\gamma_1\ge \gamma_2\ge\cdots\ge 0$.

Since our observed data live in the space $\R^p$, it is convenient to introduce the corresponding operator on the embedded signal space. Recall that $\mu_s$ is the distribution of the clean signals, supported on the embedded manifold $\mathcal S=\iota(\M)\subset\R^p$ as in Assumption~\ref{ass:model}. We define the corresponding population operator $\widetilde{\mathcal K}:L^2(\R^p,\mu_s)\to L^2(\R^p,\mu_s)$ by
\begin{align}
	(\widetilde{\mathcal K}g)(\mathbf x)
	\coloneq
	\int_{\mathcal S} \exp\!\left(-\frac{\|\mathbf x-\mathbf y\|_2^2}{\mathsf h}\right)g(\mathbf y)\,\mu_s(\dd\mathbf y),
	\qquad \mathbf x\in\mathcal S.\label{eq:pop_op_1}
\end{align}
Recall $\Rs$ in \eqref{eq:rotate} and $\Rs_r$ in \eqref{eq:transform}. For $\mathbf x\in\mathcal S$, let $\bm u$ denote the first $r$ coordinates of $\Rs^\top\mathbf x$ and define
\[
\widetilde\phi_i(\mathbf x)\coloneq\phi_i(\bm u),\qquad i\ge1.
\]
Since $\bm u\mapsto\Rs_r\bm u$ preserves distances and transports $\mu$ to $\mu_s$, the functions $\{\widetilde\phi_i\}$ are orthonormal eigenfunctions of $\widetilde{\mathcal K}$ with the same eigenvalues $\{\gamma_i\}$. For any fixed $\delta>0$, define
\begin{align}
    M(\delta)\coloneq \#\{i\ge 1:\gamma_i\ge \delta\}.
    \label{eq:M}
\end{align}
For a simple positive eigenvalue $\gamma_i$, let
\begin{equation}\label{eq:eigengap}
g_i=\begin{cases}
\gamma_1-\gamma_2,&i=1,\\
\min\{\gamma_{i-1}-\gamma_i,\gamma_i-\gamma_{i+1}\},&i>1.
\end{cases}
\end{equation}

We then turn to the empirical operator constructed from the noisy observations. This operator is built from the projected data $\{\yb_j\}_{j\in[n]}$ in \eqref{eq:projected_y}. Let $\widehat{\mu}_n$ be the empirical distribution of the noisy samples $\{\xb_j\}_{j=1}^n\subset\R^p$, that is,
$$
\widehat{\mu}_n\coloneq \frac{1}{n}\sum_{j=1}^n \delta_{\xb_j}.
$$
Recall $\Ps\in\R^{p\times\Ks}$ in \eqref{eq:P}. We define the corresponding empirical operator $\widehat{\mathcal K}_n:L^2(\R^p,\widehat{\mu}_n)\to L^2(\R^p,\widehat{\mu}_n)$ by
\begin{align}
    (\widehat{\mathcal K}_n g)(\mathbf{x})
    \coloneq
    \int \exp\!\left(-\frac{\|\Ps^\top \mathbf{x}-\Ps^\top \mathbf{y}\|_2^2}{h_n}\right)g(\mathbf{y})\,\widehat{\mu}_n(\dd\mathbf{y})
    =
    \frac{1}{n}\sum_{j=1}^n
    \exp\!\left(-\frac{\|\Ps^\top(\mathbf{x}-\xb_j)\|_2^2}{h_n}\right)g(\xb_j),
    \qquad \mathbf{x}\in\R^p\notag
\end{align}
By construction of $\widehat{\mathcal K}_n$, the nonzero eigenvalues of $\widehat{\mathcal K}_n$ coincide with those of $n^{-1}\Kb_n$. Recall from \eqref{eq:eigendecomposition} that
$$
\frac{1}{n}\Kb_n
=
{\mathbf U}{\mathbf\Lambda}{\mathbf U}^\top,
\qquad
{\mathbf\Lambda}=\diag(\lambda_1,\dots,\lambda_n),
\qquad
{\mathbf U}=(\ub_1,\dots,\ub_n).
$$
For each $\lambda_i>0$, the corresponding normalized empirical eigenfunction is given by
\begin{align}
    \widehat{\phi}_i^{(n)}(\mathbf{x})
    \coloneq
    \frac{1}{\lambda_i\sqrt n}\sum_{j=1}^n
    \exp\!\left(-\frac{\|\Ps^\top(\mathbf{x}-\xb_j)\|_2^2}{h_n}\right)u_{ij},
    \qquad \mathbf{x}\in\R^p.\label{eq:emp_op_1}
\end{align}

We now present the convergence results.  Recall $\psi_n$ in \eqref{eq:psi-proj}, $G$ in \eqref{eq:cdf}, $\mathsf{h}$ in \eqref{eq:h}, $M(\delta)$ in \eqref{eq:M} and $g_i$ in \eqref{eq:eigengap}.

\begin{theorem}[Spectral convergence]
    \label{thm:total_convergence}
    Suppose Assumptions~\ref{ass:model}, \ref{ass:signal}, and \ref{ass:lq} hold. Assume $\omega\in(0,1)$ is chosen such that $G$ has a density $\nu$ in a neighborhood of $\mathsf h$, continuous at $\mathsf h$ with $\nu(\mathsf h)>0$. Then

    $$
    \max_{i\in[n]}|\lambda_i-\gamma_i|
    =
    \OO_\prec\!\left(\psi_n+\frac{1}{\sqrt{n}}\right).
    $$
    Moreover, for any fixed $\delta>0$ and any $1\le i\le M(\delta)$  with $g_i>0$,
    there exists some $s_i\in\{-1,1\}$ such that, for any $\mathbf{x}\in\mathcal S$, we have
    $$
    \left|
    \sqrt{\lambda_i}\,\widehat{\phi}_i^{(n)}(\mathbf{x})
    -
    s_i\sqrt{\gamma_i}\,\widetilde{\phi}_i(\mathbf{x})
    \right|
    =
    \OO_\prec\!\left(
    \psi_n+\frac{1}{\sqrt{n}}
    +
    \frac1{g_i}
    (\psi_n^{1/2}+\frac{1}{\sqrt{n}})
    \right).
    $$
\end{theorem}

Theorem~\ref{thm:total_convergence} provides the methodological justification for the full project--then--embed procedure. It shows that, after the projection step denoises the pairwise geometry, the kernel matrix constructed from the projected observations still recovers the population spectral structure associated with the underlying signal manifold. Consequently, the spectral embeddings in \eqref{eq:qb} can be interpreted as reliable low-dimensional features that retain the intrinsic nonlinear geometry while remaining robust to high-dimensional noise.

\begin{remark}
	The error bound in Theorem~\ref{thm:total_convergence} has three main sources. First, $\psi_n$ is the geometric perturbation inherited from Theorem~\ref{thm:pairwise}, and measures the distortion of pairwise distances after projection by $\Ps^\top$ as in \eqref{eq:P}. Second, comparing the empirical bandwidth $h_n$ with the population bandwidth $\mathsf h$ introduces a quantile estimation error of order $n^{-1/2}$ in addition to $\psi_n$. Third, approximating the population integral operator by its empirical counterpart contributes another error of order $n^{-1/2}$.

    Similar to Remark~\ref{rmk:pairwise}, a polynomial regime guarantees consistency of both the pairwise-distance and spectral recovery results. In particular, for any $q\in(0,2)$ and $p\asymp n^\alpha$, if we have $\sigma\lesssim n^{-\mathsf c}$ for some fixed $\mathsf c>0$, $\psi_n$ in \eqref{eq:psi-proj} decays polynomially, and hence all the error terms in Theorems~\ref{thm:pairwise} and \ref{thm:total_convergence} are stochastically dominated by $n^{-\mathsf c''}$ for some fixed $\mathsf c''>0$.

	We now compare our results with those in \cite{ding2023learning}. Their theory is developed under the condition $\SNR\gg 1$, whereas our analysis also covers the regime $\SNR\lesssim 1$. For example, with $\Theta\asymp1$ and $\sigma\asymp n^{-c}$ for $0<c<\alpha/2$, the overall $\SNR\asymp n^{2c-\alpha}\ll 1$, while our consistency results continue to apply. However, this regime lies outside the high-SNR assumptions of \cite{ding2023learning}. Compared with their error bounds, our results involve two additional contributions. The first is the bias term $\psi_{n,\mathrm{bias}}$ in \eqref{eq:psi-proj-bias}, which reflects the signal energy lost in the unselected basis directions. The second is an additional bandwidth estimation error of order $n^{-1/2}$ from the fact that we use the population $\omega$-quantile $\mathsf h$ rather than the empirical $\omega$-quantile of the clean pairwise distances of $\{\sb_j\}_{j\in[n]}$, as in \cite{ding2023learning}. Furthermore, when $q\in(0,1]$, our projected-noise term $\psi_{n,\mathrm{noise}}$ in \eqref{eq:psi-proj-noise} has the same form as their bound, with the ambient noise dimension $p$ replaced by the selected dimension $\Ks$ under the conditions of Proposition~\ref{prop:K}. When $q\in(1,2)$, however, the slower decay of the basis coefficients introduces the additional factor $\Ks^{1-1/q}$ in our bound for the signal--noise interaction.
\end{remark}

Finally, we point out that the spectral convergence results extend to a broader class of kernel functions under suitable regularity conditions. We defer the precise assumptions and formal statements to Appendix~\ref{sec:proof_main}.

\section{Numerical Simulations}
\label{sec:simulation_analysis}

In this section, we conduct extensive Monte-Carlo simulations to evaluate the empirical performance of the proposed method (denoted as SPASE) on downstream classification and spectral clustering tasks. For comparison, we also include several commonly used manifold learning methods, including Isomap \cite{tenenbaum2000global} (denoted as Isomap), locally linear embedding \cite{roweis2000lle} (denoted as LLE), t-SNE \cite{van2008visualizing} (denoted as t-SNE), Diffusion Maps \cite{coifman2006diffusion} (denoted as DM), autoencoders \cite{hinton2006reducing} (denoted as AE), denoising autoencoders \cite{vincent2008extracting} (denoted as DAE), and kernel spectral embedding in \cite{ding2023learning} (denoted as KSE). We first compare downstream performance in Section~\ref{sec:sim_downstream}, then examine recovery of clean pairwise distances, sensitivity to the prescribed basis, and the choice of projection dimension in Sections~\ref{sec:distance_recovery}--\ref{sec:dimension_behavior}.

\subsection{Simulation Setup}
\label{sec:simu_setup}

We first describe the common data-generating mechanism used throughout the simulation study. For \eqref{eq:spn}, we choose the noise
$
\zb_j\iid \mathcal{N}(\mathbf 0,\sigma^2\Ib_p),
$
and construct the signal $\sb_j\in\R^p$ through \eqref{eq:rotate}. Specifically, we first generate latent coordinates $\sb_j^0=(\sb_{j,1}^0,\sb_{j,2}^0,\sb_{j,3}^0)^\top\in\R^3$ on a latent manifold $\Omega\subset\R^3$. We then randomly choose three distinct indices
$
1\le i_1<i_2<i_3\le p
$
and define $\tilde{\sb}_j=(\tilde{\sb}_{j,1},\cdots,\tilde{\sb}_{j,p})^\top\in\R^p$ by
$$
\tilde{\sb}_{j,\ell}=\begin{cases}
\sb_{j,k}^0, & \ell=i_k,\ k=1,2,3,\\
0, & \ell\notin\{i_1,i_2,i_3\}.
\end{cases}
$$
Next, we construct $\Rs\in\R^{p\times p}$ using the Haar basis induced by level-4 discrete multiresolution decomposition (see, for example, \cite{mallat2009wavelet,daubechies1992ten}), denoted by $\mathsf H=\{\mathsf h_j\}_{j=1}^p$. Recall the sparsity parameter $q$ in Assumption~\ref{ass:lq}. Larger $q$ gives slower decay of the coefficient magnitudes $\ell^{-1/q}$ in the construction below. For each $k=1,2,3$, we generate
\begin{align}
    \tilde{\rs}_{i_k}=\sum_{\ell=1}^{p}\ell^{-1/q}\mathsf h_{\pi_k(\ell)},
    \notag
\end{align}
where, for each $k=1,2,3$, $\{\pi_k(\ell)\}_{\ell\in[p]}$ is an independently generated random permutation of $[p]$. We then orthonormalize the three vectors $(\tilde{\rs}_{i_1},\tilde{\rs}_{i_2},\tilde{\rs}_{i_3})$ to obtain $(\rs_{i_1},\rs_{i_2},\rs_{i_3})$, and let $\Rs$ be an orthogonal completion whose $i_1$-th, $i_2$-th, and $i_3$-th columns are precisely these three vectors $\rs_{i_1},\rs_{i_2},\rs_{i_3}\in\R^p$. Finally, we set
$$
\sb_j=\Rs\tilde{\sb}_j\in\R^p.
$$

To assess performance across different geometric settings, we consider $\Omega$ consisting of two $2$-dimensional spheres embedded in $\R^3$, where the two spheres correspond to the two class labels. We study two choices of their locations and radii: in the first, the spheres are disjoint, both with radius $2$ and centered at $(1.5,1.5,1.5)^\top$ and $(-1.5,-1.5,-1.5)^\top$; in the second, they are concentric, centered at the origin with radii $1$ and $4$. 
% These latent structures are displayed in Figure~\ref{fig:simulated_manifolds}.

%\begin{figure}[!htbp]
%    \centering
%    \begin{subfigure}[t]{0.4\textwidth}
%        \centering
%        \includegraphics[width=\linewidth]{images/two_spheres.pdf}
%        \caption{Two Disjoint Spheres}
%    \end{subfigure}
%    \hspace{0.08\textwidth}
%    \begin{subfigure}[t]{0.4\textwidth}
%        \centering
%        \includegraphics[width=\linewidth]{images/two_compacted_spheres.pdf}
%        \caption{Two Concentric Spheres}
%    \end{subfigure}
%    \caption{Illustration of the two parameter settings of the latent two-sphere model. Different colors denote different class labels.}
%    \label{fig:simulated_manifolds}
%\end{figure}

Unless otherwise specified, we fix $q=1$, set the dimension to be $p=10000$, and generate $500$ observations uniformly from each sphere, so that the total sample size is $n=1000$. We compare the proposed method with the competing approaches on the two parameter settings of the latent two-sphere model 
% in Figure~\ref{fig:simulated_manifolds} 
under a range of noise levels $\sigma\in\{0.1,0.2,0.3,0.4,0.5\}$. For the main comparison, the prescribed basis $\Ws$ in our proposed method is taken to be the same level-4 Haar basis $\mathsf H$ used to construct $\Rs$. Sensitivity to basis misspecification is examined separately in Section~\ref{sec:basis_misspecification}. For classification, we use a $70/30$ training--testing split, train a shallow neural network with one hidden layer of $16$ ReLU units on the training embeddings, and report the classification accuracy on the test embeddings. For spectral clustering, we apply $k$-means and report the Adjusted Rand Index (ARI) \cite{hubert1985comparing}, which takes value between 0 and 1. Each experiment setup is repeated $250$ times, and the reported accuracies and ARIs are averaged over these repetitions.

\subsection[Classification and Clustering]{Classification and Clustering}\label{sec:sim_downstream}

\begin{figure}[!htbp]
    \centering
    \begin{subfigure}[t]{0.48\textwidth}
        \centering
        \includegraphics[width=\linewidth]{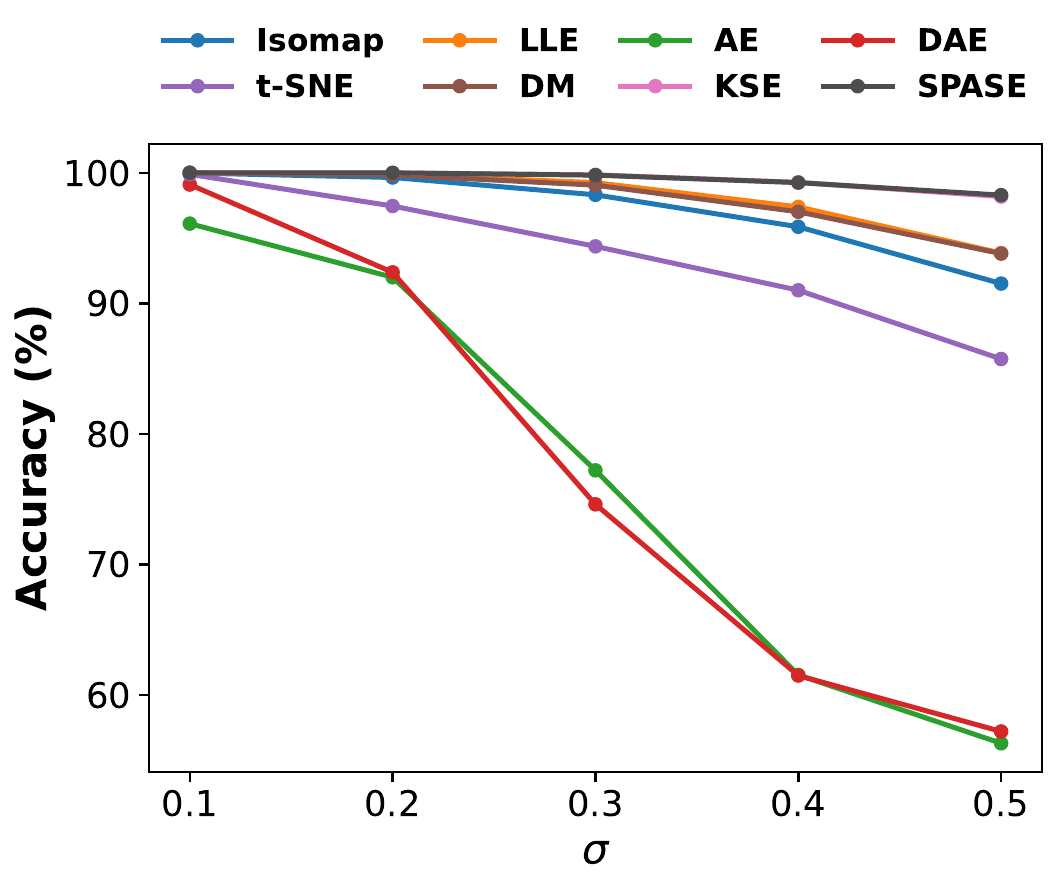}
        \caption{Classification Accuracy}
    \end{subfigure}\hfill
    \begin{subfigure}[t]{0.48\textwidth}
        \centering
        \includegraphics[width=\linewidth]{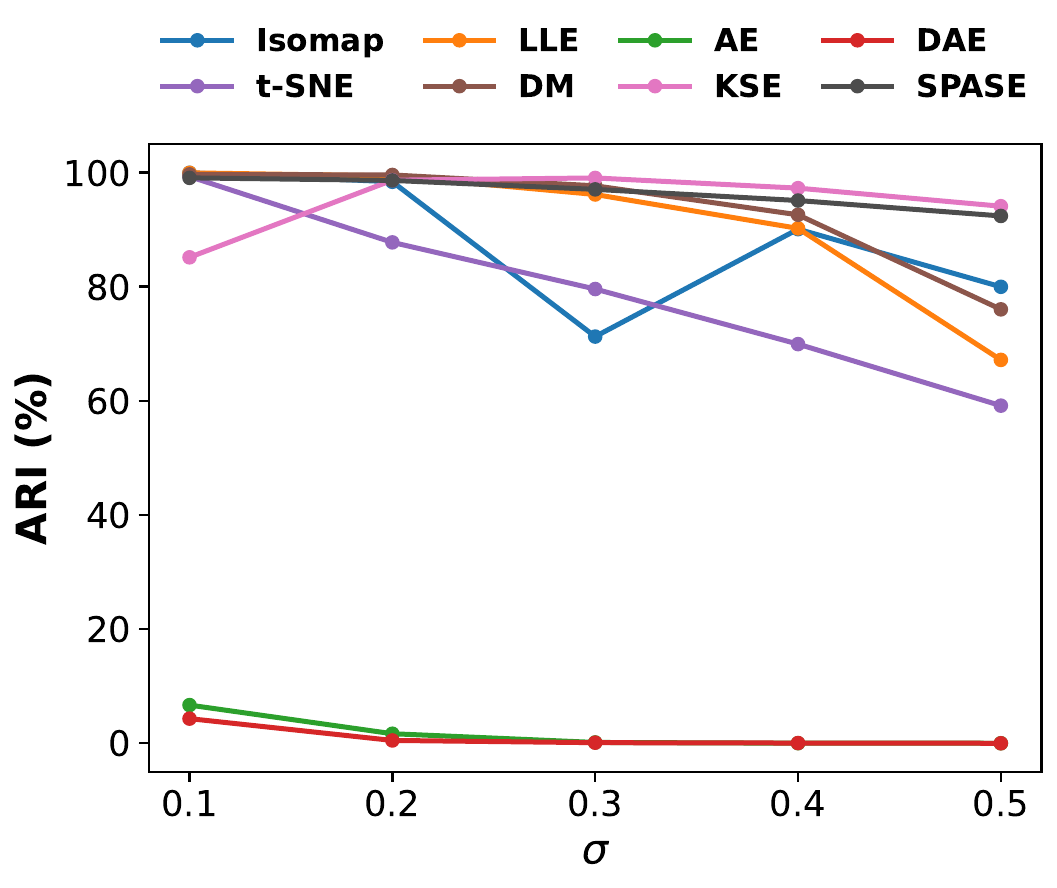}
        \caption{Clustering ARI}
    \end{subfigure}
    \caption[Classification accuracy and clustering ARI for the two disjoint spheres model across varying noise levels.]{Classification accuracy and clustering ARI for the two disjoint spheres model across varying noise levels. The reported results are based on $250$ repetitions. }
\label{fig:results_two_sphere}
\end{figure}

\begin{figure}[!htbp]
    \centering
    \begin{subfigure}[t]{0.48\textwidth}
        \centering
        \includegraphics[width=\linewidth]{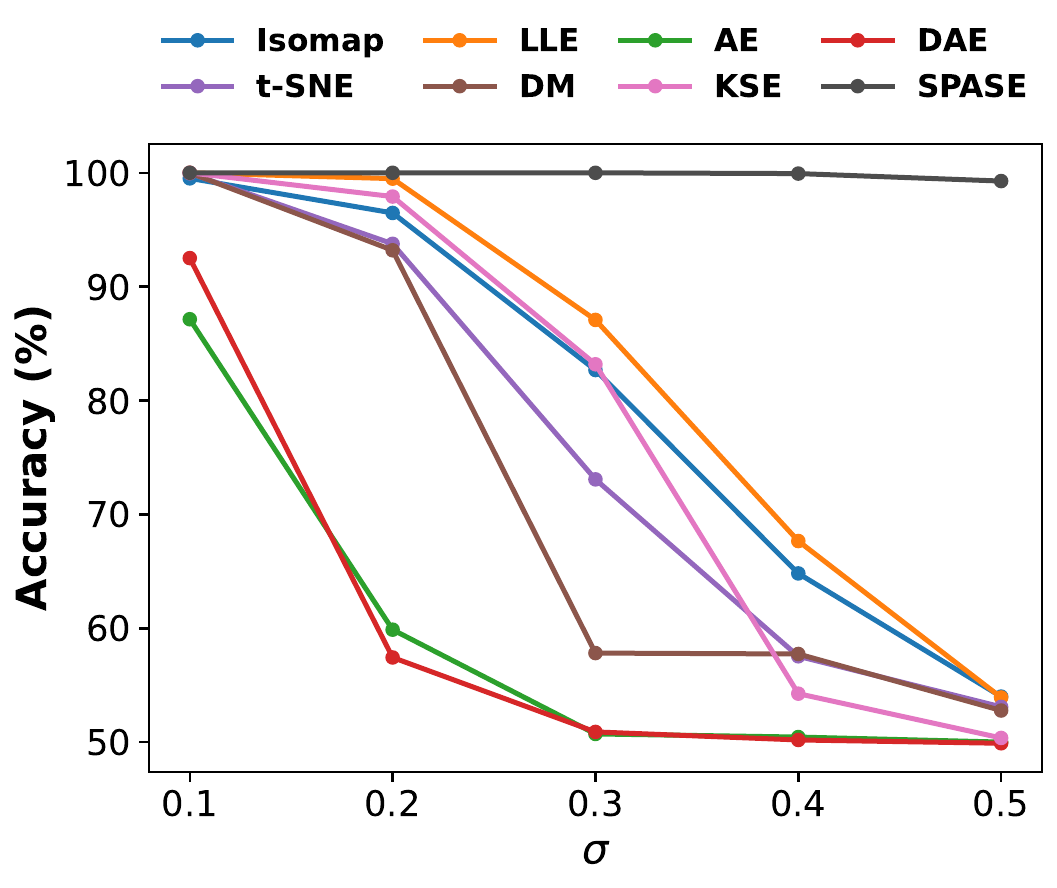}
        \caption{Classification Accuracy}
    \end{subfigure}\hfill
    \begin{subfigure}[t]{0.48\textwidth}
        \centering
        \includegraphics[width=\linewidth]{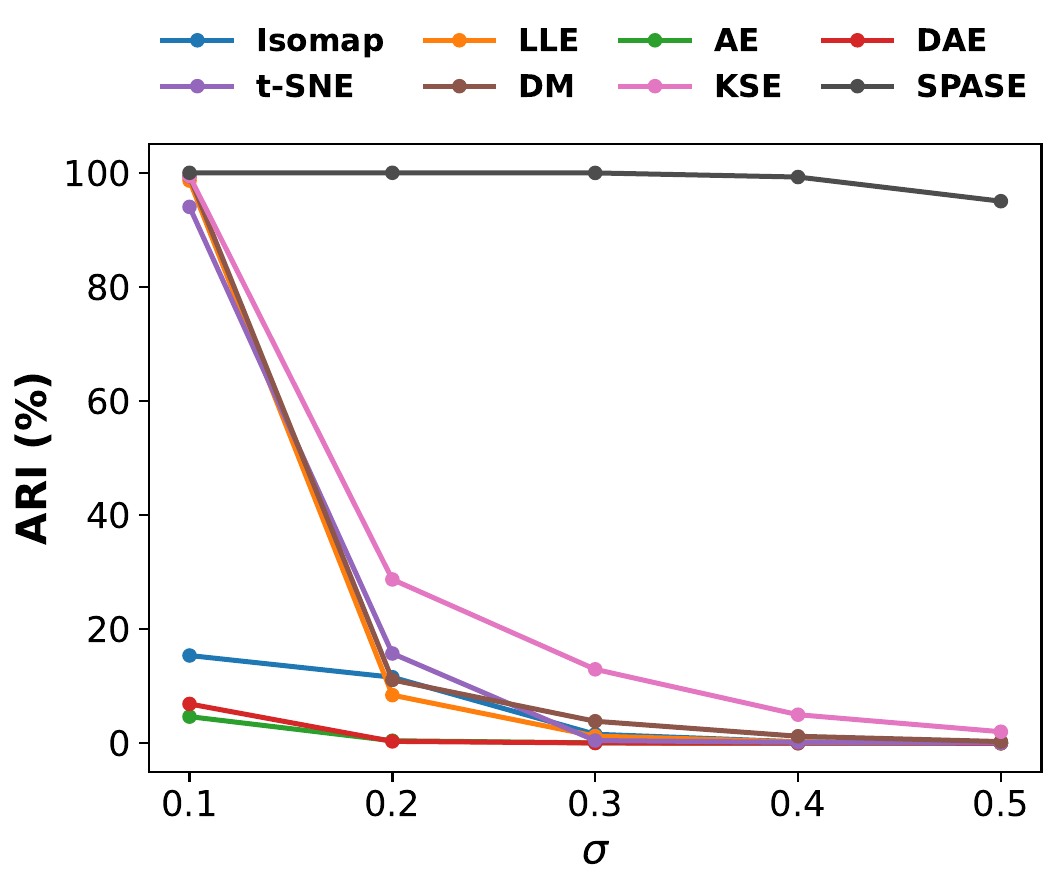}
        \caption{Clustering ARI}
    \end{subfigure}
    \caption[Classification accuracy and clustering ARI for the two-concentric-spheres model across varying noise levels.]{Classification accuracy and clustering ARI for the two-concentric-spheres model across varying noise levels. The reported results are based on $250$ repetitions. }
\label{fig:results_two_compact_sphere}
\end{figure}

The downstream classification and clustering results are reported in Figures~\ref{fig:results_two_sphere} and \ref{fig:results_two_compact_sphere}.

Figure~\ref{fig:results_two_sphere} shows that, in this relatively simple two-disjoint-spheres setting, most methods perform reasonably well when the noise level is small. For classification, SPASE, KSE, Isomap, LLE, DM, and t-SNE all achieve high accuracy at low noise levels, whereas AE and DAE deteriorate rapidly as $\sigma$ increases and approach chance-level accuracy under heavier noise. For clustering, the difference becomes more pronounced under heavier noise: SPASE and KSE remain robust, while the remaining methods show substantially larger degradation, especially AE, DAE, and t-SNE. At $\sigma=0.5$, SPASE and KSE achieve classification accuracies of $98.29\%$ and $98.18\%$ and ARIs of $0.924$ and $0.941$, respectively. Although several methods work well when the geometry is simple and the noise is mild, SPASE and KSE are the most stable at the larger noise levels.

Figure~\ref{fig:results_two_compact_sphere} presents results for the much more demanding two-concentric-spheres setting. As noise increases, most competing methods deteriorate in both classification and clustering. SPASE remains the most robust across the range of $\sigma$: its classification accuracy remains almost perfect and its clustering ARI remains high as noise increases. Among the baselines, KSE has the highest clustering ARI, but its performance is clearly below that of SPASE under moderate and high noise. At $\sigma=0.5$, SPASE attains $99.27\%$ classification accuracy and an ARI of $0.950$, compared with $50.38\%$ and $0.020$ for KSE.
KSE also performs well at lower noise levels because it incorporates kernel smoothing prior to spectral embedding, partially mitigating the effect of high-dimensional noise. However, unlike our adaptive projection strategy, it constructs the kernel directly from the noisy observations, and its performance deteriorates more substantially as the noise level increases. These results support the main motivation of SPASE: reducing noise-induced distortion of pairwise distances before spectral embedding. We examine this effect directly through pairwise-distance recovery in Section~\ref{sec:distance_recovery}.

\subsection[Recovery of Pairwise Distances]{Recovery of Pairwise Distances}
\label{sec:distance_recovery}
We next examine the recovery of clean pairwise distances using the concentric-sphere setting as in Section~\ref{sec:simu_setup}. For a projection matrix $\widehat {\bf P}\in \R^{p\times k}$, we write $\widehat{\bm u}_i=\widehat {\bf P}^\top(\xb_i-\bar\xb)\in\R^k$ for the projected observations. We compare their squared pairwise distances with those between the clean signals using the relative error
\begin{align}
	\mathcal E_{\mathrm{dist}}(\widehat {\bf P})
		=\frac{\left\{\sum_{i<j}
			\left(\|\widehat{\bm u}_i-\widehat{\bm u}_j\|_2^2-\|\sb_i-\sb_j\|_2^2\right)^2\right\}^{1/2}}
		{\left\{\sum_{i<j}\|\sb_i-\sb_j\|_2^4\right\}^{1/2}}.
	\label{eq:simdist}
\end{align}
Alongside SPASE and no projection, i.e., $\widehat {\bf P}=\mathbf I_p$, we include PCA and random orthonormal projections with dimensions $k$ matched to the SPASE-selected $\Ks$, and an oracle projection $\Rs$ as in \eqref{eq:rotate} onto the true three-dimensional signal space. All data-dependent projections are estimated from the training sample, and the relative error is evaluated on the test observations. The oracle also projects the noisy observations, so noise within the true signal space remains and its error need not be zero.

\begin{figure}[!htbp]
\centering
\includegraphics[width=.52\linewidth]{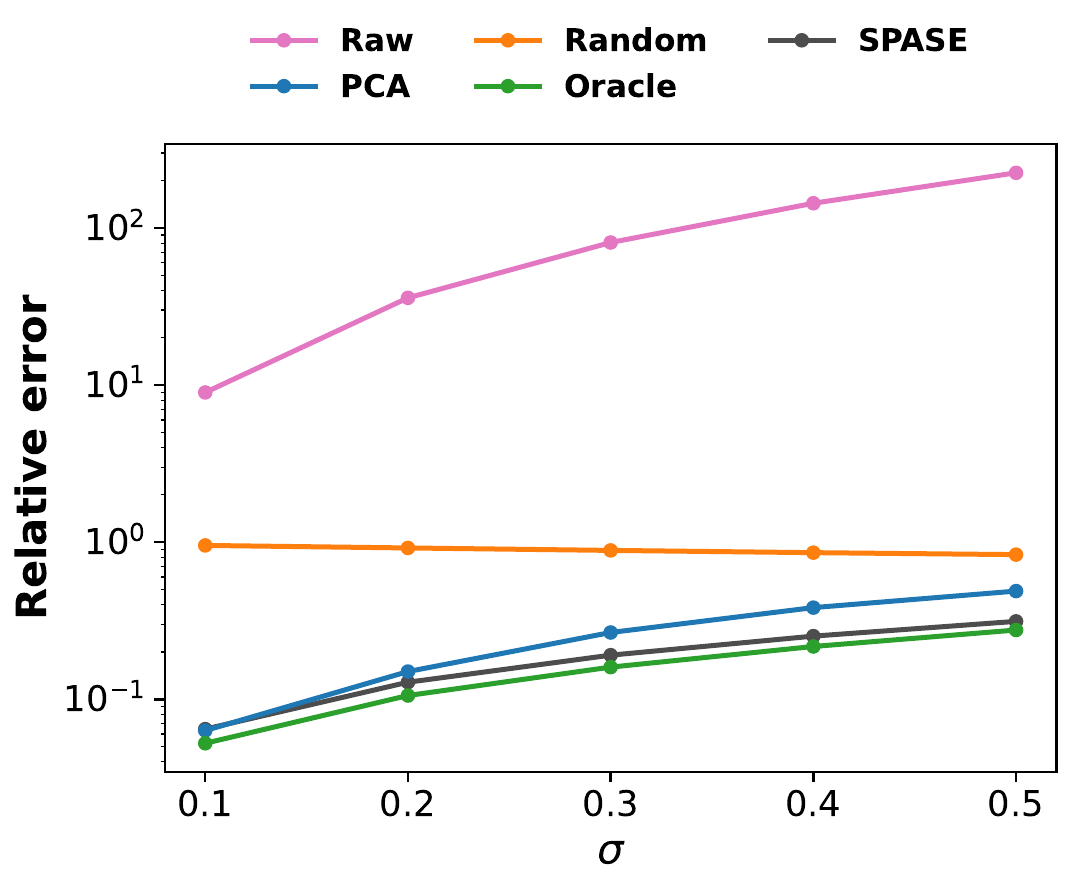}
\caption{Relative error in squared pairwise distances for concentric spheres, averaged over $250$ repetitions. Curves compare Raw (no projection), Random, SPASE, PCA, and Oracle (true signal space). Random and PCA use the SPASE-selected dimension. All projections act on noisy observations. The vertical scale is logarithmic.}
\label{fig:dist}
\end{figure}

Figure~\ref{fig:dist} shows that projection substantially reduces the distortion of pairwise distances. At $\sigma=0.5$, the mean relative errors are $224.24$ without projection, $0.833$ for random projection, $0.488$ for PCA, $0.313$ for SPASE, and $0.276$ for the oracle. SPASE achieves an error much closer to the oracle benchmark, although its projection $\Ps$ in \eqref{eq:P} is estimated from the noisy observations without knowledge of the true signal space.

This comparison illustrates the role of energy-based projection in recovering the clean pairwise geometry. In particular, at the highest noise level, the distances used by SPASE for kernel construction are much closer to the clean signal distances than those used by direct KSE. This difference is consistent with the higher classification accuracy and clustering ARI of SPASE in the concentric-sphere setting discussed in Section~\ref{sec:sim_downstream}.

\subsection[Basis Misspecification]{Basis Misspecification}
\label{sec:basis_misspecification}
In Section~\ref{sec:simu_setup}, the working basis $\Ws$ in \eqref{eq:y} is chosen as the level-4 Haar basis $\mathsf H$ used to generate the signals. In practice, however, the basis in which the signals admit a sparse representation is generally unknown. We therefore evaluate sensitivity to the choice of $\Ws$ by comparing the downstream classification performance of the matched basis with a level-2 Haar basis and level-4 Daubechies bases of orders $2$ and $4$ \cite{daubechies1992ten}. The alternative implementations are denoted as Haar L2, db2 L4, and db4 L4. We retain the same signal-generating mechanism and all other settings.

\begin{figure}[!htbp]
\centering
\includegraphics[width=\linewidth]{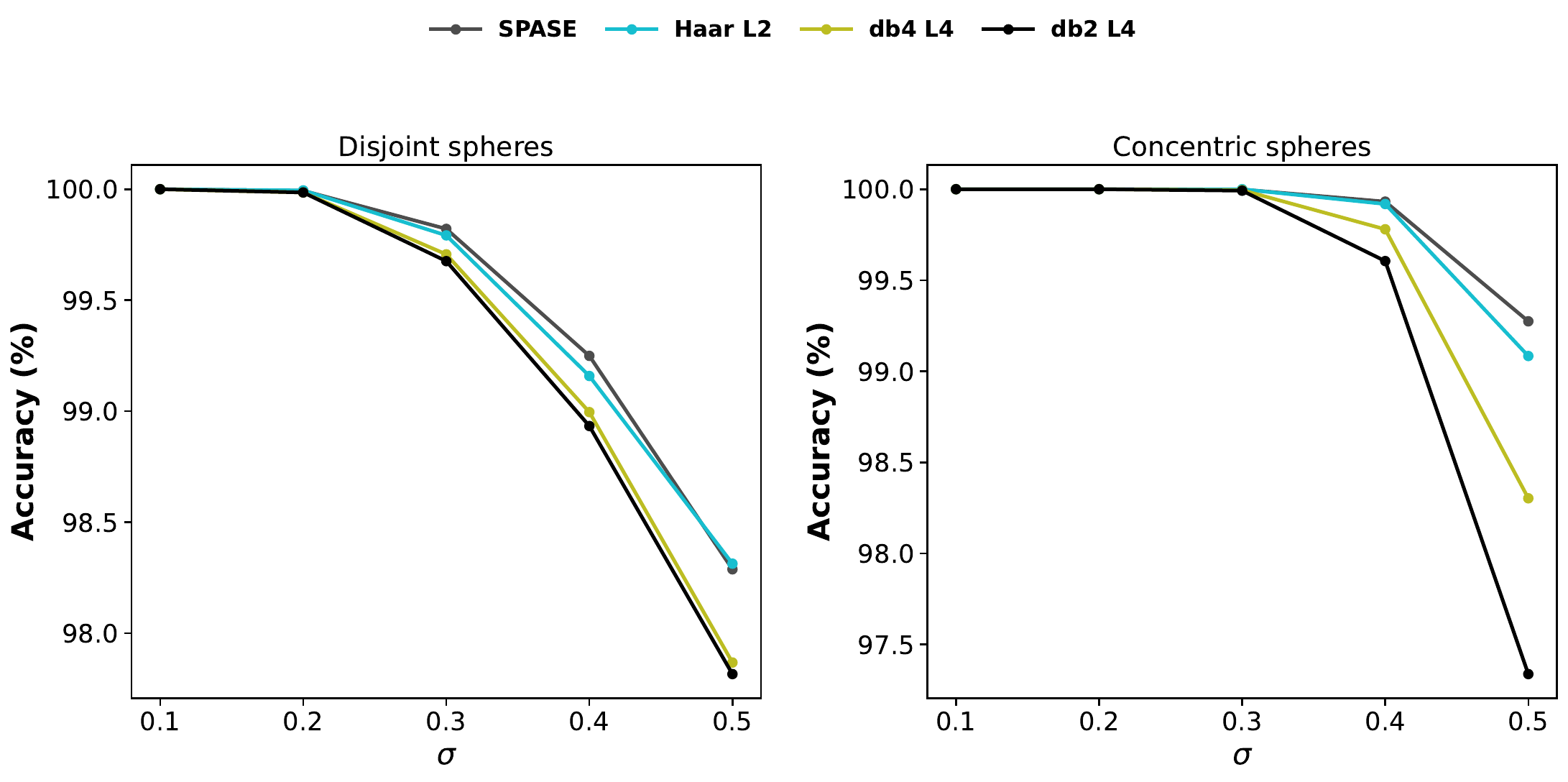}
\caption{Classification accuracy under basis misspecification for disjoint spheres (left) and concentric spheres (right). SPASE uses the matched Haar-L4 basis; Haar L2, db4 L4, and db2 L4 use the alternative working bases. The reported results are based on $250$ repetitions.}
\label{fig:basis_misspecification}
\end{figure}

Figure~\ref{fig:basis_misspecification} shows that SPASE maintains high classification accuracy with all four bases across the tested noise levels. For disjoint spheres at $\sigma=0.5$, the accuracies for Haar L4, Haar L2, db4 L4, and db2 L4 are $98.29\%$, $98.31\%$, $97.87\%$, and $97.82\%$, respectively. For comparison, KSE attains $98.18\%$, comparable to these four implementations. In the more challenging concentric-sphere setting at $\sigma=0.5$, the corresponding accuracies are $99.27\%$, $99.08\%$, $98.30\%$, and $97.34\%$, while KSE attains $50.38\%$. Thus, although changing the working basis leads to some loss of accuracy in the concentric-sphere setting, SPASE retains a substantial advantage over KSE and other methods. These results indicate that SPASE can maintain strong performance even when the working basis differs from the signal-generating basis.

\subsection[Choice of Projection Dimension]{Choice of Projection Dimension}
\label{sec:dimension_behavior}

We conduct two experiments to examine the choice of projection dimension. The first studies how the dimension $\Ks$ selected by Algorithm~\ref{alg:Kselect} adapts to sparsity and noise. The second evaluates classification accuracy across fixed projection dimensions and examines whether the adaptive selection achieves near-optimal downstream performance among them.

For the first experiment, we consider the two-disjoint-spheres setting and follow the setup in Section~\ref{sec:simu_setup}, with $q\in\{0.2,0.3,\ldots,1.6\}$ and $\sigma\in\{0.1,0.15,0.2\}$. For each configuration, we apply Algorithm~\ref{alg:Kselect} over $250$ repetitions and report the median selected dimension.

\begin{figure}[!htbp]
    \centering
    \includegraphics[width=0.55\linewidth]{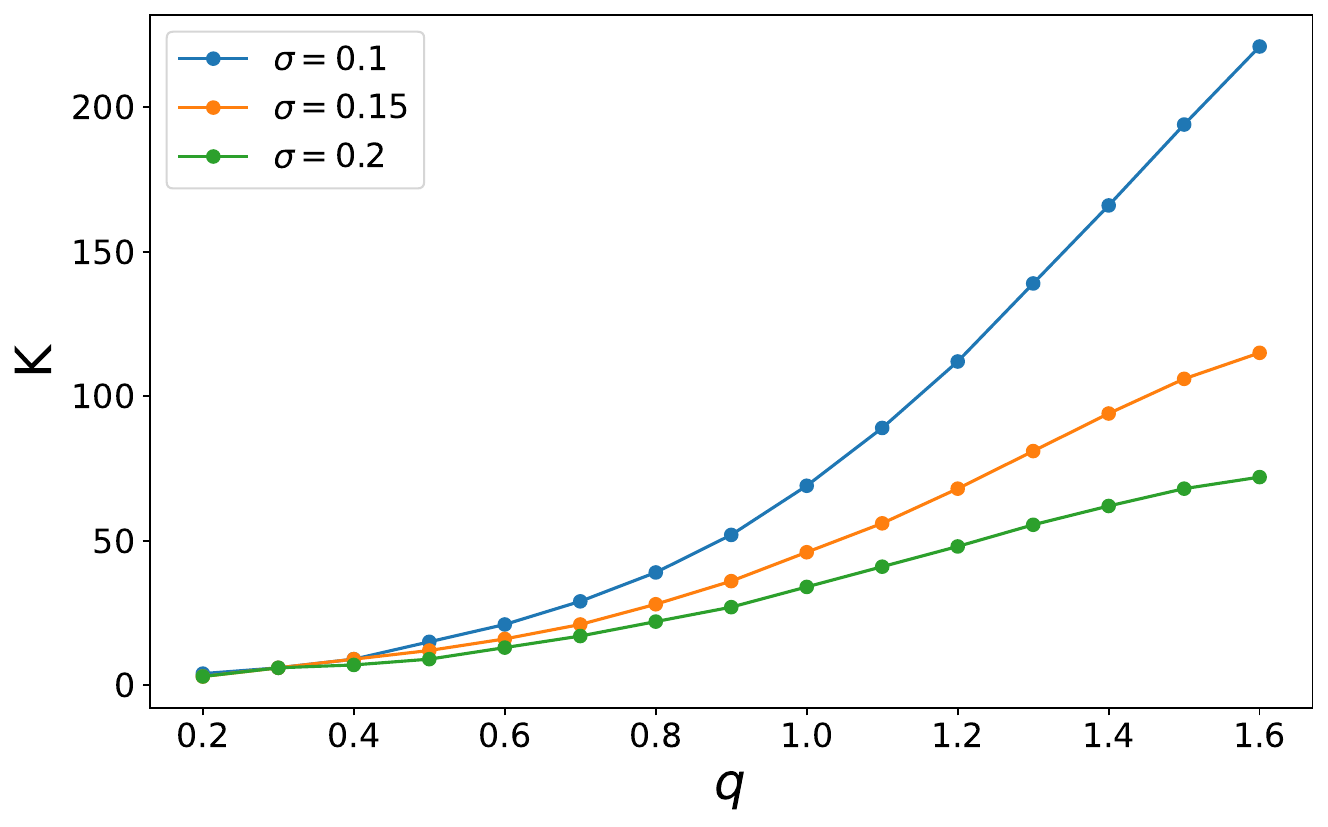}
    \caption{Selected projection dimension $\Ks$ versus sparsity parameter $q$ under different $\sigma$ in the two-disjoint-spheres model. The reported results are based on $250$ repetitions.}
\label{fig:sim_k_vs_q}
\end{figure}

Figure~\ref{fig:sim_k_vs_q} shows two clear patterns. First, the median selected dimension increases with $q$. As the coefficient magnitudes $\ell^{-1/q}$ decay more slowly, the signal energy is spread across more basis directions, and the procedure retains more directions. Second, at a given $q$, the median selected dimension generally decreases as the noise level increases. These patterns agree with Proposition~\ref{prop:K}, which characterizes how the selected dimension depends jointly on sparsity and the signal-to-noise ratio.

We next examine classification accuracy as the projection dimension varies in the concentric-sphere setting at $\sigma=0.3$ and $0.5$. We compare energy-based projection, PCA, and random orthonormal projection at fixed dimensions $k$ ranging from $1$ to $512$, each followed by KSE. At each fixed $k$, the energy-based projection retains the $k$ directions with the largest empirical energies as in \eqref{eq:Ej}. We also report the performance of SPASE, which selects $\Ks$ separately in each repetition, and use direct KSE as a reference. All methods use the same observations and training--testing splits.

\begin{figure}[!htbp]
\centering
\includegraphics[width=\linewidth]{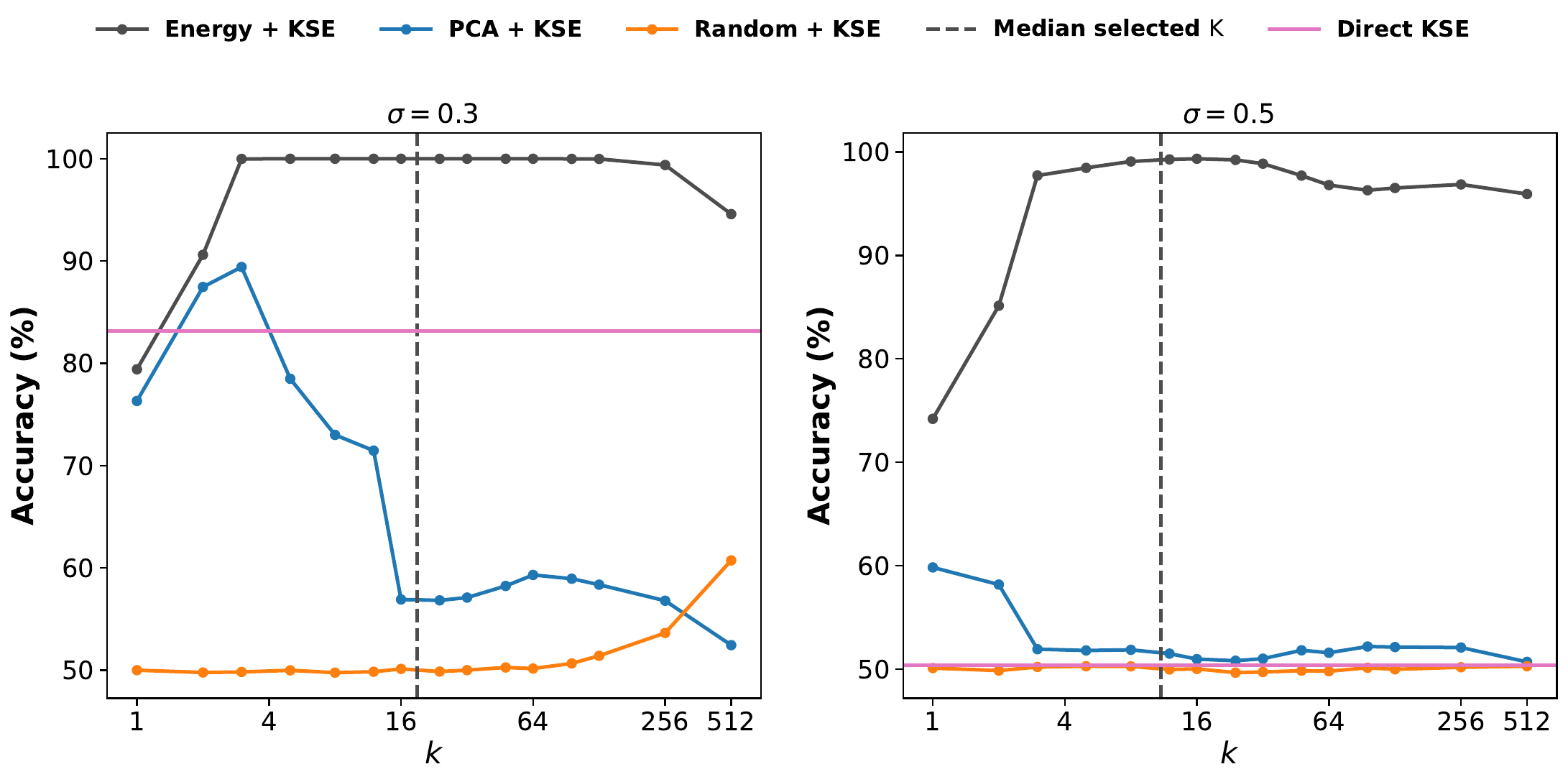}
\caption{Concentric-sphere classification accuracy versus the imposed projection dimension $k$. The vertical dashed line marks the median dimension $\Ks$ selected by SPASE ($19$ at $\sigma=0.3$ and $11$ at $\sigma=0.5$). The horizontal line gives the accuracy of direct KSE. The reported results are based on $250$ repetitions.}
\label{fig:accuracy_vs_K}
\end{figure}

Figure~\ref{fig:accuracy_vs_K} shows that energy-based projection followed by KSE achieves higher mean classification accuracy than PCA+KSE and random projection followed by KSE at every tested dimension for both noise levels. At $\sigma=0.3$, adaptive SPASE selects a median dimension of $19$ and attains nearly $100\%$ accuracy; all tested dimensions between $3$ and $128$ also yield nearly perfect accuracy. At $\sigma=0.5$, the median selected dimension is $11$, and adaptive SPASE achieves $99.27\%$ accuracy, close to the highest mean accuracy on the tested grid, $99.34\%$ at $k=16$. The tested dimensions between $8$ and $32$ yield accuracies ranging from $98.87\%$ to $99.34\%$. At both noise levels, the median dimension selected by Algorithm~\ref{alg:Kselect} lies within a broad high-accuracy region, and SPASE performs close to the best accuracy on the tested grid.

These findings indicate both the advantage of energy-based projection and the effectiveness of the adaptive dimension selection. Moreover, as discussed in Section~\ref{sec:projection}, $\tau$ as in \eqref{eq:Ks} controls the energy-based selection cutoff. The broad high-accuracy regions suggest that moderate changes in $\tau$ may have little effect on performance when the selected dimensions remain within these regions. This further demonstrates the robustness of SPASE.

\section[Data Application]{Data Application}
\label{sec:real_data}

\subsection{Data Description and Setup}
Histopathology images exhibit complex local texture patterns together with substantial variability introduced by staining conditions, tissue preparation, and imaging artifacts. Although these images are represented in a high-dimensional pixel space, their intrinsic variability is believed to be governed by a much lower-dimensional manifold reflecting tissue morphology. Consequently, histopathology provides an appropriate setting for evaluating whether adaptive projection can improve the stability of manifold estimation under controlled additive perturbations of histopathology images.

To assess the practical performance of the proposed method, we use the PathMNIST dataset from the MedMNIST collection \cite{yang2021medmnist,yang2023medmnist}. MedMNIST is an open-sourced lightweight benchmark suite designed for standardized evaluation on medical imaging tasks, and PathMNIST is one of its most widely used subsets. PathMNIST consists of hematoxylin and eosin (H\&E) stained histological images of colorectal tissue and defines a $9$-class classification problem. The class labels are: $0$ (adipose), $1$ (background), $2$ (debris), $3$ (lymphocytes), $4$ (mucus), $5$ (smooth muscle), $6$ (normal colon mucosa), $7$ (cancer-associated stroma), and $8$ (colorectal adenocarcinoma epithelium). Each image has resolution $28\times 28$ with three color channels, so that after vectorization each observation lies in $\R^{2352}$. This dataset provides a useful real-data testbed for our method: the observations are high-dimensional and strongly structured, and the downstream classification tasks remain sensitive to the quality of the learned representation. The adaptive projection is expected to suppress noise-dominated wavelet directions while preserving spatially localized morphological structures, thereby producing a more stable affinity graph for the subsequent spectral embedding. We therefore expect the proposed method to be particularly useful for this application, where robust recovery of the intrinsic data geometry is critical for downstream classification.

\begin{figure}[htbp]
  \centering
  \begin{subfigure}[t]{0.235\textwidth}
    \centering
    \includegraphics[width=\linewidth]{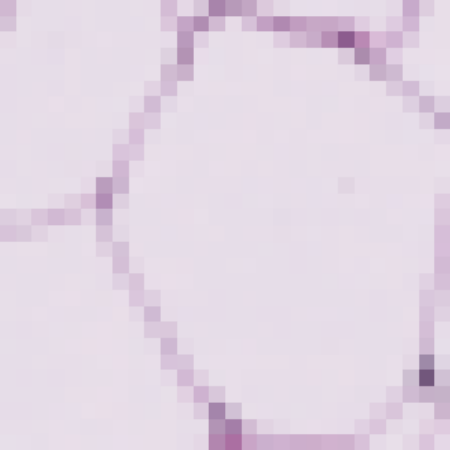}
    \caption{Class 0}
  \end{subfigure}
  \hfill
  \begin{subfigure}[t]{0.235\textwidth}
    \centering
    \includegraphics[width=\linewidth]{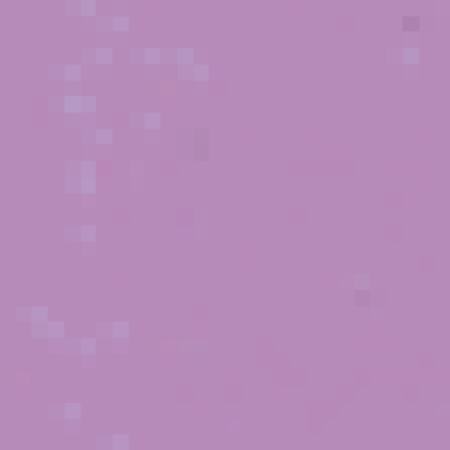}
    \caption{Class 1}
  \end{subfigure}
  \hspace{0.6em}
  \hfill
  \begin{subfigure}[t]{0.235\textwidth}
    \centering
    \includegraphics[width=\linewidth]{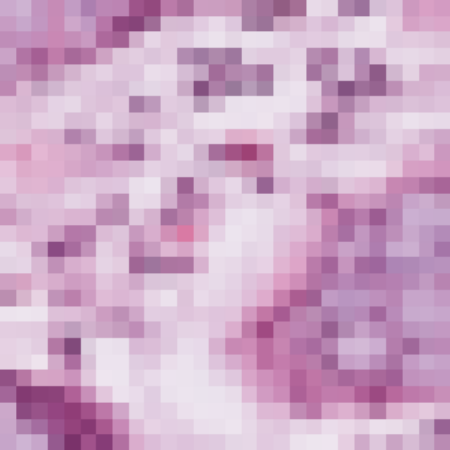}
    \caption{Class 6}
  \end{subfigure}
  \hfill
  \begin{subfigure}[t]{0.235\textwidth}
    \centering
    \includegraphics[width=\linewidth]{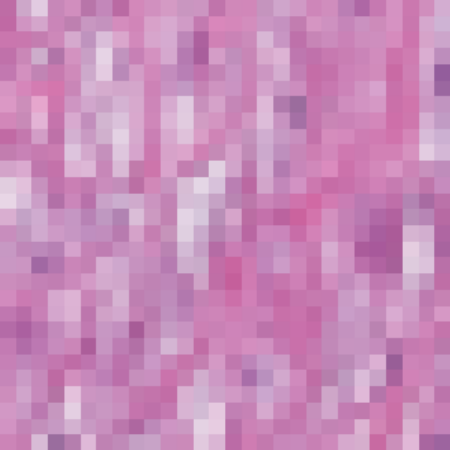}
    \caption{Class 7}
  \end{subfigure}
  \caption{Example of PathMNIST images from classes $0$ (adipose), $1$ (background), $6$ (normal colon mucosa), and $7$ (cancer-associated stroma).}
  \label{fig:pathmnist-demo}
\end{figure}

We evaluate the competing methods on downstream binary classification tasks formed by selecting two pairs of PathMNIST classes: $0$ vs.\ $1$ (adipose vs.\ background) and $6$ vs.\ $7$ (normal colon mucosa vs.\ cancer-associated stroma). The two class pairs were selected to represent different levels of classification difficulty. Classes 0 and 1 exhibit markedly different visual appearances, providing an easier discrimination task. In contrast, classes 6 and 7 share similar histological textures and differ through relatively subtle structural patterns, making them considerably more challenging and therefore a more stringent test of representation quality. Figure~\ref{fig:pathmnist-demo} shows examples from each class pair. We evaluate the proposed method against the following competing methods: Isomap, LLE, t-SNE, Diffusion Maps (DM), autoencoders (AE), denoising autoencoders (DAE), and KSE.

We consider two complementary evaluation tasks. In the first, we study robustness to ambient noise: for each class pair, we sample $500$ observations from each class, add independent Gaussian noise with varying standard deviation $\sigma$, and report the resulting classification accuracy. In the second, we study sample efficiency: we vary the number of sampled observations per class, fix the noise level at $\sigma=0.3$, and again report the resulting classification accuracy. In the two tasks, the goal is to assess whether the learned representation remains informative when the data are either increasingly corrupted or limited in number.

In all experiments, we use a $70/30$ training--testing split and use the vectorized two-dimensional discrete Haar wavelet basis induced by level-3 discrete multiresolution decomposition as the basis $\Ws$ in our proposed method. We use a downstream classifier with the same one-hidden-layer, $16$-ReLU-unit architecture as in Section~\ref{sec:simu_setup}. For each setup, we conduct the analysis over $100$ independent resamples of the data and report the mean performance across these runs.

\subsection{Accuracy of Classification}

\begin{figure}[!htbp]
    \centering
    \begin{subfigure}[t]{0.48\textwidth}
        \centering
        \includegraphics[width=\linewidth]{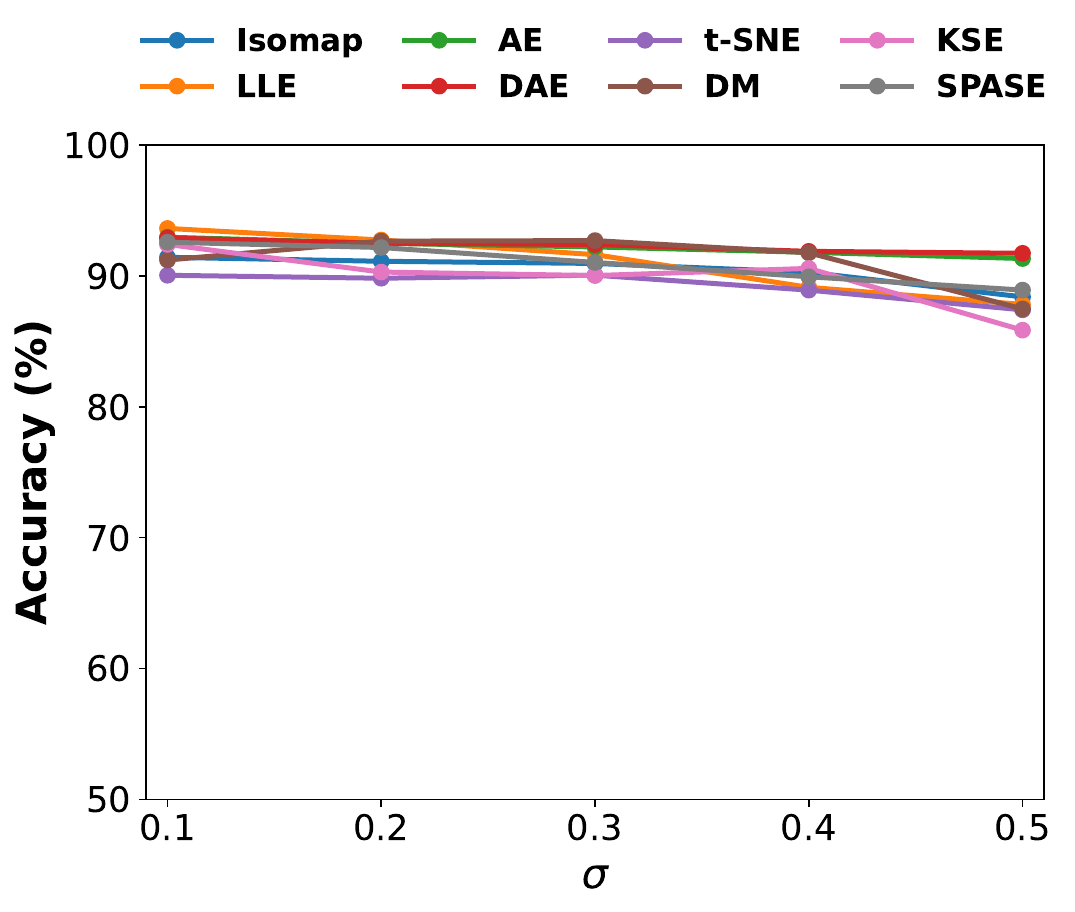}
        \caption{Accuracy vs. Noise Level}
    \end{subfigure}\hfill
    \begin{subfigure}[t]{0.48\textwidth}
        \centering
        \includegraphics[width=\linewidth]{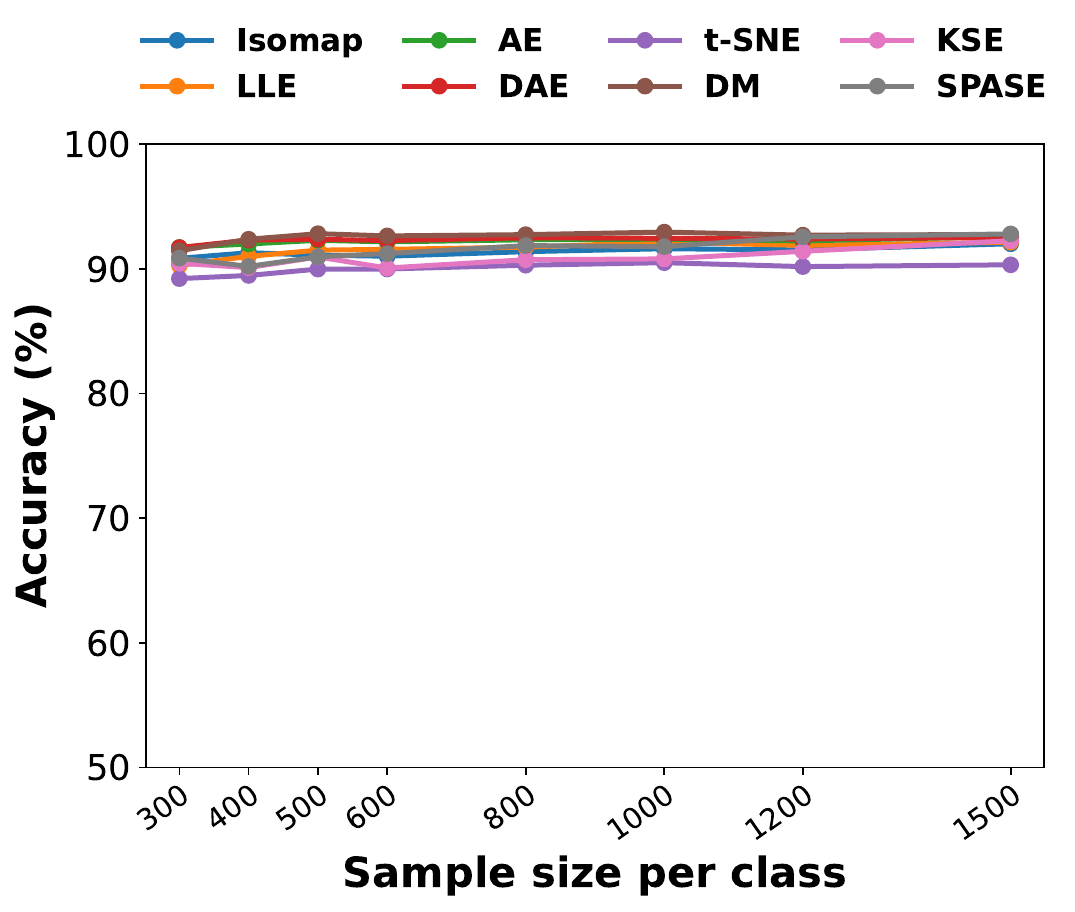}
        \caption{Accuracy vs. Sample Size}
    \end{subfigure}
    \caption{Classification accuracy for PathMNIST classes $0$ and $1$ (adipose vs.\ background). The left panel shows performance under increasing noise with $500$ samples per class; the right panel shows performance under varying sample sizes with fixed noise level $\sigma=0.3$.}
    \label{fig:pathmnist_0_1}
\end{figure}

\begin{figure}[!htbp]
    \centering
    \begin{subfigure}[t]{0.48\textwidth}
        \centering
        \includegraphics[width=\linewidth]{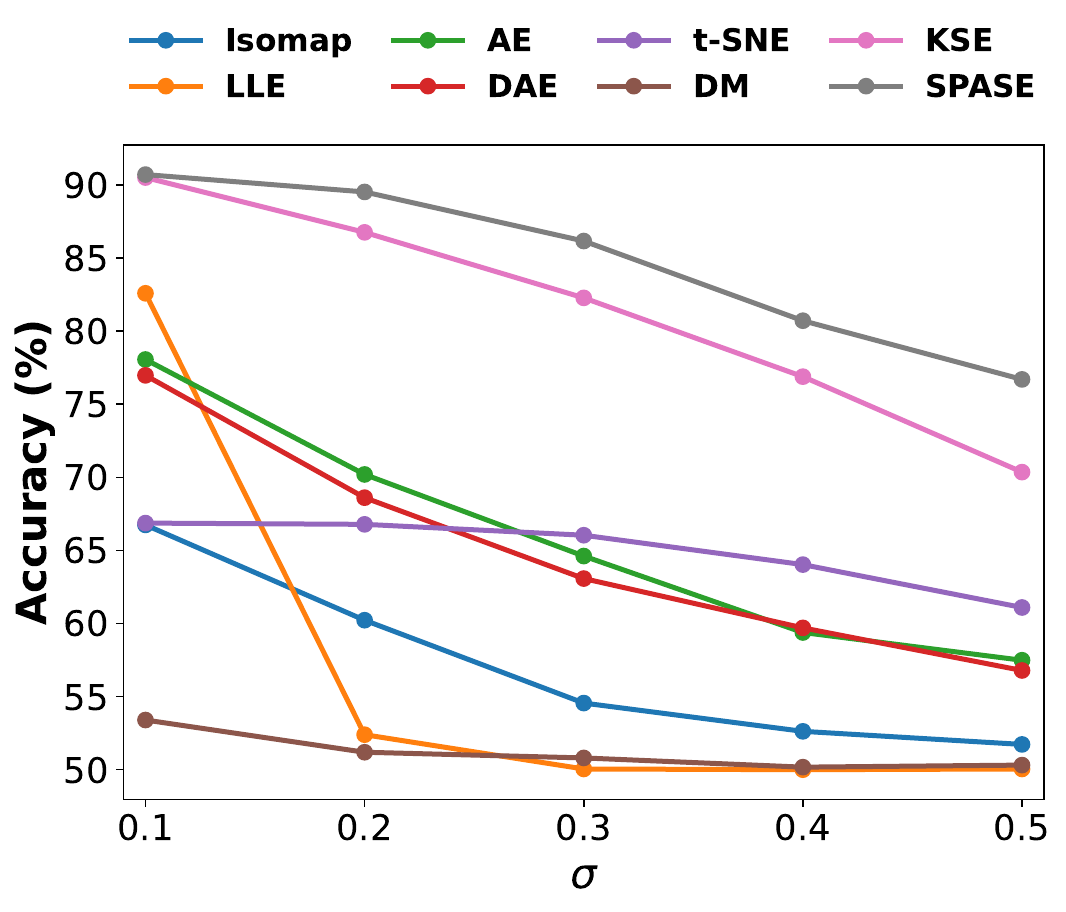}
        \caption{Accuracy vs. Noise Level}
    \end{subfigure}\hfill
    \begin{subfigure}[t]{0.48\textwidth}
        \centering
        \includegraphics[width=\linewidth]{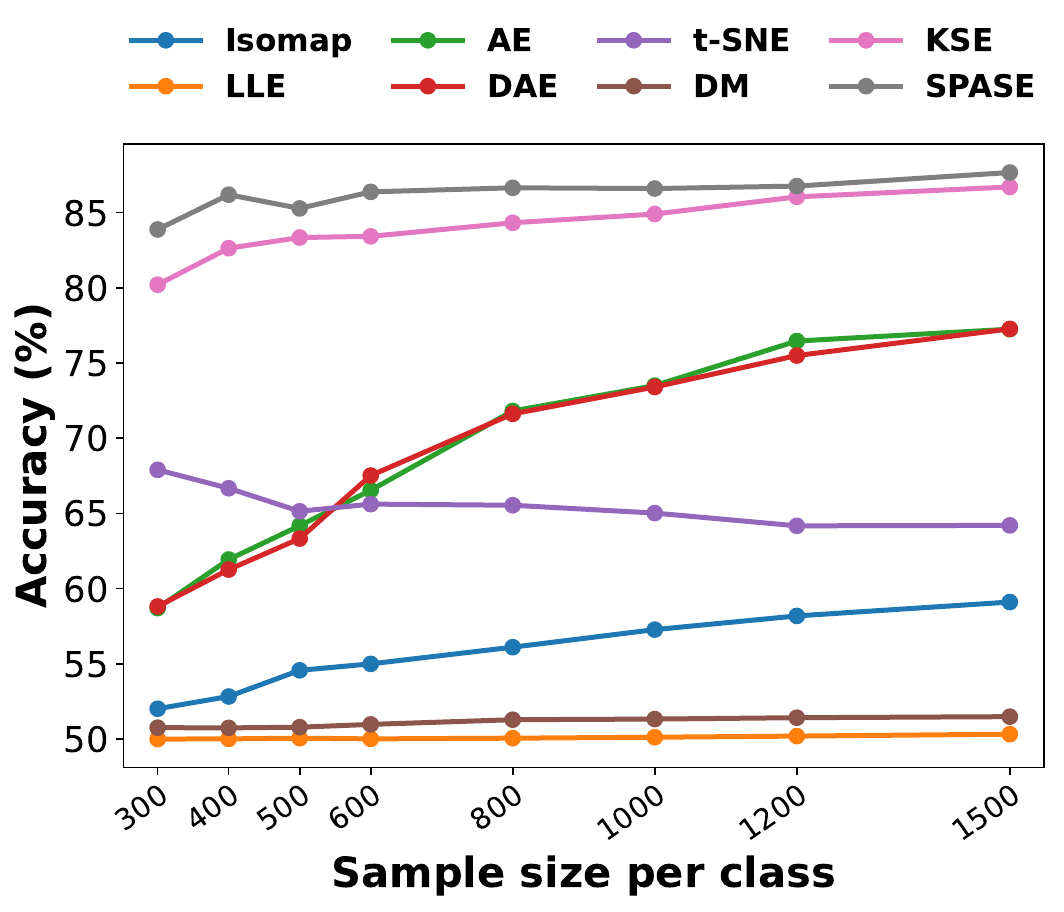}
        \caption{Accuracy vs. Sample Size}
    \end{subfigure}
    \caption{Classification accuracy for PathMNIST classes $6$ and $7$ (normal colon mucosa vs.\ cancer-associated stroma). The left panel shows performance under increasing noise with $500$ samples per class; the right panel shows performance under varying sample sizes with fixed noise level $\sigma=0.3$.}
    \label{fig:pathmnist_6_7}
\end{figure}

The results for the PathMNIST class pairs are reported in Figures~\ref{fig:pathmnist_0_1} and \ref{fig:pathmnist_6_7}. SPASE is competitive on the easier pair and has a larger advantage over KSE on the harder pair as the noise level increases.

Figure~\ref{fig:pathmnist_0_1} corresponds to the easy task classes $0$ vs.\ $1$ (adipose vs.\ background). In this case, all methods achieve relatively high classification accuracy, and SPASE remains consistently competitive and stable across both the noise-robustness and sample-size experiments.

Figure~\ref{fig:pathmnist_6_7} corresponds to the difficult task classes $6$ vs.\ $7$ (normal colon mucosa vs.\ cancer-associated stroma). In this setting, SPASE consistently outperforms the competing methods as the noise level increases. At $\sigma=0.1$, SPASE and KSE, the strongest competing method, achieve comparable classification accuracies of $90.7\%$ and $90.5\%$, respectively. At the highest noise level $\sigma=0.5$, however, SPASE maintains an accuracy of $76.7\%$, compared with $70.4\%$ for KSE, corresponding to an improvement of approximately $6.3$ percentage points.

A similar pattern is observed in the sample-size experiment. With only $300$ observations per class, SPASE achieves an accuracy of $83.9\%$, compared with $80.2\%$ for KSE, an improvement of approximately $3.7$ percentage points. As the sample size increases, the gap narrows: with $1500$ observations per class, SPASE and KSE achieve accuracies of $87.7\%$ and $86.7\%$, respectively. These comparisons indicate that the benefit of adaptive projection is particularly visible in the more challenging regimes of stronger noise and limited sample size.

For the tissue images, we do not know in advance which basis best concentrates the signal. We chose the level-3 two-dimensional Haar basis because its detail coefficients describe local intensity contrasts at several spatial scales. This gives us a way to represent tissue texture without choosing the basis from the class labels or the classification results. In Section~\ref{sec:basis_misspecification}, several working bases gave high accuracy even though only one matched the generating basis. Here the same Haar basis was used for both tissue pairs, and the energy-based rule determined which of its directions to retain.

The comparison of normal colon mucosa with cancer-associated stroma illustrates the proposed role of screening. The two tissues have similar textures. Direct KSE uses distances between the full images, so differences due to noise in many pixels can accumulate in each squared distance. SPASE first screens the Haar directions by their variation across images. If the local contrasts that distinguish the tissues vary enough to be retained, weaker directions can be discarded before their noise enters the distance calculation. The near tie under mild added noise and the gain of $6.3$ percentage points over KSE at $\sigma=0.5$ in Figure~\ref{fig:pathmnist_6_7} are consistent with this explanation.

\bibliographystyle{abbrv}

\newpage

\begin{center}
{\Large \bf Supplement to ``Adaptive Sparsity-Aware Projection-Assisted Spectral Embedding for High-Dimensional Noisy Data with Nonlinear Structures''}
\end{center}
\appendix
\renewcommand{\thealgorithm}{S.\arabic{algorithm}}
\setcounter{algorithm}{0}
\renewcommand{\theHalgorithm}{supp.\arabic{algorithm}}
\section{Additional Simulation Results}
\label{app:additional_simulation}

In this appendix, we report additional downstream classification and clustering results for the two-moons model, which are omitted from the main text for brevity. The data are generated in the same way as in Section~\ref{sec:simu_setup}, with the following two-moons latent coordinates. More specifically, the first two latent coordinates are generated from a two-moons model (see, e.g., \cite{pedregosa2011scikit}), while the third coordinate is independently sampled from a uniform distribution on $[-0.15,0.15]$.

\begin{figure}[H]
    \centering
    \begin{subfigure}[t]{0.48\textwidth}
        \centering
        \includegraphics[width=\linewidth]{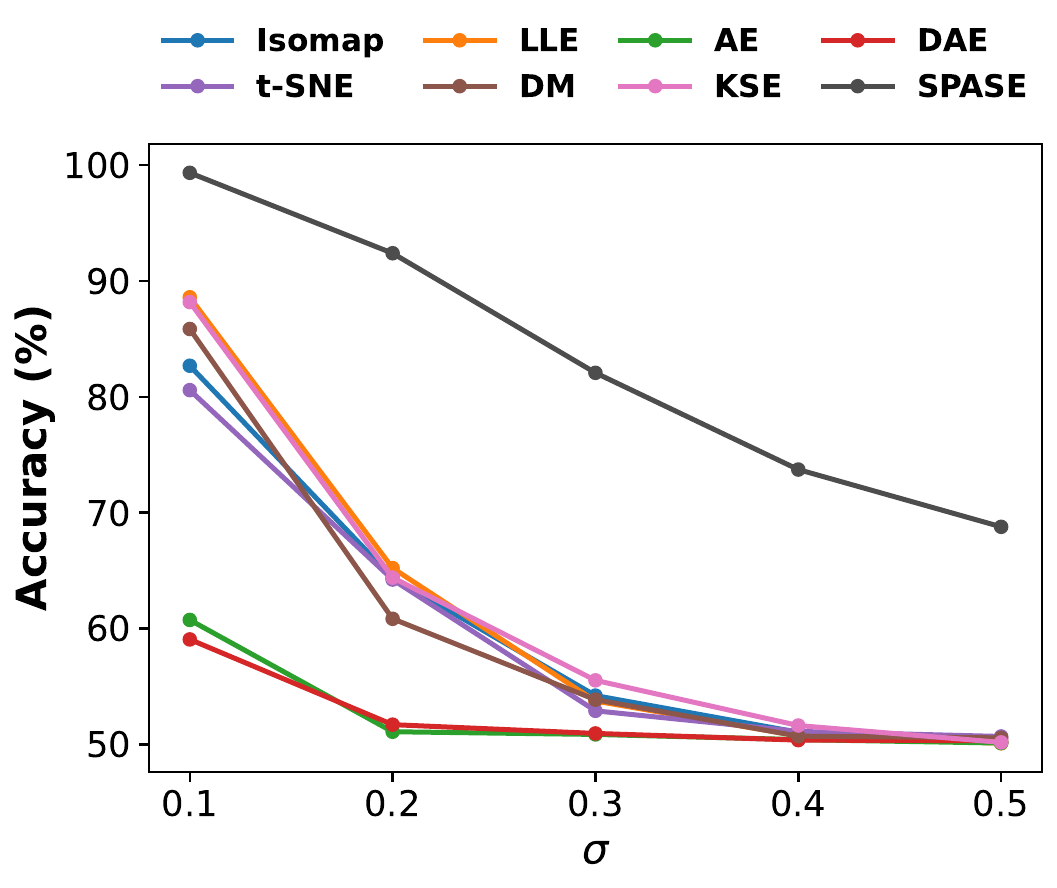}
        \caption{Classification Accuracy}
    \end{subfigure}\hfill
    \begin{subfigure}[t]{0.48\textwidth}
        \centering
        \includegraphics[width=\linewidth]{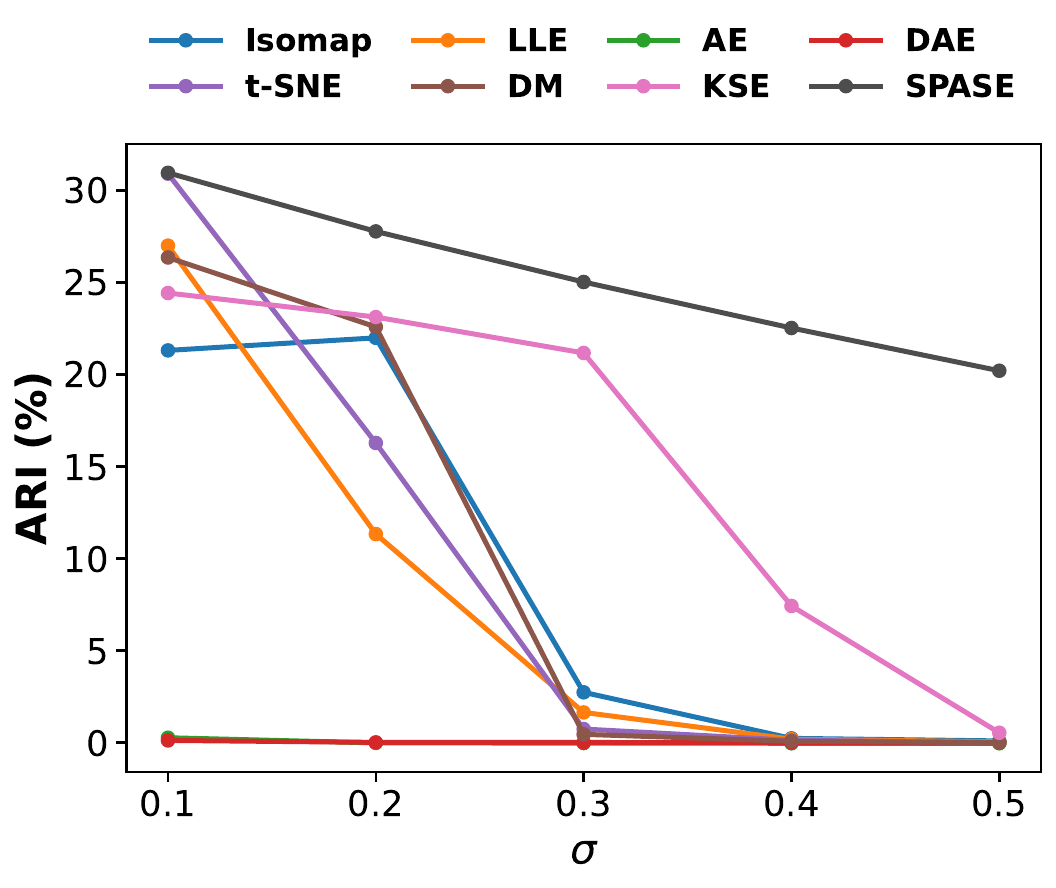}
        \caption{Clustering ARI}
    \end{subfigure}
    \caption{Classification accuracy and clustering ARI for the two-moons model across varying noise levels. The reported results are based on $250$ repetitions.}
    \label{fig:results_two_moons}
\end{figure}

Figure~\ref{fig:results_two_moons} reports the downstream classification accuracy and clustering ARI across varying noise levels. For classification, SPASE achieves the highest mean accuracy among the plotted methods at each of the five noise levels. At $\sigma=0.5$, SPASE attains approximately $68.8\%$ accuracy, compared with $50.2\%$ for KSE, while most other baselines are close to chance.

For clustering, SPASE's mean ARI decreases from $0.310$ at $\sigma=0.1$ to $0.202$ at $\sigma=0.5$. At the lowest noise level, t-SNE has a similar ARI of $0.309$, but its performance deteriorates more rapidly as noise increases. At $\sigma=0.5$, KSE has an ARI of $0.005$, while most remaining baselines are close to zero. Although the absolute ARIs are lower than in the sphere examples, SPASE retains more clustering information as the noise increases.

On the two-moons data, SPASE shows strong performance in both classification and clustering across the tested noise levels, which agrees with our results in Section~\ref{sec:sim_downstream} of the manuscript.

\section{Additional Real-Data Results}
\label{app:additional_real_data}

In this appendix, we report additional real-data results for the PathMNIST classification task formed by classes $2$ and $8$, corresponding to debris and colorectal adenocarcinoma epithelium, respectively. This class pair is omitted from the main text for brevity.

\begin{figure}[!htbp]
    \centering
    \begin{subfigure}[t]{0.48\textwidth}
        \centering
        \includegraphics[width=\linewidth]{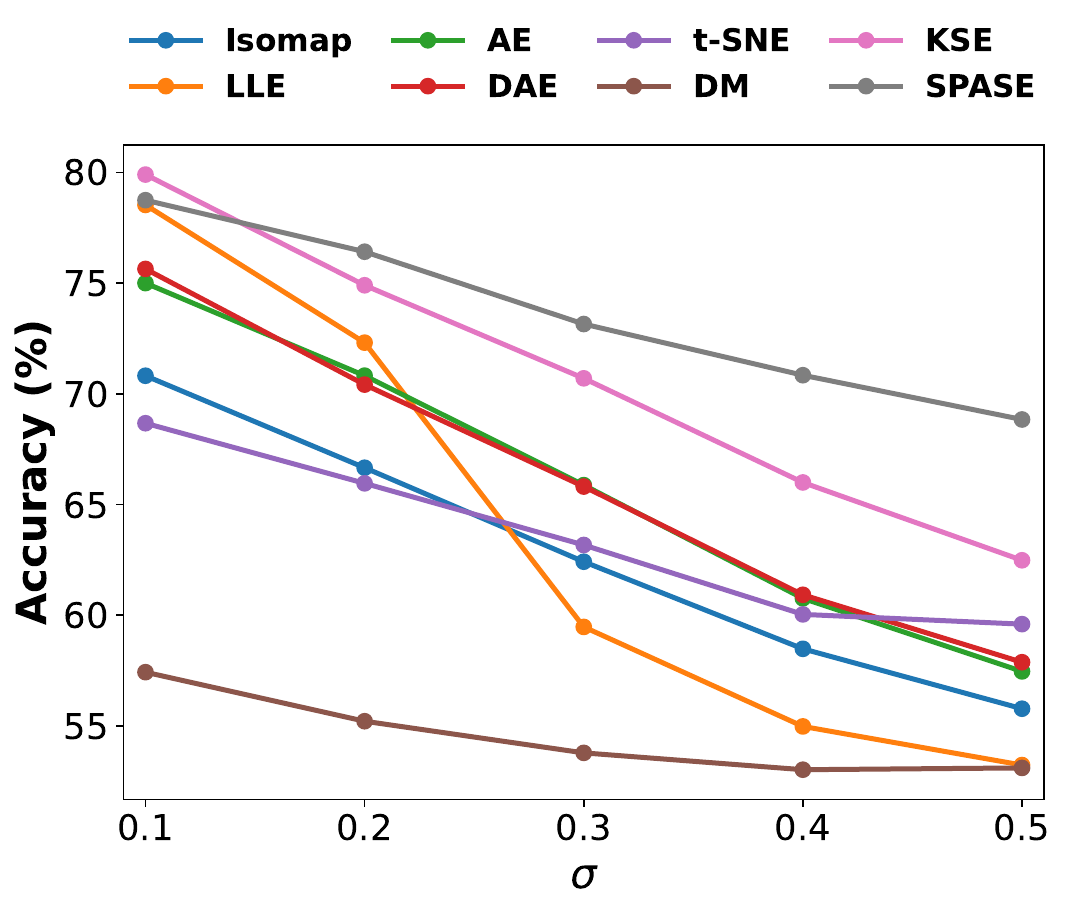}
        \caption{Accuracy vs. Noise Level}
    \end{subfigure}\hfill
    \begin{subfigure}[t]{0.48\textwidth}
        \centering
        \includegraphics[width=\linewidth]{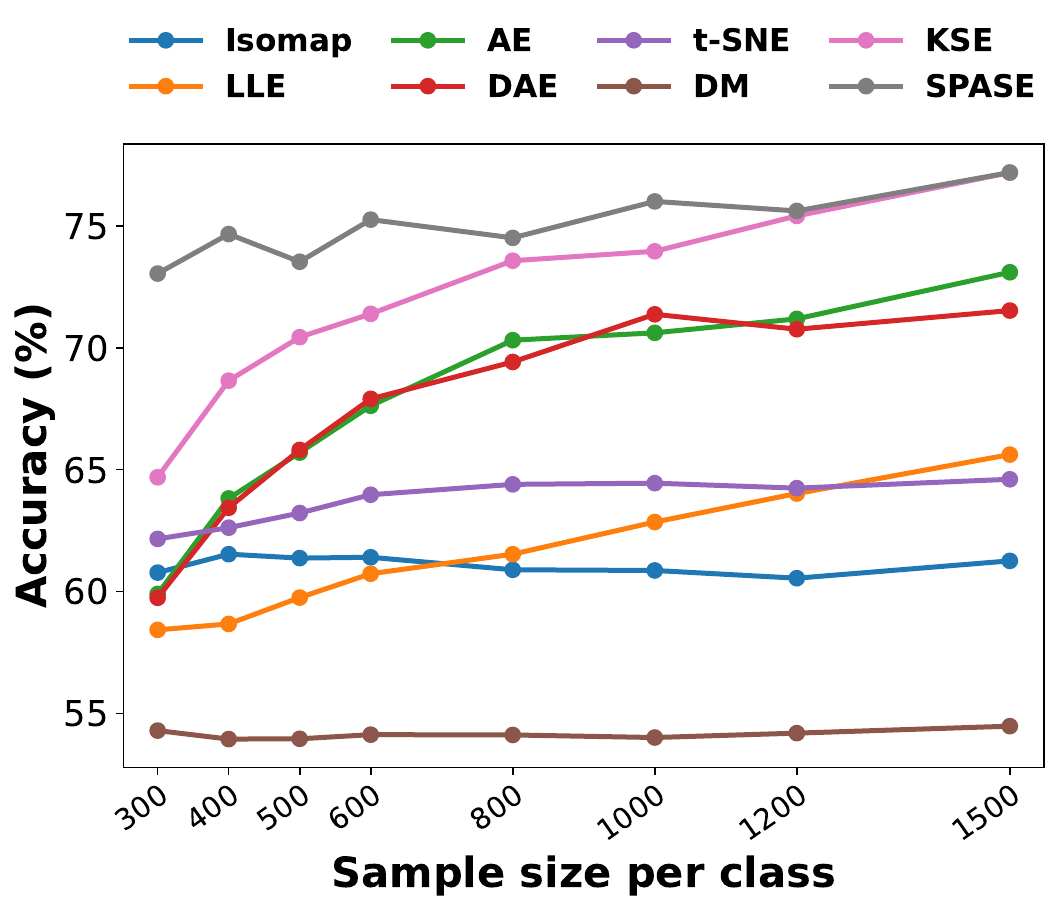}
        \caption{Accuracy vs. Sample Size}
    \end{subfigure}
    \caption{Classification accuracy for PathMNIST classes $2$ and $8$ (debris vs.\ colorectal adenocarcinoma epithelium). The left panel shows performance under increasing noise with $500$ samples per class; the right panel shows performance under varying sample sizes with fixed noise level $\sigma=0.3$.}
    \label{fig:pathmnist_2_8}
\end{figure}

Figure~\ref{fig:pathmnist_2_8} reports the results for classes $2$ vs.\ $8$ (debris vs.\ colorectal adenocarcinoma epithelium). As the noise level increases, most baseline methods deteriorate steadily, whereas the proposed method has the highest mean accuracy at the four larger noise levels. KSE is again the closest competing baseline and is slightly more accurate than SPASE at $\sigma=0.1$. The same pattern appears in the sample-size experiment: SPASE leads at smaller sample sizes, but its mean accuracy is nearly identical to that of KSE at $1500$ observations per class. Overall, SPASE is among the most accurate methods across both the noise-robustness and sample-size experiments.

\section{Technical Proofs}
\label{sec:proofs}

\subsection{Proofs for Recovery of Pairwise Distances}

In this subsection, we prove Proposition~\ref{prop:K} and Theorem~\ref{thm:pairwise}. Define
\begin{align}
    A \coloneq \Ws^\top(\rs_1,\dots,\rs_r)\in\R^{p\times r},
    \label{eq:appendix-A}
\end{align}
so that, for each $k=1,\dots,r$,
\begin{align}
    \rs_k=\sum_{j=1}^p a_{jk}\ws_j.
    \label{eq:appendix-rk}
\end{align}
Let $A_{j\cdot}$ denote the $j$-th row of $A$, and define the Euclidean row norms by
\begin{align}
    \rho_j\coloneq \|A_{j\cdot}\|_2,\qquad j=1,\dots,p.
    \label{eq:energy}
\end{align}

\begin{lemma}[Decay of row norms]
    \label{lem:row-weak-lq}
    Under Assumption~\ref{ass:lq}, there exist constants $0<c_-\le c_+<\infty$, depending only on $(c_k)_{k\in[r]}$, $(C_k)_{k\in[r]}$, $q$, and $r$, such that
    $$
    c_- m^{-1/q}\le \rho_{(m)}\le c_+ m^{-1/q},\qquad m=1,\dots,p,
    $$
    where $\rho_{(1)}\ge \rho_{(2)}\ge\cdots\ge \rho_{(p)}$ denotes the nonincreasing rearrangement of $\{\rho_j\}_{j=1}^p$.
\end{lemma}

\begin{proof}
    We first prove the upper bound. For $t>0$, define
    \begin{align}
        S_t\coloneq \{j:\rho_j\ge t\}.
        \label{eq:A.1.1}
    \end{align}
    For any $j\in S_t$, we have $\rho_j^2=\sum_{k=1}^r a_{jk}^2\ge t^2$, so there exists some $k\in[r]$ such that $|a_{jk}|\ge t/\sqrt r$. Hence
    $$
    S_t\subseteq \bigcup_{k=1}^r\{j:|a_{jk}|\ge t/\sqrt r\},
    $$
    and therefore
    \begin{align}
        |S_t|\le \sum_{k=1}^r \#\{j:|a_{jk}|\ge t/\sqrt r\}.
        \label{eq:A.1.2}
    \end{align}
    By Assumption~\ref{ass:lq}, for each $k\in[r]$ and $u>0$,
    $$
    \#\{j:|a_{jk}|\ge u\}\le (C_k/u)^q.
    $$
    Taking $u=t/\sqrt r$, we obtain
    $$
    \#\{j:|a_{jk}|\ge t/\sqrt r\}\le (C_k\sqrt r/t)^q.
    $$
    Combining this with \eqref{eq:A.1.2}, we get
    $$
    |S_t|\le \sum_{k=1}^r (C_k\sqrt r/t)^q\le (c_+/t)^q
    $$
    for some constant $c_+>0$ depending only on $(C_k)_{k\in[r]}$, $q$, and $r$. Setting $t=\rho_{(m)}$ yields
    $$
    m\le |S_{\rho_{(m)}}|\le (c_+/\rho_{(m)})^q,
    $$
    which implies
    $$
    \rho_{(m)}\le c_+ m^{-1/q}.
    $$

    We next prove the lower bound. For each $j=1,\dots,p$ and each $k\in[r]$, we have
    $$
    \rho_j=\Bigl(\sum_{\ell=1}^r a_{j\ell}^2\Bigr)^{1/2}\ge |a_{jk}|.
    $$
    Therefore, for every fixed $k\in[r]$, the decreasing rearrangements satisfy
    $$
    \rho_{(m)}\ge |a_{(m)k}|,\qquad m=1,\dots,p.
    $$
    By Assumption~\ref{ass:lq},
    $$
    |a_{(m)k}|\ge c_k m^{-1/q}.
    $$
    Hence, for each $k\in[r]$,
    $$
    \rho_{(m)}\ge c_k m^{-1/q}.
    $$
    This implies
    $$
    \rho_{(m)}\ge c_- m^{-1/q},\qquad m=1,\dots,p,
    $$
    for some constant $c_->0$. Combining the upper and lower bounds completes the proof.
\end{proof}

Lemma~\ref{lem:row-weak-lq} shows that the ordered row norms $\rho_{(m)}$ inherit the same decay. We next show that the empirical coordinate energies concentrate uniformly around their population counterparts, which allows Algorithm~\ref{alg:Kselect} to recover the correct order of the selected projection dimension. Recall that $\Sigma=\Cov(\sb_i^0)$ with $\sb_i^0$ as in \eqref{eq:rotate}. The population variance corresponding to the sample energy $E_j$ in \eqref{eq:Ej} is
\begin{align}
    \kappa_j\coloneq\operatorname{Var}(\xb_i^\top\ws_j)=\sigma^2+A_{j\cdot}\Sigma A_{j\cdot}^\top,\qquad j=1,\dots,p.
    \label{eq:muj}
\end{align}

Because the energy uses denominator $n$ after sample centering, $\mathbb E E_j=(1-1/n)\kappa_j$. If $\kappa_j=0$, then $E_j=0$ almost surely; ratios below are interpreted as zero on those coordinates.

\begin{remark}[Sub-Gaussian convention]\label{rem:noise-norm}
Throughout the asymptotic analysis, the reduced law $\mu$ and its support $\M^0$ are independent of $n$. For a scalar random variable $Z$, write
\[
\|Z\|_{\psi_2}=\inf\{t>0:\mathbb E\exp(Z^2/t^2)\le2\}.
\]
Throughout, the sub-Gaussian noise condition in Assumption~\ref{ass:signal} is understood as
$\|z_{ij}\|_{\psi_2}\le L\sigma$ for all $n,i,j$, for a fixed $L$.
This is the usual standardized sub-Gaussian control in concentration
arguments; see \cite{vershynin2018high}. It agrees with the noise scaling
used in the high-dimensional kernel analysis of \cite{ding2023learning}.
Independence and centering of the noise entries imply
\[
\|a^\top\zb_i\|_{\psi_2}\le C L\sigma\|a\|_2,
\qquad a\in\R^p,
\]
with a constant independent of $n$ and $p$. The condition includes
Gaussian noise with covariance $\sigma^2\Ib_p$ and fixes the constants
in the concentration bounds below as the noise level varies.
\end{remark}

\begin{lemma}[Concentration of empirical energies]
    \label{lem:E-conc-rel-poly}
    Under Assumptions~\ref{ass:model} and \ref{ass:signal},
    $$
    \max_{1\le j\le p}\frac{|E_j-\kappa_j|}{\kappa_j}
    =
    \OO_\prec\!\left(\sqrt{\frac{\log n}{n}}\right).
    $$
\end{lemma}

\begin{proof}

Write $\bm m_0=\mathbb E\sb_i^0$ and $\bm m=\Rs_r\bm m_0=\mathbb E\xb_i$. Compactness in Assumption~\ref{ass:signal} gives $\|\sb_i^0\|_2\le R$ and $\|\sb_i^0-\bm m_0\|_2\le2R$ for a constant $R$. The covariance eigenvalues $\lambda_1^0,\dots,\lambda_r^0$ are positive constants, so $\Theta\asymp1$.

    Fix $j\in\{1,\dots,p\}$. Recall from \eqref{eq:Ej} that
    $$
    \begin{aligned}Y_{ij}&\coloneq(\xb_i-\bm m)^\top\ws_j,\qquad \bar Y_j=\frac1n\sum_iY_{ij},\\ E_j&=\widetilde E_j-\bar Y_j^2,\qquad \widetilde E_j=\frac1n\sum_iY_{ij}^2.\end{aligned}
    $$
    Write
    $$
    Y_{ij}=S_{ij}+Z_{ij},
    \qquad
    S_{ij}\coloneq(\sb_i-\bm m)^\top\ws_j,
    \qquad
    Z_{ij}\coloneq \zb_i^\top\ws_j.
    $$
    By construction, $\mathbb E Y_{ij}=\mathbb E S_{ij}=0$.

    We first control the noise part. By Assumption~\ref{ass:signal} and the convention in Remark~\ref{rem:noise-norm}, $\zb_i$ is a centered sub-Gaussian random vector with covariance matrix $\sigma^2\Ib_p$. Hence there exists a constant $C>0$ independent of $n$ and $p$ such that, for every deterministic vector $\mathbf{a}\in\R^p$ and every $t\in\R$,
    $$
    \E\exp\bigl(t\,\mathbf{a}^\top \zb_i\bigr)
    \le
    \exp\bigl(C\sigma^2 t^2\|\mathbf{a}\|_2^2\bigr).
    $$
    Taking $\mathbf{a}=\ws_j$ and using $\|\ws_j\|_2=1$, we obtain
    \begin{align}
        \E\exp(tZ_{ij})
        =
        \E\exp(t\ws_j^\top\zb_i)
        \le
        \exp(C\sigma^2 t^2),
        \qquad t\in\R.
        \label{eq:A.3.1}
    \end{align}

    We next control the signal part. By Assumption~\ref{ass:model} and \eqref{eq:rotate}, we have $\sb_i=\sum_{k=1}^r s_{ik}\rs_k$, and therefore
    $$
    S_{ij}=(\sb_i-\bm m)^\top\ws_j=A_{j\cdot}(\sb_i^0-\bm m_0).
    $$
    Using the bound $\|\sb_i^0-\bm m_0\|_2\le2R$ above, we obtain
    \begin{align}
        |S_{ij}|
        =
        |A_{j\cdot}(\sb_i^0-\bm m_0)|
        \le
        \|A_{j\cdot}\|_2\|\sb_i^0-\bm m_0\|_2
        \le
        2R\rho_j.
        \label{eq:A.3.2}
    \end{align}
    Since $S_{ij}$ is bounded, there exists a constant $C_1\ge C>0$ independent of $n$ and $p$ such that
    \begin{align}
        \E\exp\bigl(t(S_{ij}-\E S_{ij})\bigr)
        \le
        \exp(C_1\rho_j^2 t^2),
        \qquad t\in\R.
        \label{eq:A.3.3}
    \end{align}

    Since $\sb_i$ and $\zb_i$ are independent, \eqref{eq:A.3.1} and \eqref{eq:A.3.3} yield
    \begin{align}
        \E\exp\bigl(t(Y_{ij}-\E Y_{ij})\bigr)
        &=
        \E\exp\bigl(t(S_{ij}-\E S_{ij})\bigr)\,
        \E\exp(tZ_{ij}) \notag\\
        &\le
        \exp\bigl(C_1(\rho_j^2+\sigma^2)t^2\bigr),
        \qquad t\in\R.
        \label{eq:A.3.4}
    \end{align}
    Thus $Y_{ij}-\E Y_{ij}$ is a centered sub-Gaussian random variable with variance proxy of order $\rho_j^2+\sigma^2$.

    Since the eigenvalues of $\Sigma$ are fixed positive constants, there exist constants $0<c_\Sigma<C_\Sigma<\infty$ such that
    $$
    c_\Sigma\Theta\rho_j^2
    \le
    A_{j\cdot}\Sigma A_{j\cdot}^\top
    \le
    C_\Sigma\Theta\rho_j^2.
    $$
    Therefore
    \begin{align}
        \kappa_j
        =
        \sigma^2+A_{j\cdot}\Sigma A_{j\cdot}^\top
        \asymp
        \sigma^2+\Theta\rho_j^2.
        \label{eq:A.3.5}
    \end{align}
    Since $\Theta$ is fixed, \eqref{eq:A.3.4} and \eqref{eq:A.3.5} imply that $Y_{ij}-\E Y_{ij}$ is sub-Gaussian with variance proxy of order $\kappa_j$. Hence, by the fact that the square of a sub-Gaussian random variable is sub-exponential,
    $$
    Y_{ij}^2-\kappa_j
    $$
    is sub-exponential with scale of order $\kappa_j$. By Bernstein's inequality (Lemma \ref{lemma:bernstein}), we obtain that for any $t>0$,
    \begin{align}
        \P\left(\frac{|\widetilde E_j-\kappa_j|}{\kappa_j}>t\right)
        \le
        2\exp\left(-cn\min(t^2,t)\right)
        \label{eq:A.3.6}
    \end{align}
    for some constant $c>0$ independent of $n$ and $p$.

The mean $\bar Y_j$ is sub-Gaussian with scale
$C\sqrt{\kappa_j/n}$. Hence
\[
 \mathbb P(\bar Y_j^2>\kappa_j t)\le2e^{-cnt}.
\]
For $0<t\le1$, the identity $E_j=\widetilde E_j-\bar Y_j^2$
and a union bound give
\[
 \mathbb P(|E_j-\kappa_j|>2\kappa_j t)
 \le2e^{-cnt^2}+2e^{-cnt}.
\]
Thus the sample-centering correction is of smaller order and the
same concentration conclusion follows.

    Taking
    $$
    t=n^\varepsilon\sqrt{\frac{\log n}{n}}
    $$
    for any fixed $0<\varepsilon<1/2$, and using that $p\asymp n^\alpha$, a union bound over $j=1,\dots,p$ yields
    $$
    \max_{1\le j\le p}\frac{|E_j-\kappa_j|}{\kappa_j}
    =
    \OO_\prec\!\left(\sqrt{\frac{\log n}{n}}\right).
    $$
\end{proof}

We now prove Proposition~\ref{prop:K}.

\begin{proof}[\bf{Proof of Proposition~\ref{prop:K}}]
    Let
    $$
    d_j\coloneq A_{j\cdot}\Sigma A_{j\cdot}^\top,
    \qquad
    \kappa_j\coloneq \sigma^2+d_j,
    $$
    and let
    $$
    d_{(1)}\ge\cdots\ge d_{(p)},
    \qquad
    \kappa_{(1)}\ge\cdots\ge \kappa_{(p)}
    $$
    denote the corresponding decreasing rearrangements. Since the eigenvalues of $\Sigma$ are fixed positive constants, there exist constants $0<c_\Sigma<C_\Sigma<\infty$ such that
    \begin{align}
        c_\Sigma \Theta \rho_j^2
        \le
        A_{j\cdot}\Sigma A_{j\cdot}^\top
        \le
        C_\Sigma \Theta \rho_j^2,
        \qquad j=1,\dots,p.
        \label{eq:A.K.1}
    \end{align}
    By Lemma~\ref{lem:row-weak-lq}, there exist constants $0<c_-<c_+<\infty$ such that
    $$
    c_- m^{-1/q}\le \rho_{(m)}\le c_+ m^{-1/q},
    \qquad m=1,\dots,p.
    $$
    Combining this with \eqref{eq:A.K.1}, we obtain constants $0<a_1<a_2<\infty$ such that
    \begin{align}
        a_1\Theta m^{-2/q}\le d_{(m)}\le a_2\Theta m^{-2/q},
        \qquad m=1,\dots,p.
        \label{eq:A.K.2}
    \end{align}
    Hence
    \begin{align}
        \kappa_{(m)}=\sigma^2+d_{(m)}\asymp \sigma^2+\Theta m^{-2/q},
        \qquad m=1,\dots,p.
        \label{eq:A.K.3}
    \end{align}

    Define the ordered population tail average
    $$
    \bar\kappa_{\mathrm{tail}}
    \coloneq
    \frac{1}{s}\sum_{\ell=p+1-s}^{p}\kappa_{(\ell)}.
    $$
    Write $\bar E_{\mathrm{tail}}=s^{-1}\sum_{\ell=p+1-s}^pE_{(\ell)}=\hat\sigma^2$.
    Since $s$ is fixed, \eqref{eq:A.K.2} implies
    \begin{align}
        \bar\kappa_{\mathrm{tail}}
        =
        \sigma^2+\frac{1}{s}\sum_{\ell=p+1-s}^{p} d_{(\ell)}
        \asymp
        \sigma^2+\Theta p^{-2/q}.
        \label{eq:A.K.4}
    \end{align}
    By Lemma~\ref{lem:E-conc-rel-poly}, we have
    $$
    \max_{1\le j\le p}\frac{|E_j-\kappa_j|}{\kappa_j}
    =
    \OO_\prec\!\left(\frac{1}{\sqrt n}\right).
    $$
    Hence, for any fixed $\delta>0$, with high probability, for sufficiently large $n$,
    $$
    (1-\delta)\kappa_j\le E_j\le (1+\delta)\kappa_j,\qquad j=1,\dots,p.
    $$
    Rearranging these inequalities yields
    $$
    (1-\delta)\kappa_{(m)}\le E_{(m)}\le (1+\delta)\kappa_{(m)},
    \qquad m=1,\dots,p,
    $$
    and averaging over the last $s$ ordered terms gives
    \begin{align}
        (1-\delta)\bar\kappa_{\mathrm{tail}}\le \bar E_{\mathrm{tail}}\le (1+\delta)\bar\kappa_{\mathrm{tail}}.
        \label{eq:A.K.5}
    \end{align}

    Choose $\delta>0$ small enough so that
    $$
    u_1\coloneq \frac{(1+\tau)(1+\delta)}{1-\delta}>1,
    \qquad
    u_2\coloneq \frac{(1+\tau)(1-\delta)}{1+\delta}>1.
    $$
    Define
    $$
    N_1\coloneq \max\{m:\kappa_{(m)}\ge u_1\bar\kappa_{\mathrm{tail}}\},
    \qquad
    N_2\coloneq \max\{m:\kappa_{(m)}\ge u_2\bar\kappa_{\mathrm{tail}}\}.
    $$
    Use zero for the maximum of an empty index set.
    On the simultaneous coordinate-concentration event above, on which \eqref{eq:A.K.5} also holds, if $\kappa_{(m)}\ge u_1\bar\kappa_{\mathrm{tail}}$, then
    $$
    E_{(m)}
    \ge
    (1-\delta)\kappa_{(m)}
    \ge
    (1+\tau)(1+\delta)\bar\kappa_{\mathrm{tail}}
    \ge
    (1+\tau)\bar E_{\mathrm{tail}},
    $$
    which implies $m\le \Ks$. Hence
    \begin{align}
        N_1\le \Ks.
        \label{eq:A.K.6}
    \end{align}
    Similarly, if $\kappa_{(m)}<u_2\bar\kappa_{\mathrm{tail}}$, then
    $$
    E_{(m)}
    \le
    (1+\delta)\kappa_{(m)}
    <
    (1+\tau)(1-\delta)\bar\kappa_{\mathrm{tail}}
    \le
    (1+\tau)\bar E_{\mathrm{tail}},
    $$
    so $m>\Ks$. Therefore
    \begin{align}
        \Ks\le N_2.
        \label{eq:A.K.7}
    \end{align}

    	For each fixed $u>1$, the inequality
    	$$
    	\kappa_{(m)}\ge u\bar\kappa_{\mathrm{tail}}
    	$$
    	holds if and only if
    	$$
    	d_{(m)}\ge (u-1)\sigma^2+\frac{u}{s}\sum_{\ell=p+1-s}^p d_{(\ell)}.
    	$$
    	By \eqref{eq:A.K.2}, the expression on the right is bounded above and below by constant multiples of $B_n\coloneq\sigma^2+\Theta p^{-2/q}$. Define $L_n\coloneq(\Theta/B_n)^{q/2}\le p$.
    	We now apply this observation to the thresholds $u_1$ and $u_2$. By the lower bound in \eqref{eq:A.K.2}, we can choose a fixed $b_0\in(0,1)$ sufficiently small so that $\kappa_{(m)}\ge u_1\bar\kappa_{\mathrm{tail}}$ whenever $m\le b_0L_n$. Similarly, the upper bound gives a fixed $C>0$ sufficiently large so that $\kappa_{(m)}<u_2\bar\kappa_{\mathrm{tail}}$ whenever $m>CL_n$. It follows from the definitions of $N_1$ and $N_2$ that
    	$$
    	\lfloor b_0L_n\rfloor\le N_1\le N_2\le CL_n.
    	$$
    	The condition $\sigma^2/\Theta=\oo(1)$ in the proposition implies $L_n\to\infty$. Thus, for all sufficiently large $n$, we have $L_n\ge2/b_0$ and hence $\lfloor b_0L_n\rfloor\ge b_0L_n/2$. Taking $b_1=b_0/2$ and $b_2>C$, we obtain

    \begin{align}
        b_1\left(\frac{\Theta}{\sigma^2+\Theta p^{-2/q}}\right)^{q/2}
        \le
        N_1\le N_2
        \le
        b_2\left(\frac{\Theta}{\sigma^2+\Theta p^{-2/q}}\right)^{q/2}.
        \label{eq:A.K.8}
    \end{align}
    Combining \eqref{eq:A.K.6}, \eqref{eq:A.K.7}, and \eqref{eq:A.K.8}, we obtain that, with probability tending to one,
    \begin{align}
        \Ks\asymp \left(\frac{\Theta}{\sigma^2+\Theta p^{-2/q}}\right)^{q/2}.
        \label{eq:A.K.9}
    \end{align}

    If $\SNR\lesssim p^{2/q-1}$, then $\Theta p^{-2/q}\lesssim \sigma^2$, so
    $$
    \sigma^2+\Theta p^{-2/q}\asymp \sigma^2.
    $$
    Substituting this into \eqref{eq:A.K.9} yields
    $$
    \Ks\asymp (\Theta/\sigma^2)^{q/2}.
    $$
    If $\SNR\gtrsim p^{2/q-1}$, then $\sigma^2\lesssim \Theta p^{-2/q}$, so
    $$
    \sigma^2+\Theta p^{-2/q}\asymp \Theta p^{-2/q}.
    $$
    Substituting this into \eqref{eq:A.K.9} yields
    $$
    \Ks\asymp p.
    $$
    The same counting bound will also be used in the recovery proofs
below. It only requires $L_n\ge2/b_0$. Set $c_*=(b_0/2)^{2/q}/2$.
For sufficiently large $p$, the condition $\sigma^2/\Theta\le c_*$ gives
$L_n\ge(2c_*)^{-q/2}=2/b_0$. Thus, on the same concentration event,
\begin{equation}\label{eq:K-small-noise}
 \Ks\asymp
 \left(\frac{\Theta}{\sigma^2+\Theta p^{-2/q}}\right)^{q/2},
 \qquad \sigma^2/\Theta\le c_*.
\end{equation}
The comparison constants are uniform in this range.
    This completes the proof.
\end{proof}

With Proposition~\ref{prop:K} in hand, we next study how accurately the selected projection reconstructs the underlying signal subspace. This subspace reconstruction error will then be translated into a uniform bound for the projected clean pairwise distances. Define
$$
\mathcal U\coloneq \operatorname{span}\{\rs_1,\dots,\rs_r\}\subset\R^p,
$$
and let $\Pi_{\mathcal U}$ denote the orthogonal projector onto $\mathcal U$.
Recall that $\Pi_\Ps=\Ps\Ps^\top$ projects onto the selected basis directions. The error in the following lemma is the largest norm of the discarded component $(\Ib-\Pi_\Ps)v$ over unit vectors $v\in\mathcal U$.

\begin{lemma}[Signal-subspace reconstruction error]
    \label{lem:proj-gap-from-residual}
    Let $\mathcal I_{\Ks}$ be the index set selected by Algorithm~\ref{alg:Kselect}. Then
    $$
    \|(\Ib-\Pi_\Ps)\Pi_{\mathcal U}\|_{\op}
    \le
    \Bigl(\sum_{j\notin\mathcal I_{\Ks}}\rho_j^2\Bigr)^{1/2}.
    $$
\end{lemma}

\begin{proof}
    Let $\Rs_r\coloneq (\rs_1,\dots,\rs_r)\in\R^{p\times r}$, so that $\mathcal U=\Col(\Rs_r)$ and $\Pi_{\mathcal U}=\Rs_r\Rs_r^\top$. Since the columns of $\Rs_r$ are orthonormal, we have
    $$
    \|(\Ib-\Pi_\Ps)\Pi_{\mathcal U}\|_{\op}
    \le
    \|(\Ib-\Pi_\Ps)\Pi_{\mathcal U}\|_F
    =
    \|(\Ib-\Pi_\Ps)\Rs_r\Rs_r^\top\|_F
    \le
    \|(\Ib-\Pi_\Ps)\Rs_r\|_F.
    $$
    By \eqref{eq:appendix-A}, we have $\Rs_r=\Ws A$. Since $\Pi_\Ps$ is the orthogonal projector onto $\Col(\Ws_{\mathcal I_{\Ks}})$,
    $$
    (\Ib-\Pi_\Ps)\Rs_r
    =
    (\Ib-\Pi_\Ps)\Ws A
    =
    \Ws_{\mathcal I_{\Ks}^c}A_{\mathcal I_{\Ks}^c,\cdot}.
    $$
    Using the orthonormality of $\Ws$, we obtain
    $$
    \|(\Ib-\Pi_\Ps)\Rs_r\|_F^2
    =
    \|A_{\mathcal I_{\Ks}^c,\cdot}\|_F^2
    =
    \sum_{j\notin\mathcal I_{\Ks}}\|A_{j\cdot}\|_2^2
    =
    \sum_{j\notin\mathcal I_{\Ks}}\rho_j^2.
    $$
    Taking square roots completes the proof.
\end{proof}

\begin{theorem}[Convergence of signal-subspace reconstruction]
    \label{thm:proj-gap-final-poly}
    Suppose Assumptions~\ref{ass:model}, \ref{ass:signal}, and \ref{ass:lq} hold. Then
    $$
    \|(\Ib-\Pi_\Ps)\Pi_{\mathcal U}\|_{\op}^2
    =
    \begin{cases}
        \OO_\prec\!\left(\left(\dfrac{\Theta}{\sigma^2}\right)^{q/2-1}\right), & \text{if } \SNR\lesssim p^{2/q-1}, \\[1ex]
        \OO_\prec\!\left(p^{1-2/q}\right), & \text{if } \SNR\gtrsim p^{2/q-1}.
    \end{cases}
    $$
\end{theorem}

\begin{proof}
    Let $\mathcal I_{\Ks}\subset[p]$ be the index set selected by Algorithm~\ref{alg:Kselect}. By definition, $|\mathcal I_{\Ks}|=\Ks$. If $\Ks=p$, the residual is zero and the claimed bound holds. We may therefore assume $\Ks<p$. Let $t_n=\sigma^2/\Theta$, and recall $c_*$ from \eqref{eq:K-small-noise}. If $t_n>c_*$, then for all sufficiently large $n$, we are in the first $\SNR$ regime and have $t_n^{1-q/2}\ge c_*^{1-q/2}>0$. Since we also have $\|(\Ib-\Pi_\Ps)\Pi_{\mathcal U}\|_{\op}\le1$, we obtain the desired bound in this case.

    It remains to consider $t_n\le c_*$. On the concentration event associated with \eqref{eq:K-small-noise}, we have $\Ks\ge1$ for all sufficiently large $n$. We work on this event below, so that $1\le\Ks<p$ and the counting bound \eqref{eq:K-small-noise} applies. When defining the population top-$\Ks$ index set below, we break ties by index.

    With this convention, we define the population top-$\Ks$ index set by
    $$
    \mathcal I^*
    \coloneq
    \{j\in[p]: \rho_j \text{ is among the $\Ks$ largest of } (\rho_1,\dots,\rho_p)\},
    $$
    and recall that $\rho_{(1)}\ge \rho_{(2)}\ge\cdots\ge \rho_{(p)}$ denotes the nonincreasing rearrangement of $\{\rho_j\}_{j=1}^p$.

    By Lemma~\ref{lem:proj-gap-from-residual}, we bound the squared residual by the sum of squared row norms over the discarded directions. Splitting this sum according to $\mathcal I^*$, we obtain
    \begin{align}
        \|(\Ib-\Pi_\Ps)\Pi_{\mathcal U}\|_{\op}^2
        \le
        \sum_{j\notin\mathcal I_{\Ks}}\rho_j^2
        \le
        \sum_{j\notin\mathcal I^*}\rho_j^2
        +
        \sum_{j\in\mathcal I^*\setminus\mathcal I_{\Ks}}\rho_j^2.
        \label{eq:thm.1.1}
    \end{align}
    We first control $\sum_{j\notin\mathcal I^*}\rho_j^2$. By Lemma~\ref{lem:row-weak-lq}, $\rho_{(m)}\le C' m^{-1/q}$ for some constant $C'>0$. Hence
    \begin{align}
        \sum_{j\notin\mathcal I^*}\rho_j^2
        =
        \sum_{m=\Ks+1}^p \rho_{(m)}^2
        \le
        \sum_{m=\Ks+1}^\infty (C'm^{-1/q})^2
        \lesssim
        \Ks^{1-2/q}.
        \label{eq:thm.1.2}
    \end{align}

    To bound the second term in \eqref{eq:thm.1.1}, we define
    $$
    S_{\mathrm{miss}}\coloneq \mathcal I^*\setminus\mathcal I_{\Ks},
    \qquad
    S_{\mathrm{add}}\coloneq \mathcal I_{\Ks}\setminus\mathcal I^*.
    $$
    Then $|S_{\mathrm{miss}}|=|S_{\mathrm{add}}|\le \Ks$. By Lemma~\ref{lem:E-conc-rel-poly}, we have
    $$
    \max_{1\le j\le p}\frac{|E_j-\kappa_j|}{\kappa_j}
    =
    \OO_\prec\!\left(\sqrt{\frac{\log n}{n}}\right).
    $$
    Fix $0<\varepsilon<1/2$ and set
    $$
    \delta_n\coloneq n^\varepsilon\sqrt{\frac{\log n}{n}}.
    $$
    For all sufficiently large $n$, we have $\delta_n\le1/3$. The preceding concentration bound gives, with high probability,
    $$
    \max_{1\le j\le p}\frac{|E_j-\kappa_j|}{\kappa_j}\le \delta_n.
    $$
    For each $j\in S_{\mathrm{miss}}$, the equality $|S_{\mathrm{miss}}|=|S_{\mathrm{add}}|$ allows us to choose $l\in S_{\mathrm{add}}$. Since $l$ is selected and $j$ is not, the empirical ordering gives $E_l\ge E_j$. We therefore have
    $$
    \kappa_l(1+\delta_n)\ge E_l \ge E_j \ge \kappa_j(1-\delta_n),
    $$
    which implies
    $$
    \kappa_j\le \kappa_l\frac{1+\delta_n}{1-\delta_n}\le (1+3\delta_n)\kappa_l.
    $$
    Equivalently,
    \begin{align}
        \kappa_j-\sigma^2 \le (\kappa_l-\sigma^2) + 3\delta_n\kappa_l.
        \label{eq:thm.1.3}
    \end{align}
    Since $\kappa_j=\sigma^2+A_{j\cdot}\Sigma A_{j\cdot}^\top$ and the eigenvalues of $\Sigma$ are fixed positive constants, there exist constants $0<c_\Sigma<C_\Sigma<\infty$ such that
    \begin{align}
        c_\Sigma\Theta \rho_j^2 \le \kappa_j-\sigma^2 \le C_\Sigma\Theta \rho_j^2,
        \qquad
        \kappa_l \le \sigma^2+C_\Sigma\Theta \rho_l^2.
        \label{eq:thm.1.4}
    \end{align}
    Combining \eqref{eq:thm.1.3} and \eqref{eq:thm.1.4} yields
    $$
    c_\Sigma\Theta \rho_j^2
    \le
    C_\Sigma\Theta \rho_l^2 + 3\delta_n(\sigma^2+C_\Sigma\Theta \rho_l^2).
    $$
    Hence, for some constant $C>0$,
    $$
    \rho_j^2 \le C(1+\delta_n)\rho_l^2 + C\delta_n\frac{\sigma^2}{\Theta}.
    $$
    Since $l\in S_{\mathrm{add}}\subset(\mathcal I^*)^c$, we have $\rho_l\le \rho_{(\Ks+1)}$. Therefore,
    $$
    \rho_j^2 \le C(1+\delta_n)\rho_{(\Ks+1)}^2 + C\delta_n\frac{\sigma^2}{\Theta},
    \qquad
    j\in S_{\mathrm{miss}}.
    $$
    Summing over $j\in S_{\mathrm{miss}}$ gives
    $$
    \sum_{j\in S_{\mathrm{miss}}}\rho_j^2
    \le
    C\Ks(1+\delta_n)\rho_{(\Ks+1)}^2 + C\Ks\delta_n\frac{\sigma^2}{\Theta}.
    $$
    Since $\rho_{(\Ks+1)}^2\lesssim \Ks^{-2/q}$ by Lemma~\ref{lem:row-weak-lq}, we obtain
    \begin{align}
        \sum_{j\in S_{\mathrm{miss}}}\rho_j^2
        \lesssim
        (1+\delta_n)\Ks^{1-2/q} + \Ks\delta_n\frac{\sigma^2}{\Theta}.
        \label{eq:thm.1.5}
    \end{align}

    Combining the tail bound \eqref{eq:thm.1.2} and the missed-direction bound \eqref{eq:thm.1.5} with \eqref{eq:thm.1.1}, and using $\delta_n\le1/3$, we obtain
    \begin{align}
        \|(\Ib-\Pi_\Ps)\Pi_{\mathcal U}\|_{\op}^2
        \lesssim
        \Ks^{1-2/q} + \Ks\delta_n\frac{\sigma^2}{\Theta}.
        \label{eq:thm.1.6}
    \end{align}

    We now apply the counting bound \eqref{eq:K-small-noise} in the two $\SNR$ regimes. If $\SNR\lesssim p^{2/q-1}$, we have $\Ks\asymp (\Theta/\sigma^2)^{q/2}$. Substituting this order into the two terms in \eqref{eq:thm.1.6}, we obtain
    $$
    \Ks^{1-2/q}\asymp \left(\frac{\Theta}{\sigma^2}\right)^{q/2-1},
    \qquad
    \Ks\frac{\sigma^2}{\Theta}\delta_n
    \asymp
    \left(\frac{\Theta}{\sigma^2}\right)^{q/2-1}\delta_n.
    $$
    Since $\delta_n=\oo(1)$, we obtain
    $$
    \|(\Ib-\Pi_\Ps)\Pi_{\mathcal U}\|_{\op}^2
    =
    \OO_\prec\!\left(\left(\frac{\Theta}{\sigma^2}\right)^{q/2-1}\right).
    $$
    If $\SNR\gtrsim p^{2/q-1}$, the same counting bound gives $\Ks\asymp p$. In this regime, we also have $\sigma^2\lesssim \Theta p^{-2/q}$, and hence
    $$
    \Ks\frac{\sigma^2}{\Theta}\delta_n\lesssim p^{1-2/q}\delta_n.
    $$
    Substituting these bounds into \eqref{eq:thm.1.6}, we obtain
    $$
    \|(\Ib-\Pi_\Ps)\Pi_{\mathcal U}\|_{\op}^2
    =
    \OO_\prec\!\left(p^{1-2/q}\right).
    $$
    This completes the proof.
\end{proof}

\begin{lemma}[Clean pairwise-distance bound]
    \label{lem:pairwise-clean}
    Under Assumptions~\ref{ass:model}, \ref{ass:signal}, and \ref{ass:lq},
    $$
    \max_{1\le i<j\le n}\Bigl|\|\Ps^\top(\sb_i-\sb_j)\|_2^2-\|\sb_i-\sb_j\|_2^2\Bigr|
    =
    \begin{cases}
        \OO_\prec\left(\left(\dfrac{\Theta}{\sigma^2}\right)^{q/2-1}\right), & \text{if } \SNR\lesssim p^{2/q-1}, \\[1ex]
        \OO_\prec(p^{1-2/q}), & \text{if } \SNR\gtrsim p^{2/q-1}.
    \end{cases}
    $$
\end{lemma}

\begin{proof}
    Let
    $$
    \Delta\sb_{ij}\coloneq \sb_i-\sb_j.
    $$
    Since $\sb_i,\sb_j\in\mathcal U$, we have $\Delta\sb_{ij}\in\mathcal U$. Therefore,
    \begin{align}
        \Bigl|\|\Ps^\top(\sb_i-\sb_j)\|_2^2-\|\sb_i-\sb_j\|_2^2\Bigr|
        &=
        \Bigl|\|\Pi_\Ps\Delta\sb_{ij}\|_2^2-\|\Delta\sb_{ij}\|_2^2\Bigr| \notag\\
        &=
        \|(\Ib-\Pi_\Ps)\Delta\sb_{ij}\|_2^2 \notag\\
        &\le
        \|(\Ib-\Pi_\Ps)\Pi_{\mathcal U}\|_{\op}^2\|\Delta\sb_{ij}\|_2^2.
        \label{eq:pairwise-clean-proof}
    \end{align}
    By Assumption~\ref{ass:signal}, the latent vectors $\sb_i^0$ have compact support, so there exists a constant $D>0$ such that
    $$
    \|\sb_i-\sb_j\|_2=\|\sb_i^0-\sb_j^0\|_2\le D,
    \qquad 1\le i,j\le n.
    $$
    Hence \eqref{eq:pairwise-clean-proof} yields
    $$
    \max_{1\le i<j\le n}\Bigl|\|\Ps^\top(\sb_i-\sb_j)\|_2^2-\|\sb_i-\sb_j\|_2^2\Bigr|
    \le
    D^2\|(\Ib-\Pi_\Ps)\Pi_{\mathcal U}\|_{\op}^2.
    $$
    By Theorem~\ref{thm:proj-gap-final-poly}, this gives the desired result.
\end{proof}

Lemma~\ref{lem:pairwise-clean} controls the signal-only projection bias. To complete the pairwise-distance analysis, it remains to bound the stochastic contribution from the projected noise and the signal--noise interaction term.

\begin{lemma}[Uniform control of projected noise terms]
    \label{lem:noise-terms}
    Under Assumptions~\ref{ass:model}, \ref{ass:signal} and \ref{ass:lq}, we have
    $$
    \max_{1\le i<j\le n}\|\Ps^\top(\zb_i-\zb_j)\|_2^2
    =
    \OO_\prec(\sigma^2\Ks)
    $$
    and
    $$
    \max_{1\le i<j\le n}\left|(\Ps^\top(\sb_i-\sb_j))^\top\Ps^\top(\zb_i-\zb_j)\right|
    =
    \OO_\prec\!\bigl(\sigma\Ks^{(1-1/q)_+}\bigr).
    $$
\end{lemma}

\begin{proof}
    For $1\le i<j\le n$, write
    $$
    \Delta \zb_{ij}\coloneq \zb_i-\zb_j,
    \qquad
    \Delta \sb_{ij}\coloneq \sb_i-\sb_j.
    $$

    We first prove the bound for the pure noise term. By the definition of
    the selected index set $\mathcal I_{\Ks}$ and the projector
    $\Pi_\Ps=\Ps\Ps^\top$, we have
    $$
    \Pi_\Ps=\sum_{\ell\in\mathcal I_{\Ks}} \ws_\ell\ws_\ell^\top,
    $$
    and therefore
    $$
    \|\Ps^\top\Delta \zb_{ij}\|_2^2
    =
    \Delta \zb_{ij}^\top \Pi_\Ps \Delta \zb_{ij}
    =
    \sum_{\ell\in\mathcal I_{\Ks}} (\ws_\ell^\top\Delta \zb_{ij})^2
    \le
    \Ks \max_{1\le \ell\le p} (\ws_\ell^\top\Delta \zb_{ij})^2.
    $$
    By Assumption~\ref{ass:signal}, each $\Delta \zb_{ij}$ is a centered
    sub-Gaussian random vector with independent entries and covariance
    matrix $2\sigma^2\Ib_p$. Hence, by the tail bound for linear forms of
    centered sub-Gaussian random vectors, there exists a constant $c>0$ independent of $n$ and $p$ such that for every nonzero deterministic vector
    $\mathbf a\in\R^p$,
    $$
    \P\bigl(|\mathbf a^\top \Delta \zb_{ij}|>u\bigr)
    \le
    2\exp\!\left(-\frac{cu^2}{\sigma^2\|\mathbf a\|_2^2}\right),
    \qquad u>0.
    $$
    Taking $\mathbf a=\ws_\ell$ and using $\|\ws_\ell\|_2=1$, we obtain
    $$
    \P\left(|\ws_\ell^\top\Delta \zb_{ij}|>u\right)
    \le
    2\exp\!\left(-\frac{cu^2}{\sigma^2}\right),
    \qquad u>0.
    $$
    Fix $\varepsilon>0$ and $D>0$, and set
    $$
    u\coloneq M\sigma n^\varepsilon,
    $$
    where $M>0$ will be chosen later. Since the number of pairs $(i,j)$
    is of order $n^2$ and $p\asymp n^\alpha$, a union bound over all $(i,j)$
    and $\ell\in[p]$ gives
    $$
    \P\left(
    \max_{1\le i<j\le n}\max_{1\le \ell\le p}
    |\ws_\ell^\top(\zb_i-\zb_j)|>M\sigma n^\varepsilon
    \right)
    \le
    C n^{\alpha+2}\exp(-cM^2n^{2\varepsilon}).
    $$
    Choosing $M>0$ large enough yields
    $$
    \P\left(
    \max_{1\le i<j\le n}\max_{1\le \ell\le p}
    |\ws_\ell^\top(\zb_i-\zb_j)|>M\sigma n^\varepsilon
    \right)
    \le n^{-D}
    $$
    for all sufficiently large $n$. Hence
    $$
    \max_{1\le i<j\le n}\max_{1\le \ell\le p}
    |\ws_\ell^\top(\zb_i-\zb_j)|
    =
    \OO_\prec(\sigma).
    $$
    Therefore,
    $$
    \max_{1\le i<j\le n}\|\Ps^\top(\zb_i-\zb_j)\|_2^2
    =
    \OO_\prec(\sigma^2\Ks).
    $$

    We next prove the bound for the interaction term. Write
    \begin{align}
        \left|(\Ps^\top\Delta \sb_{ij})^\top\Ps^\top\Delta \zb_{ij}\right|
        &=
        \left|\sum_{\ell\in\mathcal I_{\Ks}}
        (\ws_\ell^\top\Delta \sb_{ij})(\ws_\ell^\top\Delta \zb_{ij})\right|
        \notag\\
        &\le
        \max_{1\le \ell\le p}|\ws_\ell^\top\Delta \zb_{ij}|
        \sum_{\ell\in\mathcal I_{\Ks}}|\ws_\ell^\top\Delta \sb_{ij}|.
        \label{eq:interaction-split}
    \end{align}
    By \eqref{eq:rotate}, we have
    $$
    \ws_\ell^\top\Delta \sb_{ij}
    =
    A_{\ell\cdot}(\sb_i^0-\sb_j^0).
    $$
    Since $\{\sb_i^0\}$ have compact support, there exists a constant
    $D_0>0$ such that
    $$
    \|\sb_i^0-\sb_j^0\|_2\le D_0,
    \qquad 1\le i,j\le n.
    $$
    Hence
    $$
    |\ws_\ell^\top\Delta \sb_{ij}|
    \le
    \|A_{\ell\cdot}\|_2\,\|\sb_i^0-\sb_j^0\|_2
    \le
    D_0\rho_\ell,
    $$
    where $\rho_\ell=\|A_{\ell\cdot}\|_2$. Therefore,
    \begin{align}
        \sum_{\ell\in\mathcal I_{\Ks}}|\ws_\ell^\top\Delta \sb_{ij}|
        \le
        D_0\sum_{\ell\in\mathcal I_{\Ks}}\rho_\ell
        \le
        D_0\sum_{m=1}^{\Ks}\rho_{(m)}.
        \label{eq:interaction-signal-sum}
    \end{align}
    By Lemma~\ref{lem:row-weak-lq}, $\rho_{(m)}\lesssim m^{-1/q}$, so
    \begin{align}
        \sum_{m=1}^{\Ks}\rho_{(m)}
        \lesssim
        \begin{cases}
            1, & 0<q<1, \\[0.5ex]
            1+\log \Ks, & q=1, \\[0.5ex]
            \Ks^{1-1/q}, & 1<q<2.
        \end{cases}
        \label{eq:interaction-row-sum}
    \end{align}
    Since $\Ks\le p$ and $p\asymp n^\alpha$, the logarithmic factor in the
    case $q=1$ is absorbed by $\prec$-notation. Thus
    \eqref{eq:interaction-row-sum} implies
    $$
    \sum_{m=1}^{\Ks}\rho_{(m)}
    =
    \OO_\prec\!\bigl(\Ks^{(1-1/q)_+}\bigr).
    $$
    Combining this with \eqref{eq:interaction-split},
    \eqref{eq:interaction-signal-sum}, and the first part of the proof,
    we conclude that
    $$
    \max_{1\le i<j\le n}\left|(\Ps^\top(\sb_i-\sb_j))^\top\Ps^\top(\zb_i-\zb_j)\right|
    =
    \OO_\prec\!\bigl(\sigma\Ks^{(1-1/q)_+}\bigr),
    $$
    which completes the proof.
\end{proof}

We are now ready to prove Theorem~\ref{thm:pairwise}. The argument combines the clean projection bias from Lemma~\ref{lem:pairwise-clean} with the projected-noise bounds in Lemma~\ref{lem:noise-terms}.
\begin{proof}[\bf{Proof of Theorem~\ref{thm:pairwise}}]
Let $t_n=\sigma^2/\Theta$, and recall $c_*$ from \eqref{eq:K-small-noise}. Fix $0<\delta<\tau/[2(2+\tau)]$ and work on the energy-concentration event in Lemma~\ref{lem:E-conc-rel-poly}. Recall the signal variances $d_j=A_{j\cdot}\Sigma A_{j\cdot}^\top$ from the proof of Proposition~\ref{prop:K}. For each selected coordinate, we have $E_j\ge(1+\tau)\hat\sigma^2$ and $E_j\le(1+\delta)(\sigma^2+d_j)$. Since the population tail average is at least $\sigma^2$, we also have $\hat\sigma^2\ge(1-\delta)\sigma^2$. Combining these inequalities, we obtain
\[
 d_j\ge a_-\sigma^2,
 \qquad a_-=\frac{(1+\tau)(1-\delta)}{1+\delta}-1>0.
\]
By the upper bound $d_{(j)}\le C\Theta j^{-2/q}$ in \eqref{eq:A.K.2}, we then have $\Ks\lesssim t_n^{-q/2}$ whenever $t_n>0$.

We first consider $t_n>c_*$. Set $T=\max\{1,2/a_-\}$. If $c_*<t_n<T$, the preceding bound gives $\Ks\lesssim1$, and we also have $\sigma^2<T\Theta$. Using the bounded signal support and Lemma~\ref{lem:noise-terms}, we therefore obtain a pairwise-distance error of order $\OO_\prec(1)$.

If $t_n\ge T$, we have $a_-\sigma^2\ge2\Theta$, whereas $d_j\le\sum_\ell d_\ell=\Theta$. Thus, no coordinate can satisfy the selection inequality above, and $\Ks=0$ on the concentration event. The projected distances are then zero and the clean distances are bounded, so we again obtain an error of order $\OO_\prec(1)$. In both cases, for all sufficiently large $n$, we are in the first $\SNR$ regime and have $\psi_{n,\mathrm{bias}}=t_n^{1-q/2}\ge c_*^{1-q/2}>0$. We therefore obtain the claimed rate for $t_n>c_*$. By the standardized noise convention in Remark~\ref{rem:noise-norm}, the concentration event used here holds with high probability uniformly in $\sigma$. For any fixed $D>0$, its failure probability is at most $n^{-D}$ for all sufficiently large $n$.

It remains to consider $t_n\le c_*$. We work on the concentration event associated with \eqref{eq:K-small-noise} and use its counting bound for the selected dimension in the argument below.

    For this remaining case, we write, for $1\le i<j\le n$,
    $$
    \Delta \xb_{ij}\coloneq \xb_i-\xb_j,
    \qquad
    \Delta \sb_{ij}\coloneq \sb_i-\sb_j,
    \qquad
    \Delta \zb_{ij}\coloneq \zb_i-\zb_j.
    $$
    Since $\yb_i=\Ps^\top(\xb_i-\bar\xb)$, the training mean cancels and we have
    $$
    \yb_i-\yb_j
    =
    \Ps^\top\Delta \xb_{ij}
    =
    \Ps^\top\Delta \sb_{ij}+\Ps^\top\Delta \zb_{ij}.
    $$
    Therefore,
    \begin{align}
        \|\yb_i-\yb_j\|_2^2-\|\sb_i-\sb_j\|_2^2
        &=
        (\|\Ps^\top\Delta \sb_{ij}\|_2^2-\|\Delta \sb_{ij}\|_2^2)
        +
        \|\Ps^\top\Delta \zb_{ij}\|_2^2
        +
        2(\Ps^\top\Delta \sb_{ij})^\top\Ps^\top\Delta \zb_{ij}.
        \label{eq:pairwise-decomp}
    \end{align}
    We bound the three terms on the right-hand side
    separately. The first term is the projection bias,
    while the latter two terms represent the stochastic projected-noise
    contribution.
    Taking absolute values and maxima over $1\le i<j\le n$, we obtain
    \begin{align}
        \max_{1\le i<j\le n}\left|\|\yb_i-\yb_j\|_2^2-\|\sb_i-\sb_j\|_2^2\right|
        &\le
        \max_{1\le i<j\le n}\left|\|\Ps^\top(\sb_i-\sb_j)\|_2^2-\|\sb_i-\sb_j\|_2^2\right| \notag\\
        &\quad+
        \max_{1\le i<j\le n}\|\Ps^\top(\zb_i-\zb_j)\|_2^2 \notag\\
        &\quad+
        2\max_{1\le i<j\le n}\left|(\Ps^\top(\sb_i-\sb_j))^\top\Ps^\top(\zb_i-\zb_j)\right|.
        \label{eq:pairwise-union}
    \end{align}

    We first control the projection bias term. By Lemma~\ref{lem:pairwise-clean},
    \begin{align}
        \max_{1\le i<j\le n}\left|\|\Ps^\top(\sb_i-\sb_j)\|_2^2-\|\sb_i-\sb_j\|_2^2\right|
        =
        \OO_\prec(\psi_{n,\mathrm{bias}}).
        \label{eq:pairwise-bias}
    \end{align}

    We next control the two projected-noise terms. By Lemma~\ref{lem:noise-terms},
    \begin{align}
        \max_{1\le i<j\le n}\|\Ps^\top(\zb_i-\zb_j)\|_2^2
        =
        \OO_\prec(\sigma^2\Ks),
        \label{eq:pairwise-noise1}
    \end{align}
    and
    \begin{align}
        \max_{1\le i<j\le n}\left|(\Ps^\top(\sb_i-\sb_j))^\top\Ps^\top(\zb_i-\zb_j)\right|
        =
        \OO_\prec\!\bigl(\sigma\Ks^{(1-1/q)_+}\bigr).
        \label{eq:pairwise-noise2}
    \end{align}
    Hence
    \begin{align}
        \max_{1\le i<j\le n}
        \left|
        \|\Ps^\top\Delta \zb_{ij}\|_2^2
        +
        2(\Ps^\top\Delta \sb_{ij})^\top\Ps^\top\Delta \zb_{ij}
        \right|
        =
        \OO_\prec\!\bigl(\sigma^2\Ks+\sigma\Ks^{(1-1/q)_+}\bigr).
        \label{eq:pairwise-noise3}
    \end{align}

    By the counting bound \eqref{eq:K-small-noise}, the selected projection dimension $\Ks$ admits the asymptotic characterization stated in \eqref{eq:K-cases}. Substituting that characterization into \eqref{eq:pairwise-noise3} yields the same order as the rate $\psi_{n,\mathrm{noise}}$ defined in \eqref{eq:psi-proj-noise}. Thus
    \begin{align}
        \max_{1\le i<j\le n}
        \left|
        \|\Ps^\top\Delta \zb_{ij}\|_2^2
        +
        2(\Ps^\top\Delta \sb_{ij})^\top\Ps^\top\Delta \zb_{ij}
        \right|
        =
        \OO_\prec(\psi_{n,\mathrm{noise}}).
        \label{eq:pairwise-noise4}
    \end{align}

    Finally, combining \eqref{eq:pairwise-union} with the projection-bias bound \eqref{eq:pairwise-bias} and the noise bound \eqref{eq:pairwise-noise4}, we obtain
    $$
    \max_{1\le i<j\le n}\left|\|\yb_i-\yb_j\|_2^2-\|\sb_i-\sb_j\|_2^2\right|
    =
    \OO_\prec\!\left(\psi_{n,\mathrm{bias}}+\psi_{n,\mathrm{noise}}\right)
    =
    \OO_\prec(\psi_n),
    $$
    which proves Theorem~\ref{thm:pairwise}.
\end{proof}

\subsection{Proofs for Spectral Convergence}
\label{sec:proof_main}

In this subsection, we bound the error of the empirical bandwidth $h_n$ in \eqref{eq:hn-alg} relative to the population bandwidth $\mathsf h$ in \eqref{eq:h}, and then prove Theorem~\ref{thm:total_convergence}.

We define the clean empirical distribution function
$$
G_n^0(t)\coloneq \frac{2}{n(n-1)}\sum_{1\le i<j\le n} 1_{\{\|\sb_i-\sb_j\|_2^2\le t\}},\qquad t\ge 0,
$$
and let
$$
h\coloneq \inf\{t\ge 0:G_n^0(t)\ge \omega\}.
$$
All three bandwidths are $\omega$-quantiles of squared distances, computed from the projected observations for $h_n$, the clean sample for $h$, and the clean population law for $\mathsf h$. Recall \eqref{eq:hn-alg} and \eqref{eq:cdf}.
The following proposition gives the bandwidth error bound.
\begin{proposition}[Bandwidth consistency]
    \label{prop:bandwidth}
    Suppose Assumptions~\ref{ass:model}, \ref{ass:signal},
    and \ref{ass:lq} hold. For any prespecified $\omega\in(0,1)$ satisfying the local density condition in Theorem~\ref{thm:total_convergence},
    $$
    \left|\frac{h_n}{\mathsf h}-1\right|
    =
    \OO_\prec\!\left(\psi_n+\frac1{\sqrt n}\right).
    $$
\end{proposition}

\begin{remark}
    Proposition~\ref{prop:bandwidth} separates the projection error $|h_n-h|=\OO_\prec(\psi_n)$ from the clean quantile error $|h-\mathsf h|=\OO_\prec(n^{-1/2})$. Their sum gives the error relative to the population bandwidth, which is needed in the spectral comparison below.
\end{remark}

\begin{remark}
    \label{rem:bandwidth-regularity}
The local density condition in Theorem~\ref{thm:total_convergence} is a
standard sufficient condition for quantile stability; see, for example,
\cite{serfling1980approximation,vandervaart1998asymptotic}.
It ensures that $G$ crosses the level $\omega$ at a rate bounded away
from zero near $\mathsf h$, as shown in \eqref{eq:bw.local} below.
This local property is the regularity needed in the bandwidth proof.
Algorithm~\ref{alg:omega} provides a practical way to select this
quantile parameter.
\end{remark}

\begin{proof}
    We first compare $h_n$ with the clean empirical bandwidth $h$. Let
    $$
    \Delta_n\coloneq
    \max_{1\le i<j\le n}
    \left|\|\yb_i-\yb_j\|_2^2-\|\sb_i-\sb_j\|_2^2\right|.
    $$
    By Theorem~\ref{thm:pairwise},
    \begin{align}
        \Delta_n=\OO_\prec(\psi_n).
        \label{eq:bw.1}
    \end{align}
    By definition of $\Delta_n$, for every $t\ge 0$,
    $$
    G_n^0(t-\Delta_n)\le G_n(t)\le G_n^0(t+\Delta_n),
    $$
    where we use the convention $G_n^0(s)=0$ for $s<0$. Recall that
    $$
    h=\inf\{t\ge 0:G_n^0(t)\ge \omega\},
    \qquad
    h_n=\inf\{t\ge 0:G_n(t)\ge \omega\}.
    $$
    Hence
    $$
    h-\Delta_n\le h_n\le h+\Delta_n.
    $$
    Combining this with \eqref{eq:bw.1}, we obtain
    \begin{align}
        |h_n-h|\le \Delta_n=\OO_\prec(\psi_n).
        \label{eq:bw.2}
    \end{align}

    We next compare $h$ with the population bandwidth $\mathsf h$. Recall \eqref{eq:rotate}. Since $\Rs$ is orthogonal, we have
    $$
    \|\sb_i-\sb_j\|_2^2=\|\sb_i^0-\sb_j^0\|_2^2.
    $$
    Hence
    $$
    G_n^0(t)
    =
    \frac{2}{n(n-1)}\sum_{1\le i<j\le n}
    1_{\{\|\sb_i^0-\sb_j^0\|_2^2\le t\}},
    $$
    so $G_n^0(t)$ is a function of the independent latent variables $\sb_1^0,\dots,\sb_n^0$. Moreover, since $\sb_1^0,\sb_2^0\iid \mu$,
    \begin{align}
        \E G_n^0(t)
        =
        \P\bigl(\|\sb_1^0-\sb_2^0\|_2^2\le t\bigr)
        =
        G(t).
        \label{eq:bw.3}
    \end{align}

    Since $\mu\ll V_0$ and the Riemannian volume measure $V_0$ is nonatomic,
    $$
    G(0)=\P(\bm U=\bm V)=0.
    $$
    Moreover, since $\M^0$ is compact, there exists a constant $C_0>0$ such that $\bm D<C_0$. Thus $G(C_0)=1$. Since $\omega\in(0,1)$ and $G$ is right-continuous at $0$, it follows from the definition of $\mathsf h$ that
    $$
    0<\mathsf h<C_0.
    $$

By the density condition in Theorem~\ref{thm:total_convergence}, $G$ is continuous
near $\mathsf h$. The definition of its quantile gives
$G(\mathsf h)\ge\omega$. If the inequality were strict, continuity
would give $G(\mathsf h-s)>\omega$ for some $s>0$, contradicting
the definition of $\mathsf h$. Therefore $G(\mathsf h)=\omega$.
Since $\nu$ is continuous at $\mathsf h$ and $\nu(\mathsf h)>0$,
we can choose constants $0<\delta_0<\mathsf h/2$ and $c_0>0$ such
that $G$ has density $\nu$ on
$[\mathsf h-\delta_0,\mathsf h+\delta_0]$ and
$\nu(t)\ge c_0$ throughout this interval. Integration then gives
\begin{align*}
G(\mathsf h+s)-\omega
&=\int_{\mathsf h}^{\mathsf h+s}\nu(t)\,\dd t\ge c_0s,\\
\omega-G(\mathsf h-s)
&=\int_{\mathsf h-s}^{\mathsf h}\nu(t)\,\dd t\ge c_0s.
\end{align*}
Thus
\begin{align}
    G(\mathsf h+s)\ge \omega+c_0s,
    \qquad
    G(\mathsf h-s)\le \omega-c_0s,
    \qquad 0<s\le\delta_0.
    \label{eq:bw.local}
\end{align}

    We next establish concentration for the clean empirical distribution function. Fix $t\ge 0$ and define
    $$
    f_t(\sb_1^0,\dots,\sb_n^0)
    \coloneq
    G_n^0(t)
    =
    \frac{2}{n(n-1)}\sum_{1\le i<j\le n}
    1_{\{\|\sb_i^0-\sb_j^0\|_2^2\le t\}}.
    $$
    If we replace $\sb_i^0$ by another value ${\sb_i^0}'$, then at most $(n-1)$ summands in the above sum can change, and each changes by at most $1$. Hence
    $$
    \sup_{\sb_1^0,\dots,\sb_n^0,\,{\sb_i^0}'}
    \left|
    f_t(\sb_1^0,\dots,\sb_i^0,\dots,\sb_n^0)
    -
    f_t(\sb_1^0,\dots,{\sb_i^0}',\dots,\sb_n^0)
    \right|
    \le
    \frac{2}{n}.
    $$
    By Lemma~\ref{lemma:McDiarmid}, for any $u>0$,
    \begin{align}
        \P\bigl(|G_n^0(t)-G(t)|>u\bigr)
        \le
        2\exp\!\left(-\frac{2u^2}{n(2/n)^2}\right)
        =
        2\exp\!\left(-\frac{n u^2}{2}\right).
        \label{eq:bw.4}
    \end{align}
    Let $\varepsilon\in(0,1/2)$ and $D>0$ be arbitrary, and define
    $$
    r_n\coloneq \frac{n^\varepsilon}{\sqrt n},
    $$
    Since $r_n=\oo(1)$, for sufficiently large $n$ we have
    $$
    r_n\le \min\{\delta_0,\mathsf h/2\}.
    $$
    Define
    $$
    t_n^+\coloneq \mathsf h+r_n,
    \qquad
    t_n^-\coloneq \mathsf h-r_n.
    $$
    Since $t_n^+$ and $t_n^-$ are deterministic for each fixed $n$, \eqref{eq:bw.4} and \eqref{eq:bw.local} give
    \begin{align}
        \P\bigl(G_n^0(t_n^+)<\omega\bigr)
        &\le
        \P\bigl(G_n^0(t_n^+)-G(t_n^+)<-c_0r_n\bigr)
        \le
        2\exp\!\left(-\frac{n c_0^2r_n^2}{2}\right),
        \label{eq:bw.5}
        \\
        \P\bigl(G_n^0(t_n^-)\ge \omega\bigr)
        &\le
        \P\bigl(G_n^0(t_n^-)-G(t_n^-)\ge c_0r_n\bigr)
        \le
        2\exp\!\left(-\frac{n c_0^2r_n^2}{2}\right).
        \label{eq:bw.6}
    \end{align}
    Since $nr_n^2=n^{2\varepsilon}$, for sufficiently large $n$, we have that
    $$
    2\exp\!\left(-\frac{n c_0^2r_n^2}{2}\right)\le n^{-D}
    $$
    Hence, by a union bound applied to \eqref{eq:bw.5} and \eqref{eq:bw.6},
    $$
    \P\bigl(G_n^0(t_n^+)\ge \omega,\ G_n^0(t_n^-)<\omega\bigr)
    \ge 1-2n^{-D}.
    $$
    On this event, by the definition of $h$,
    $$
    \mathsf h-r_n\le h\le \mathsf h+r_n,
    $$
    and therefore
    $$
    |h-\mathsf h|\le r_n.
    $$
    Since $\varepsilon>0$ and $D>0$ were arbitrary, this proves
    \begin{align}
        |h-\mathsf h|
        =
        \OO_\prec\!\left(\frac{1}{\sqrt n}\right).
        \label{eq:bw.7}
    \end{align}

    Combining \eqref{eq:bw.2} and \eqref{eq:bw.7}, we obtain
    \begin{align}
        |h_n-\mathsf h|
        \le
        |h_n-h|+|h-\mathsf h|
        =
        \OO_\prec\!\left(\psi_n+\frac{1}{\sqrt n}\right).
        \label{eq:bw.8}
    \end{align}
    Since $\mathsf h$ is a positive constant, \eqref{eq:bw.8} yields
    $$
    \left|\frac{h_n}{\mathsf h}-1\right|
    =
    \OO_\prec\!\left(\psi_n+\frac1{\sqrt n}\right),
    $$
    which completes the proof.
\end{proof}

We now use Proposition~\ref{prop:bandwidth} to prove the spectral convergence result for a general kernel function $f$.

\begin{assumption}[Kernel regularity]
    \label{ass:f-reg}
    The kernel function $f:[0,\infty)\to\R$ is bounded, that is,
    $$
    \sup_{x\ge 0}|f(x)|\le B_f
    $$
    for some constant $B_f>0$. Moreover, there exist two measurable functions
    $
    \zeta:[0,\infty)\to\R,
    $
    and
    $
    \tau:[0,\infty)\times[0,\infty)\to(0,1],
    $
    such that for all $s,t\ge 0$,
    \begin{align}
        |f(t)-f(s)|\le L_f\bigl(t^{\zeta(t)}+s^{\zeta(s)}\bigr)|t-s|^{\tau(s,t)},
        \label{eq:f-reg-general}
    \end{align}
    where $L_f>0$ is a constant, $\zeta(x)\ge 0$ for $0\le x\le 1$, and $\zeta(x)\le 0$ for $x>1$.
    Finally, for every $h>0$ and every dimension $m\in\N$, the kernel
    $$
    \Phi_h(u,v)\coloneq f\!\left(\frac{\|u-v\|_2}{h^{1/2}}\right), \qquad u,v\in\R^m,
    $$
    is positive semidefinite. Assume also that $\tau_0=\inf_{s,t\ge0}\tau(s,t)>0$, and interpret $0^0=1$ when needed in \eqref{eq:f-reg-general}. All regularity constants are uniform in $n$.
\end{assumption}
\begin{remark}
    For $f(x)=\exp(-x^2)$ used in \eqref{eq:Kn}, one may take $\tau(s,t)\equiv1$, $\zeta(x)=1$ for $0\le x\le1$, $\zeta(x)=0$ for $x>1$, and $L_f=2$.
\end{remark}

Under Assumption~\ref{ass:f-reg}, we define the corresponding kernel matrix, empirical eigenfunctions, and population integral operator. Recall \eqref{eq:Kn}, \eqref{eq:emp_op_1}, and \eqref{eq:pop_op_1}. Using the same notation, define $\Kb_n=(K(i,j))$ by
\begin{align}
    K(i,j)=f\!\left(\frac{\|\yb_i-\yb_j\|_2}{h_n^{1/2}}
\right),
\quad 1\le i,j\le n,\notag
\end{align}
with eigenvalues $\lambda_1\ge\cdots\ge\lambda_n$ and eigenvectors $\ub_1,\cdots,\ub_n$ of the normalized matrix $n^{-1}\Kb_n$. Denote the empirical eigenfunctions for $\lambda_i>0$ by
\begin{align}
    \widehat{\phi}_i^{(n)}(\mathbf{x})
    \coloneq
    \frac{1}{\lambda_i\sqrt n}\sum_{j=1}^n
    f\!\left(\frac{\|\Ps^\top(\mathbf{x}-\xb_j)\|_2}{h_n^{1/2}}\right)u_{ij},
    \qquad \mathbf{x}\in\R^p,\notag
\end{align}
and the population integral operator $\widetilde{\mathcal K}:L^2(\R^p,\mu_s)\to L^2(\R^p,\mu_s)$ by
\begin{align}
    (\widetilde{\mathcal K}g)(\mathbf{x})\coloneq
    \int f\!\left(\frac{\|\mathbf{x}-\mathbf{y}\|_2}{\mathsf{h}^{1/2}}\right)g(\mathbf{y})\, \mu_s(\dd\mathbf{y}),
    \qquad \mathbf{x}\in\mathcal S,\notag
\end{align}
with eigenvalue-eigenfunction pairs $\{(\gamma_i,\widetilde{\phi}_i)\}_{i\ge 1}$. Zero-bandwidth kernels and zero-eigenvalue weighted coordinates are defined to be zero.

Define $\zeta_0\coloneq \inf_{0\le x\le 1}\zeta(x)$, $\tau_0\coloneq \inf_{s,t\ge 0}\tau(s,t)$, and
$$
\xi\coloneq
\begin{cases}
\tau_0, & \text{if } \zeta_0\ge \tau_0,\\
(\tau_0+\zeta_0)/2, & \text{if } 0\le \zeta_0<\tau_0.
\end{cases}
$$
Recall $G$ in \eqref{eq:cdf} and $\psi_n$ in \eqref{eq:psi-proj}. We have the following spectral convergence results.

\begin{theorem}[Spectral convergence with general $f$]
    \label{thm:total_convergence_f}
    Suppose Assumptions~\ref{ass:model}, \ref{ass:signal}, \ref{ass:lq} and \ref{ass:f-reg} hold. For $\omega$ satisfying the density condition in Theorem~\ref{thm:total_convergence}, we have
    $$
    \max_{i\in[n]}|\lambda_i-\gamma_i|
    =
    \OO_\prec\!\left(\psi_n^\xi+n^{-\tau_0/2}\right).
    $$
    Moreover, for any fixed $\delta>0$ and any $1\le i\le M(\delta)$ with $g_i>0$ as in \eqref{eq:eigengap},
    there exists a sign $s_i\in\{-1,1\}$ such that, for any $\mathbf{x}\in\mathcal S$, we have
    $$
    \left|
    \sqrt{\lambda_i}\,\widehat{\phi}_i^{(n)}(\mathbf{x})
    -
    s_i\sqrt{\gamma_i}\,\widetilde{\phi}_i(\mathbf{x})
    \right|
    =
    \OO_\prec\!\left(
    \psi_n^\xi+n^{-\tau_0/2}
    +
    \frac1{g_i}
    \left(
    \psi_n^{\xi/2}
    +
    n^{-\tau_0/2}
    \right)\right).
    $$
\end{theorem}
\begin{remark}
    For the Gaussian kernel, $\xi=\tau_0=1$, so the stated rates reduce to Theorem~\ref{thm:total_convergence}.
\end{remark}
\begin{proof}
    Recall $\Phi_h$ from Assumption~\ref{ass:f-reg}. We define the normalized kernel matrices
    $$
        \Kb_x=\frac1n\Kb_n,\qquad
        \Kb_y(i,j)=\frac1n\Phi_h(\sb_i,\sb_j),\qquad
        \Kb_0(i,j)=\frac1n\Phi_{\mathsf h}(\sb_i,\sb_j), \qquad 1\le i,j\le n.
    $$
    Denote the diameter of $\mathcal S$ by $D$. By \eqref{eq:bw.7}, for sufficiently large $n$, we have $|h-\mathsf h|\le\mathsf h/4$ with high probability.

    For a clean squared distance $a\le D^2$ and its projected counterpart $b$, we have $|b/h_n-a/h|\le C(|b-a|+|h_n-h|)$ if $h_n\ge\mathsf h/2$. In the entrywise comparison leading to Eqs.~(109)--(111) of \cite{ding2023learning}, the kernel regularity bound gives $|f(t)-f(s)|\le C|t^2-s^2|^\xi$. Applying this step with $t=\sqrt{b/h_n}$ and $s=\sqrt{a/h}$, we obtain
    \begin{equation}\label{eq:c13-projected-entry}
        \left|f\!\left(\sqrt{b/h_n}\right)-f\!\left(\sqrt{a/h}\right)\right|
        \le C\bigl(|b-a|+|h_n-h|\bigr)^\xi.
    \end{equation}
    For $h_n<\mathsf h/2$ this follows from boundedness and $|h_n-h|\ge\mathsf h/4$. To compare the two clean kernels, take $s=\sqrt{a/h}$ and $t=\sqrt{a/\mathsf h}$. Then $|s-t|\le C|h-\mathsf h|\le1$ with high probability by \eqref{eq:bw.7}. Since $s^{\zeta(s)}$ and $t^{\zeta(t)}$ in \eqref{eq:f-reg-general} are each at most one and $\tau(s,t)\ge\tau_0$, we have
    \begin{equation}\label{eq:c13-bandwidth-entry}
        \sup_{0\le a\le D^2}\left|f\!\left(\sqrt{a/h}\right)-f\!\left(\sqrt{a/\mathsf h}\right)\right|
        \le C|h-\mathsf h|^{\tau_0}=\OO_\prec(n^{-\tau_0/2}).
    \end{equation}

    To evaluate the eigenfunctions, we fix $\mathbf x\in\mathcal S$ and define three vectors in $\R^n$ by their entries
    $$
        (k_x)_j=\Phi_{h_n}(\Ps^\top\mathbf x,\Ps^\top\xb_j),\qquad
        (k_y)_j=\Phi_h(\mathbf x,\sb_j),\qquad
        (k_0)_j=\Phi_{\mathsf h}(\mathbf x,\sb_j), \qquad j=1,\cdots,n.
    $$
    Thus $\sqrt{\lambda_i}\widehat\phi_i^{(n)}(\mathbf x)=k_x^\top\ub_i/\sqrt{n\lambda_i}$ for $\lambda_i>0$. Write $M_n=\max_{j,\ell}|\ws_\ell^\top\zb_j|$. The sub-Gaussian bound \eqref{eq:A.3.1} gives $\P(|\ws_\ell^\top\zb_j|>\sigma t)\le2\exp(-ct^2)$ for $t>0$. A union bound gives $\P(M_n>\sigma t)\le2np\exp(-ct^2)$, hence $M_n=\OO_\prec(\sigma)$ since $p\asymp n^\alpha$.

    Recall from \eqref{eq:rotate} that $\mathcal U=\operatorname{span}\{\rs_1,\ldots,\rs_r\}$ is the signal subspace, so $\mathbf d_j=\mathbf x-\sb_j\in\mathcal U$. By compactness in Assumption~\ref{ass:signal} and Lemma~\ref{lem:row-weak-lq}, we have $\|\mathbf d_j\|_2\le D$ and $|\ws_\ell^\top\mathbf d_j|\le D\rho_\ell$. Expanding the squared distance, we have
    $$
        Q_j\coloneq\|\Ps^\top(\mathbf x-\xb_j)\|_2^2-\|\mathbf d_j\|_2^2
        =-\|(\Ib-\Pi_\Ps)\mathbf d_j\|_2^2+\|\Ps^\top\zb_j\|_2^2-2\mathbf d_j^\top\Pi_\Ps\zb_j.
    $$
    Bounding the last two terms over the $\Ks$ selected directions, we obtain
    \begin{equation}\label{eq:c13-query-distance}
        \max_j|Q_j|\le D^2\|(\Ib-\Pi_\Ps)\Pi_{\mathcal U}\|_{\op}^2
        +\Ks M_n^2+2DM_n\sum_{m\le\Ks}\rho_{(m)}=\OO_\prec(\psi_n),
    \end{equation}
    where in the last step we used Theorem~\ref{thm:proj-gap-final-poly} for the bias and \eqref{eq:interaction-row-sum} for the interaction. For the noise terms in \eqref{eq:pairwise-noise3}, recall the cutoff $c_*$ in \eqref{eq:K-small-noise}. We use \eqref{eq:K-small-noise} when $\sigma^2/\Theta\le c_*$ and the bound $\Ks\lesssim(\Theta/\sigma^2)^{q/2}$ from the proof of Theorem~\ref{thm:pairwise} when $\sigma^2/\Theta>c_*$.

    By \eqref{eq:c13-projected-entry},
    \begin{equation}\label{eq:c13-query-kernel}
        n^{-1/2}\|k_x-k_y\|_2\le\max_j|(k_x)_j-(k_y)_j|
        \le C\bigl(\max_j|Q_j|+|h_n-h|\bigr)^\xi=\OO_\prec(\psi_n^\xi),
    \end{equation}
    where in the last step we used \eqref{eq:c13-query-distance} and the bandwidth bound $|h_n-h|\le\Delta_n=\OO_\prec(\psi_n)$ in \eqref{eq:bw.2}. Moreover, we also have $|b-a|\le\Delta_n$. By \eqref{eq:c13-projected-entry}--\eqref{eq:c13-bandwidth-entry} and $\|A/n\|_{\op}\le\max_{j,k}|A_{jk}|$, we obtain
    \begin{equation}\label{eq:c13-kernel-errors}
    \begin{aligned}
        e_{\rm p}\coloneq\|\Kb_x-\Kb_y\|_{\op}&=\OO_\prec(\psi_n^\xi),\\
        e_{\rm h}\coloneq\|\Kb_y-\Kb_0\|_{\op}&=\OO_\prec(n^{-\tau_0/2}),\\
        n^{-1/2}\|k_y-k_0\|_2&=\OO_\prec(n^{-\tau_0/2}).
    \end{aligned}
    \end{equation}

    In the reproducing kernel Hilbert space $\mathcal H$ of $\Phi_{\mathsf h}$, we write $k_s=\Phi_{\mathsf h}(\cdot,s)$. For $g\in\mathcal H$, define the empirical and population operators by
    $$
        \mathcal C_n g=\frac1n\sum_{j=1}^n g(\sb_j)k_{\sb_j},\qquad
        \mathcal C g=\int_{\mathcal S}g(s)k_s\,\mu_s(\dd s).
    $$
    Their nonzero eigenvalues are those of $\Kb_0$ and $\widetilde{\mathcal K}$, respectively. Since the clean samples are i.i.d.\ and the kernel is bounded, \cite[Lemma~5]{ding2023learning} gives $\eta_n\coloneq\|\mathcal C_n-\mathcal C\|_{\op}=\OO_\prec(n^{-1/2})$. By Weyl's inequality and \eqref{eq:c13-kernel-errors}, using $n^{-1/2}\le n^{-\tau_0/2}$ for $0<\tau_0\le1$, we have
    \begin{equation}\label{eq:c13-eigenvalues}
        \max_{i\le n}|\lambda_i-\gamma_i|\le e_{\rm p}+e_{\rm h}+\eta_n
        =\OO_\prec(\psi_n^\xi+n^{-\tau_0/2}).
    \end{equation}

    Write $\varepsilon=e_{\rm p}+e_{\rm h}+\eta_n$. Fix $i\le M(\delta)$ with $g_i>0$ and suppose first that $\varepsilon<\min(g_i,\delta)/16$. Let $(\beta_j,\vb_j)$ and $(\beta_j^0,\vb_j^0)$ be the $j$th eigenpairs of $\Kb_y$ and $\Kb_0$, respectively. The Weyl bounds used in \eqref{eq:c13-eigenvalues} give
    $$
        \max\{|\lambda_j-\gamma_j|,\,|\beta_j-\gamma_j|,\,|\beta_j^0-\gamma_j|\}\le\varepsilon,\qquad j\in[n].
    $$
    Since $\gamma_i\ge\delta$ by \eqref{eq:M}, all three $i$th eigenvalues exceed $\delta/2$. Recall from \eqref{eq:eigengap} that $g_i$ is the distance from $\gamma_i$ to its nearest neighbor in the population spectrum. Each end of an adjacent gap moves by at most $\varepsilon$, so the gap at the $i$th eigenvalue of each matrix is at least $g_i-2\varepsilon>7g_i/8$. We now define the clean weighted coordinates at $\mathbf x$ by
    $$
        F_{i,y}=\frac{k_y^\top\vb_i}{\sqrt{n\beta_i}},\qquad
        F_{i,0}=\frac{k_0^\top\vb_i^0}{\sqrt{n\beta_i^0}}.
    $$
    To compare with $F_{i,y}$, we first change the eigenvalue factor and then the kernel vector. Since $n^{-1/2}\|k_x\|_2\le B_f$ and $\|\ub_i\|_2=1$, we have
    $$
    \begin{aligned}
        \left|\frac1{\sqrt{\lambda_i}}-\frac1{\sqrt{\beta_i}}\right|
        \frac{|k_x^\top\ub_i|}{\sqrt n}
        &\le C_{\delta,B_f}|\lambda_i-\beta_i|\le C_{\delta,B_f} e_{\rm p}
        =\OO_\prec(\psi_n^\xi),\\
        \frac{|(k_x-k_y)^\top\ub_i|}{\sqrt{n\beta_i}}
        &\le C_\delta\,n^{-1/2}\|k_x-k_y\|_2=\OO_\prec(\psi_n^\xi),
    \end{aligned}
    $$
    where we used \eqref{eq:c13-kernel-errors} and \eqref{eq:c13-query-kernel}, respectively. For the remaining term, by Cauchy--Schwarz we have that, for either sign $s_1$,
    $$
        \frac{\bigl|k_y^\top(\ub_i-s_1\vb_i)\bigr|}{\sqrt{n\beta_i}}
        \le \frac{\|k_y\|_2}{\sqrt{n\beta_i}}\,\|\ub_i-s_1\vb_i\|_2
        \le C_{\delta,B_f}\|\ub_i-s_1\vb_i\|_2.
    $$
    Moreover, by the squared alignment estimate in \cite[Eqs.~(113)--(116)]{ding2023learning}, we can choose $s_1$ such that
    $$
        \|\ub_i-s_1\vb_i\|_2^2=\OO_\prec\!\left(\frac{\psi_n^\xi}{g_i^2}\right).
    $$
    Combining the three terms as in \cite[Eqs.~(117)--(119)]{ding2023learning}, we obtain
    \begin{equation}\label{eq:c13-squared-coordinate}
        \bigl|\sqrt{\lambda_i}\widehat\phi_i^{(n)}(\mathbf x)-s_1F_{i,y}\bigr|^2
        =\OO_\prec\!\left(\psi_n^{2\xi}+\psi_n^\xi/g_i^2\right).
    \end{equation}

    \noindent Next, by Weyl's inequality applied to the clean kernel matrices $\Kb_y$ and $\Kb_0$, we have $|\beta_i-\beta_i^0|\le\|\Kb_y-\Kb_0\|_{\op}=e_{\rm h}$. Since the gap at $\beta_i^0$ exceeds $7g_i/8$, the eigenvector perturbation bound in \cite[Lemma~14]{ding2023learning} gives $\|\vb_i-s_2\vb_i^0\|_2\le Ce_{\rm h}/g_i$ for a sign $s_2$. Substituting these bounds and \eqref{eq:c13-kernel-errors} into the same decomposition, we have
    \begin{equation}\label{eq:c13-bandwidth-coordinate}
        |F_{i,y}-s_2F_{i,0}|
        \le C_{\delta,B_f}\bigl(e_{\rm h}+n^{-1/2}\|k_y-k_0\|_2+e_{\rm h}/g_i\bigr)
        =\OO_\prec\!\left(n^{-\tau_0/2}(1+g_i^{-1})\right).
    \end{equation}

    \noindent Finally, $F_{i,0}$ and $\sqrt{\gamma_i}\widetilde\phi_i(\mathbf x)$ are values of unit eigenfunctions of $\mathcal C_n$ and $\mathcal C$ at $\mathbf x$. The same perturbation bound gives an $\mathcal H$-norm error of at most $C\eta_n/g_i$ after alignment by a sign $s_3$. Since $|v(\mathbf x)|\le\sqrt{f(0)}\|v\|_{\mathcal H}$, we get
    \begin{equation}\label{eq:c13-population-coordinate}
        |F_{i,0}-s_3\sqrt{\gamma_i}\widetilde\phi_i(\mathbf x)|=\OO_\prec(n^{-1/2}/g_i).
    \end{equation}

    It remains to consider $\varepsilon\ge\min(g_i,\delta)/16$. This case can occur at high noise. As shown in the proof of Theorem~\ref{thm:pairwise}, for sufficiently large $\sigma^2/\Theta$ we have $\Ks=0$ with high probability. Then $h_n=0$ and $\lambda_i=0$ by the zero-bandwidth convention, so \eqref{eq:c13-eigenvalues} gives $\varepsilon\ge\gamma_i\ge\delta$. To bound the coordinates throughout this case, let $G_{\mathbf x}$ be the Gram matrix of $\Phi_{h_n}$ on $\Ps^\top\xb_1,\ldots,\Ps^\top\xb_n,\Ps^\top\mathbf x$, in that order. Let $a_i=(\ub_i^\top,0)^\top$ and $e_{n+1}=(0,\ldots,0,1)^\top$ in $\R^{n+1}$. The first $n$ entries of $G_{\mathbf x}e_{n+1}$ form $k_x$. For $1\le j,k\le n$, recall from \eqref{eq:projected_y} that $\yb_j-\yb_k=\Ps^\top(\xb_j-\xb_k)$, so
    $$
        G_{\mathbf x}(j,k)=\Phi_{h_n}(\Ps^\top\xb_j,\Ps^\top\xb_k)
        =\Phi_{h_n}(\yb_j,\yb_k)=\Kb_n(j,k).
    $$
    Since the last entry of $a_i$ is zero, $a_i^\top G_{\mathbf x}a_i=\ub_i^\top\Kb_n\ub_i$. Applying Cauchy--Schwarz for the positive semidefinite matrix $G_{\mathbf x}$ to $a_i$ and $e_{n+1}$, we obtain
    $$
        |k_x^\top\ub_i|^2\le(\ub_i^\top\Kb_n\ub_i)(e_{n+1}^\top G_{\mathbf x}e_{n+1})
        \le(\ub_i^\top\Kb_n\ub_i)f(0)=n\lambda_i f(0),
    $$
    where in the last inequality we used $e_{n+1}^\top G_{\mathbf x}e_{n+1}\le f(0)$, and in the last equality we used $n^{-1}\Kb_n\ub_i=\lambda_i\ub_i$ and $\|\ub_i\|_2=1$. Thus the empirical weighted coordinate is bounded by $\sqrt{B_f}$, as is the population coordinate by the reproducing property. Since $e_{\rm p}\le2B_f$ and $g_i\le\gamma_i\le B_f$, the condition on $e_{\rm p}+e_{\rm h}+\eta_n$ in this case gives, for either sign $s_i$,
    $$
        \left|\sqrt{\lambda_i}\,\widehat\phi_i^{(n)}(\mathbf x)-s_i\sqrt{\gamma_i}\,\widetilde\phi_i(\mathbf x)\right|
        \le 2\sqrt{B_f}
        \le C_{\delta,B_f}\,\frac{\sqrt{e_{\rm p}}+e_{\rm h}+\eta_n}{g_i}.
    $$
    In the first case we choose $s_i=s_1s_2s_3$ and add the square root of \eqref{eq:c13-squared-coordinate} to the bounds in \eqref{eq:c13-bandwidth-coordinate} and \eqref{eq:c13-population-coordinate}. In both cases we obtain
    \begin{equation}\label{eq:c13-eigenfunctions}
        \left|\sqrt{\lambda_i}\,\widehat\phi_i^{(n)}(\mathbf x)-s_i\sqrt{\gamma_i}\,\widetilde\phi_i(\mathbf x)\right|
        =\OO_\prec\!\left(\psi_n^\xi+n^{-\tau_0/2}
        +\frac{\psi_n^{\xi/2}+n^{-\tau_0/2}}{g_i}\right).
    \end{equation}
    For the Gaussian kernel, $\xi=\tau_0=1$, so \eqref{eq:c13-eigenvalues} and \eqref{eq:c13-eigenfunctions} give the eigenvalue and eigenfunction rates in Theorem~\ref{thm:total_convergence}, respectively. This completes the proof.
\end{proof}

\section{Supplementary Algorithms}
\label{sec:supp_algo}

\subsection[Selection of omega]{Selection of $\omega$}
We now specify how to choose $\omega$ in \eqref{eq:hn-alg}. We first generate $m$ candidate quantile levels from the trimmed interval $(\varepsilon,1-\varepsilon)$, and for each candidate we identify the corresponding location in the ordered pairwise squared distances. We then compare the local spacing of neighboring ordered distances within a window of size $s$ around each candidate and select the one with the smallest such spacing. Throughout the paper, we use the default values $\varepsilon=0.05$, $s=5$, and $m=100$. This procedure is summarized in Algorithm~\ref{alg:omega}.

\begin{algorithm}[!htb]
    \caption{\bf Selection of the quantile level $\omega$}
    \label{alg:omega}
    \normalsize
    \setlength{\abovedisplayskip}{5pt}
    \setlength{\belowdisplayskip}{5pt}
    \setlength{\abovedisplayshortskip}{3pt}
    \setlength{\belowdisplayshortskip}{3pt}
    \begin{flushleft}
        \vspace{0pt}
        \noindent{\bf Inputs:} Ordered pairwise squared distances
        $\{d_{(j)}\}_{j=1}^{M}$ with $M=\binom{n}{2}$, a trimming constant
        $\varepsilon\in(0,1/2)$, a small fixed integer $s$ with $1\le s\le M\varepsilon$, and a positive
        integer $m$.

        \noindent{\bf Step one:}
        Generate $m$ candidate values
        $$
        \omega^{(1)},\dots,\omega^{(m)}
        \stackrel{\mathrm{iid}}{\sim}\mathrm{Unif}(\varepsilon,1-\varepsilon).
        $$

        \noindent{\bf Step two:}
        For each $\ell=1,\dots,m$, compute
        $$
        j_\ell=\lceil M\omega^{(\ell)}\rceil.
        $$
        Define
        $$
        \mathcal I_s:=\{\ell\in[m]: s+1\le j_\ell\le M-s\}.
        $$

        \noindent{\bf Step three:}
        Let
        \begin{align}
            \ell^\star
            \in
            \arg\min_{\ell\in\mathcal I_s}
            \left\{d_{(j_\ell+s)}-d_{(j_\ell-s)}\right\},
            \notag
        \end{align}
        where ties are broken by choosing the smallest $\ell$. Set
        \begin{align}
            \omega=\omega^{(\ell^\star)}.
            \notag
        \end{align}

        \noindent{\bf Output:} The selected quantile level $\omega$.
    \end{flushleft}
\end{algorithm}

\subsection{Selection of the Number of Embeddings Used}
We specify how to choose the number of embeddings used in \eqref{eq:qb}. We select $\mathsf d$ by an eigenvalue-ratio rule in the spirit of \cite{ding2022impact}: among the empirical eigenvalues of $n^{-1}\Kb_n$, we identify the last index at which the ratio of two successive eigenvalues exceeds a prescribed fixed threshold. We use the default threshold $\alpha=1.05$. The resulting procedure is summarized in Algorithm~\ref{alg:d}.

\begin{algorithm}[!htb]
    \caption{\bf Selection of number of embeddings}
    \label{alg:d}
    \normalsize
    \setlength{\abovedisplayskip}{5pt}
    \setlength{\belowdisplayskip}{5pt}
    \setlength{\abovedisplayshortskip}{3pt}
    \setlength{\belowdisplayshortskip}{3pt}
    \begin{flushleft}
        \vspace{0pt}
        \noindent{\bf Inputs:} Sorted eigenvalues
        $\{\hat\lambda_i\}_{i=1}^n$ of $n^{-1}\Kb_n$ with
        $\hat\lambda_1\ge\cdots\ge\hat\lambda_n$, and threshold $\alpha>1$.

        \noindent{\bf Action:}
        Set
        $$
        {\mathsf d}=\max\left(\left\{1\le i\le n-1:\hat\lambda_i>\alpha\hat\lambda_{i+1}\right\}\cup\{1\}\right).
        $$

        \noindent{\bf Output:} ${\mathsf d}$, the number of embeddings used.
    \end{flushleft}
\end{algorithm}

\section{Concentration Inequalities}
In this section, we introduce the following standard concentration inequalities used in the proofs.

\begin{lemma}[Bernstein's Inequality for Sub-Exponential Variables]
    \label{lemma:bernstein}
    Let $X_1, \dots, X_N$ be independent mean-zero sub-exponential random variables. Then for every $t \ge 0$,
    $$
    \P\left( \left| \sum_{i=1}^N X_i \right| \ge t \right) \le 2 \exp \left( -c \min \left( \frac{t^2}{\sum_{i=1}^N \|X_i\|_{\psi_1}^2}, \frac{t}{\max_i \|X_i\|_{\psi_1}} \right) \right),
    $$
    where $c > 0$ is some universal constant and $\norm{\cdot}_{\psi_1}$ denotes the sub-exponential norm.
\end{lemma}
\begin{proof}
    See Theorem 2.8.1 in \cite{vershynin2018high}.
\end{proof}

\begin{lemma}[Bounded differences inequality]
    \label{lemma:McDiarmid}
    Let $X_1,\dots,X_n$ be independent random vectors with $X_i \in \mathcal{X}$ for $i=1,\dots,n$ and let $f:\mathcal{X}\times\cdots\times\mathcal{X}\to\mathbb{R}$ be measurable. Assume that there exist constants $c_1,\dots,c_n\ge 0$ such that, for every
    $i=1,\dots,n$,
    $$
    \sup_{\substack{x_j\in\mathcal{X}\ (1\le j\le n)\\ x_i'\in\mathcal{X}}}
    \left|
    f(x_1,\dots,x_n)
    -
    f(x_1,\dots,x_{i-1},x_i',x_{i+1},\dots,x_n)
    \right|
    \le c_i.
    $$
    Define
    \[
    Z=f(X_1,\dots,X_n),
    \qquad
    v=\frac14\sum_{i=1}^n c_i^2.
    \]
    Then for every $t>0$,
    $$
    \mathbb{P}\!\left(|Z-\mathbb{E}Z|>t\right)
    \le
    2\exp\!\left(-\frac{2t^2}{\sum_{i=1}^n c_i^2}\right).
    $$
\end{lemma}
\begin{proof}
    See Theorem 6.2 in \cite{BLM2013}.
\end{proof}

\end{document}

%% file: template.tex
\usepackage[utf8]{inputenc}

\usepackage{amsfonts,epsfig,epstopdf,titling,url,array}
\usepackage{amssymb,amsmath, amsthm, mathabx,mathtools}
\usepackage{dsfont}
\usepackage{bm}
\usepackage{mathrsfs}
\usepackage{stmaryrd}
\usepackage{amsmath} %
\usepackage{amssymb}
\usepackage{amsthm}

\usepackage{caption}
\usepackage{graphicx}
\usepackage{pgfplots}
\usepackage{subcaption}
\usepackage{tikz}
\usetikzlibrary{positioning}
\usetikzlibrary{shapes.geometric, arrows.meta, positioning, calc, decorations.pathmorphing}
\usepackage{wrapfig}

\usepackage{array}
\usepackage{longtable}

\usepackage{color, enumerate}
\usepackage{color}
\usepackage{enumitem}
\usepackage{enumerate}

\usepackage{algorithm}
\usepackage[noend]{algpseudocode}
\usepackage{algpseudocode}

\usepackage{comment, authblk}
\usepackage{pifont}
\usepackage{lipsum}
\usepackage{titlesec}

\usepackage[inline]{showlabels}

\usepackage{a4wide}
\usepackage[a4paper, total={6in, 8in}]{geometry}

\usepackage{hyperref}
\hypersetup{
	colorlinks,
	citecolor=blue,
	linkcolor=blue,
	urlcolor=blue}

\titleformat*{\subsection}{\large\bfseries}
\numberwithin{equation}{section}

\pgfplotsset{compat=newest}
\pgfplotsset{plot coordinates/math parser=false}
\newlength\figureheight
\newlength\figurewidth

\newcommand{\zb}{\mathbf{z}}

\DeclareMathOperator{\Cov}{Cov}

\numberwithin{equation}{section}

\newcommand{\xb}{\mathbf{x}}
\newcommand{\yb}{\mathbf{y}}

\newcommand{\Kb}{\mathbf{K}}

\newcommand{\Ib}{\mathbf{I}}

\newcommand{\beq}{\begin{equation}}
	\newcommand{\bEq}{\end{equation}}

\newcommand{\be}{\begin{equation}}
	\newcommand{\ee}{\end{equation}}

\newcommand{\dd}{\mathrm d}

\renewcommand{\epsilon}{\varepsilon}
\renewcommand{\leq}{\leqslant}
\renewcommand{\geq}{\geqslant}
\newcommand{\floor}[1] {\lfloor {#1} \rfloor}

\renewcommand{\le}{\leq}
\renewcommand{\ge}{\geq}

\renewcommand{\P}{\mathbb{P}}
\newcommand{\E}{\mathbb{E}}
\newcommand{\R}{\mathbb{R}}

\newcommand{\N}{\mathbb{N}}

\newcommand{\qb}[1]{\bigl[{#1}\bigr]}

\newcommand{\norm}[1]{\lVert #1 \rVert}

\DeclareMathOperator{\diag}{diag}

\DeclareMathOperator{\OO}{O}
\DeclareMathOperator{\oo}{o}

\DeclareMathOperator{\OOp}{\OO_{\mathbb{P}}}
\DeclareMathOperator{\oop}{\oo_{\mathbb{P}}}

\makeatletter
\theoremstyle{plain} %
\@ifundefined{theorem}{\newtheorem{theorem}{Theorem}[section]}{}
\@ifundefined{theorem*}{\newtheorem*{theorem*}{Theorem}}{}
\@ifundefined{lemma}{\newtheorem{lemma}[theorem]{Lemma}}{}
\@ifundefined{assumption}{\newtheorem{assumption}[theorem]{Assumption}}{}
\@ifundefined{lemma*}{\newtheorem*{lemma*}{Lemma}}{}
\@ifundefined{corollary}{}{}
\@ifundefined{corollary*}{\newtheorem*{corollary*}{Corollary}}{}
\@ifundefined{proposition}{\newtheorem{proposition}[theorem]{Proposition}}{}
\@ifundefined{proposition*}{\newtheorem*{proposition*}{Proposition}}{}
\@ifundefined{claim}{}{}

\@ifundefined{definition}{}{}
\@ifundefined{definition*}{\newtheorem*{definition*}{Definition}}{}
\theoremstyle{plain}
\@ifundefined{example}{}{}
\@ifundefined{example*}{\newtheorem*{example*}{Example}}{}
\@ifundefined{remark}{\newtheorem{remark}[theorem]{Remark}}{}
\@ifundefined{remarks}{}{}
\@ifundefined{remark*}{\newtheorem*{remark*}{Remark}}{}
\@ifundefined{remarks*}{\newtheorem*{remarks*}{Remarks}}{}
\@ifundefined{notation}{}{}
\makeatother

\newcommand{\vb}{\mathbf{v}}

\makeatletter
\def\env@dmatrix{\hskip -\arraycolsep
	\let\@ifnextchar\new@ifnextchar
	\extrarowheight=2ex
	\array{*\c@MaxMatrixCols{>{\displaystyle}c}}}

\allowdisplaybreaks